%% file: ANA-STDM-2021-18-PAPER.tex
\documentclass[PAPER, atlasdraft=false, UKenglish, cernpreprint, texmf, orcidlogo]{atlasdoc}
\usepackage{placeins}

\usepackage{atlaspackage}
\usepackage{atlasbiblatex}

\usepackage{atlasphysics}
\graphicspath{{logos/}{figures/}}

\usepackage{ANA-STDM-2021-18-PAPER-defs}
\usepackage{atlasjetetmiss}

\AtlasTitle{Measurement of inclusive dijet cross-sections in proton--proton collisions at $\sqrt{s} = 13~\text{TeV}\xspace$ with the ATLAS detector}

\AtlasAbstract{%
Inclusive dijet cross-sections have been measured in proton--proton collisions at a centre-of-mass energy of 13~TeV using data with an integrated luminosity of 140~fb$^{-1}$, recorded by the ATLAS detector at the Large Hadron Collider during 2015--2018. Jets are identified using the anti-$k_{t}$ algorithm with a radius parameter of $R = 0.4$. The inclusive dijet double-differential cross-sections are measured first as a function of the invariant dijet mass and the half absolute rapidity separation between the two leading jets, $(m_{\mathrm{jj}}$, $y^{\ast})$, and second as a function of the invariant dijet mass and the total longitudinal boost of the dijet system, $(m_{\mathrm{jj}}$, $y_{\mathrm{boost}})$. The measured dijet system covers the invariant mass range from 240~GeV to almost 10~TeV, with dijet separation $y^{\ast} < 3.0$ and dijet boost $y_{\mathrm{boost}} < 3.0$. The results are unfolded to the particle level and compared with state-of-the-art next-to-next-to-leading-order full colour perturbative QCD calculations, corrected for non-perturbative and electroweak effects.
}

\author{The ATLAS Collaboration}

\AtlasRefCode{STDM-2021-18}

\PreprintIdNumber{CERN-EP-2025-283}

\AtlasJournal{EPJC}
\AtlasCoverEgroupAnalysisTeam{atlas-stdm-2021-18-analysis-team@cern.ch}

\hypersetup{pdftitle={ATLAS document},pdfauthor={The ATLAS Collaboration}}

\begin{document}

\maketitle
\tableofcontents

%
%

%
%

\FloatBarrier
\newpage

\section{Introduction}
Dijet cross-section measurements provide a fundamental test of Quantum Chromodynamics (QCD), in particular its perturbative regime (pQCD), by probing the dynamics of hard parton--parton scattering. Using data collected with the ATLAS detector~\cite{PERF-2007-01} at the Large Hadron Collider (LHC)~\cite{Evans:2008zzb} at a centre-of-mass energy of $13$~\TeV, a measurement of the inclusive dijet double-differential cross-section is presented as a function of key kinematic observables. This analysis offers detailed insights into parton distribution functions (PDFs) and higher-order QCD effects,
providing robust constraints on theoretical predictions.

In this context, the dijet final state provides a particularly suitable environment, benefiting from the copious production of jets in hadron-induced processes over a broad phase space. Several measurements were already performed at the LHC: by CMS at centre-of-mass energies of $7$, $8$, and $13$ \TeV~\cite{CMS-QCD-10-025,CMS-SMP-16-011,CMS-SMP-21-008} and by ATLAS at $7$ and $13$ \TeV~\cite{STDM-2012-03,STDM-2016-03}. The most recent 13~\TeV ATLAS (CMS) measurements were based on partial LHC Run-2 data samples corresponding to $3.2$~\ifb  ($36.3$~\ifb).

In this article, measurements of the dijet cross-section in proton--proton ($pp$) collisions at $\sqrt{s}=13~\TeV$ using \antikt~\cite{Cacciari:2008gp} jets with a radius parameter $R=0.4$ are presented.
The double-differential cross-sections are measured as functions of the invariant dijet mass, \mjj, and one of two angular variables, \ystar or \yboost, defined as
\begin{equation*}
\mjj = \sqrt{(p_1+p_2)^2}, \quad \ystar = \frac{1}{2}|y_1 - y_2| , \quad \yboost = \frac{1}{2}|y_1 + y_2|,
\end{equation*}
using the jet four-momenta, $p_1, p_2$, and the rapidities, $y_1, y_2$, of the \pt-ordered leading and subleading jets.
The invariant mass \mjj provides a direct handle on the energy scale of the hard interaction, while the angular variables offer insight into the dynamics of the dijet system.

The variable \ystar is directly related to the scattering angle $\thetastar$ in the partonic centre-of-mass frame through $\tanh (\ystar) = \cos(\thetastar) $, and the commonly used dijet angular variable $\chi$ defined as $\chi = e^{2\ystar} = \frac{1 + \cos(\thetastar)}{1 - \cos(\thetastar)}$. The variable \yboost characterises the overall rapidity of the dijet system in the laboratory frame. Both angular variables are directly related to the longitudinal momentum fractions of the incoming partons $x_{1,2} = \frac{\mjj}{\sqrt{s}} e^{\pm \yboost}$ with $\mjj = 2p_{\text{T,avg}} \cosh(\ystar)$, where the formulas are derived for back-to-back jets at the Born level. The positive (negative) sign corresponds to the leading (subleading) parton, and $p_{\text{T,avg}}$ denotes their average transverse momentum. Small \yboost values correspond to $x_1 \approx x_2$, whereas large \yboost selects strongly asymmetric momentum fractions. All three variables, \mjj, \ystar, and \yboost provide sensitivity to the parton momentum fractions $x_{1,2}$ and, consequently, to the PDFs~\cite{PhysRevLett.123.102001}.

This measurement builds on previous dijet cross-section results by using a significantly larger data sample collected by ATLAS with an integrated luminosity of $140$~\ifb. The \yboost variable is used for the first time in the dijet cross-section measurement by ATLAS\@. The increased data sample enhances statistical precision, particularly in high-mass and forward regions where earlier measurements were limited by the sample size. Additionally, an improved jet energy calibration is implemented, incorporating updated systematic uncertainties~\cite{JETM-2018-05,JETM-2022-06}. These updates reduce the experimental uncertainties in the jet energy scale by a factor of two to three for $\mjj$ above a few hundred~\GeV, enabling more precise constraints on theoretical predictions.
The phase space of the measurement extends from $\mjj=240\, \GeV$ to $10\,\TeV$, covering a wide kinematic range where the two highest-\pt jets are required to be within $|y|<3.0$. This broad coverage allows stringent tests of pQCD across a wide range of the relevant hard-scattering scale, including regions where higher-order effects and the running of the strong coupling become increasingly significant.

To fully exploit these improvements, the results of this measurement are compared with state-of-the-art theoretical predictions, including next-to-next-to-leading-order (\NNLO) QCD calculations with full colour for jet production~\cite{Chen_2022,} and the most recent PDFs~\cite{Hou:2019efy,Bailey:2020ooq,NNPDF:2017mvq,NNPDF:2021njg,STDM-2020-32}. These comparisons highlight the sensitivity to \NNLO pQCD corrections, particularly in kinematic regions where higher-order corrections could become increasingly relevant. %

Recently, \alphas was extracted from dijet cross-section measurements~\cite{Ahmadova:2024emn}, using the invariant dijet mass as the characteristic energy scale. This approach extends renormalisation group evolution (RGE) studies up to scales of approximately $7$~\TeV, probing the running of \alphas at an unprecedented energy.
The present measurement, with its extended \mjj range and reduced experimental uncertainties, provides an opportunity to further test the validity of this scale choice and its implications for high-energy QCD dynamics.

After the introduction and a brief description of the ATLAS detector in \Cref{sec:ATLASdetector}, the definition of the cross-sections is given in \Cref{sec:crossSectionDefinitions}. Data and Monte Carlo (MC) samples, event selection, trigger strategy, and jet calibration are described in~\Cref{sec:DatasetSelection}.
The unfolding procedure is explained in~\Cref{sec:unfolding}.
In~\Cref{sec:experimentalUncertainties}, sources of experimental uncertainties are presented.
\Cref{sec:theoreticalPrediction} describes the details and settings of the calculation of the pQCD prediction, the non-perturbative corrections and electroweak corrections.
In \Cref{sec:Results}, the final results and theory are compared and interpreted using statistical $\chi^2$ tests.

\FloatBarrier


%
%
%

\newcommand{\AtlasCoordFootnote}{%
ATLAS uses a right-handed coordinate system with its origin at the nominal interaction point (IP)
in the centre of the detector and the \(z\)-axis along the beam pipe.
The \(x\)-axis points from the IP to the centre of the LHC ring,
and the \(y\)-axis points upwards.
Polar coordinates \((r,\phi)\) are used in the transverse plane,
\(\phi\) being the azimuthal angle around the \(z\)-axis.
The pseudorapidity is defined in terms of the polar angle \(\theta\) as \(\eta = -\ln \tan(\theta/2)\) and is equal to the rapidity
$ y = \frac{1}{2} \ln \left( \frac{E + p_z}{E - p_z} \right) $ in the relativistic limit.
Angular distance is measured in units of \(\Delta R \equiv \sqrt{(\Delta y)^{2} + (\Delta\phi)^{2}}\).}

\FloatBarrier
\section{ATLAS detector}
\label{sec:ATLASdetector}

The ATLAS detector~\cite{PERF-2007-01} at the LHC covers nearly the entire solid angle around the collision point.\footnote{\AtlasCoordFootnote}
It consists of an inner tracking detector surrounded by a thin superconducting solenoid, electromagnetic and hadronic calorimeters,
and a muon spectrometer incorporating three large superconducting air-core toroidal magnets.

The inner-detector system (ID) is immersed in a \qty{2}{\tesla} axial magnetic field
and provides charged-particle tracking in the range \(|\eta| < 2.5\).
The high-granularity silicon pixel detector covers the vertex region and typically provides four measurements per track,
the first hit generally being in the insertable B-layer (IBL) installed before Run~2~\cite{ATLAS-TDR-19,PIX-2018-001}.
It is followed by the SemiConductor Tracker (SCT), which usually provides eight measurements per track.
These silicon detectors are complemented by the transition radiation tracker (TRT),
which enables radially extended track reconstruction up to \(|\eta| = 2.0\).
The TRT also provides electron identification information
based on the fraction of hits (typically 30 in total) above a higher energy-deposit threshold corresponding to transition radiation.

The calorimeter system covers the pseudorapidity range \(|\eta| < 4.9\).
Within the region \(|\eta|< 3.2\), electromagnetic calorimetry is provided by barrel and
endcap high-granularity lead/liquid-argon (LAr) calorimeters,
with an additional thin LAr presampler covering \(|\eta| < 1.8\)
to correct for energy loss in material upstream of the calorimeters.
Hadronic calorimetry is provided by the steel/scintillator-tile calorimeter,
segmented into three barrel structures within \(|\eta| < 1.7\), and two copper/LAr hadronic endcap calorimeters.
The solid angle coverage is completed by forward copper/LAr and tungsten/LAr calorimeter modules
optimised for electromagnetic and hadronic energy measurements, respectively.

The muon spectrometer (MS) comprises separate trigger and
high-precision tracking chambers measuring the deflection of muons in a magnetic field generated by the superconducting air-core toroidal magnets.
The field integral of the toroids ranges between \num{2.0} and \qty{6.0}{\tesla\metre}
across most of the detector.
Three layers of precision chambers, each consisting of layers of monitored drift tubes, cover the region \(|\eta| < 2.7\),
complemented by cathode-strip chambers in the forward region, where the background is highest.
The muon trigger system covers the range \(|\eta| < 2.4\) with resistive-plate chambers in the barrel, and thin-gap chambers in the endcap regions.

The luminosity is measured mainly by the LUCID--2~\cite{LUCID2} detector that records Cherenkov light produced in the quartz windows of photomultipliers located close to the beampipe.

Events were selected by the first-level trigger system implemented in custom hardware,
followed by selections made by algorithms implemented in software in the high-level trigger~\cite{TRIG-2016-01}.
The first-level trigger accepted events from the \qty{40}{\MHz} bunch crossings at a rate close to \qty{100}{\kHz},
which the high-level trigger further reduced in order to record complete events to disk at about \qty{1.25}{\kHz}.

A software suite~\cite{SOFT-2022-02} is used in data simulation, in the reconstruction
and analysis of real and simulated data, in detector operations, and in the trigger and data acquisition
systems of the experiment.

\FloatBarrier


\FloatBarrier

\section{Dijet cross-section definition}
\label{sec:crossSectionDefinitions}
The measured cross-section is corrected to particle level using simulated samples with jets constructed directly from generated stable particles with lifetimes satisfying $c\tau > 10$~mm, excluding muons and neutrinos. The jets are clustered using the collinear and infrared-safe \antikt algorithm~\cite{Cacciari:2008gp} with a radius parameter of $R=0.4$, as implemented in the \Fastjet~\cite{Fastjet} package.

The event selection requires at least two jets with transverse momentum $\pT > 75$~\GeV and rapidity $|y| < 3.0$. Additionally, the scalar sum of the two leading jet \pt, $\HTtwo = \ptone + \pttwo$, is required to be above $200$~\GeV, ensuring the stability of the pQCD calculations~\cite{Frixione_1997,Alioli_2011}.

The inclusive dijet double-differential cross-section is calculated for (\mjj, \ystar) and (\mjj, \yboost) at the particle level as a ratio of dijet event yields $N^\text{part}$ to the integrated luminosity \Lint and the bin widths of $\Delta \mjj$, and $\Delta \ystar$ or $\Delta \yboost$, respectively,
\begin{equation*}
\frac{ \mathrm{d}^2 \sigma}{ \mathrm{d} \mjj \mathrm{d} \ystar}
=
\frac{1}{\Lint }
\frac{ N^\text{part}}{ \Delta \mjj \Delta \ystar},
\end{equation*}
\begin{equation*}
\frac{ \mathrm{d}^2 \sigma}{ \mathrm{d} \mjj \mathrm{d} \yboost}
=
\frac{1}{\Lint }
\frac{N^\text{part}}{ \Delta \mjj \Delta \yboost}.
\end{equation*}

The dijet event topologies in corresponding analysis bins of \ystar and \yboost are illustrated in \Cref{fig:crossSection:dijetTopology}. The \ystar and \yboost equidistant binning from $0.0$ to $3.0$ and the 0.5 bin width are chosen for two reasons: first, to ensure consistency with the previous ATLAS dijet cross-section measurement~\cite{STDM-2016-03} performed in \ystar bins; and second, to satisfy the kinematic constraints between \ystar and \yboost variables.
\begin{figure}[htbp]
\centering
\includegraphics[width=0.55\linewidth]{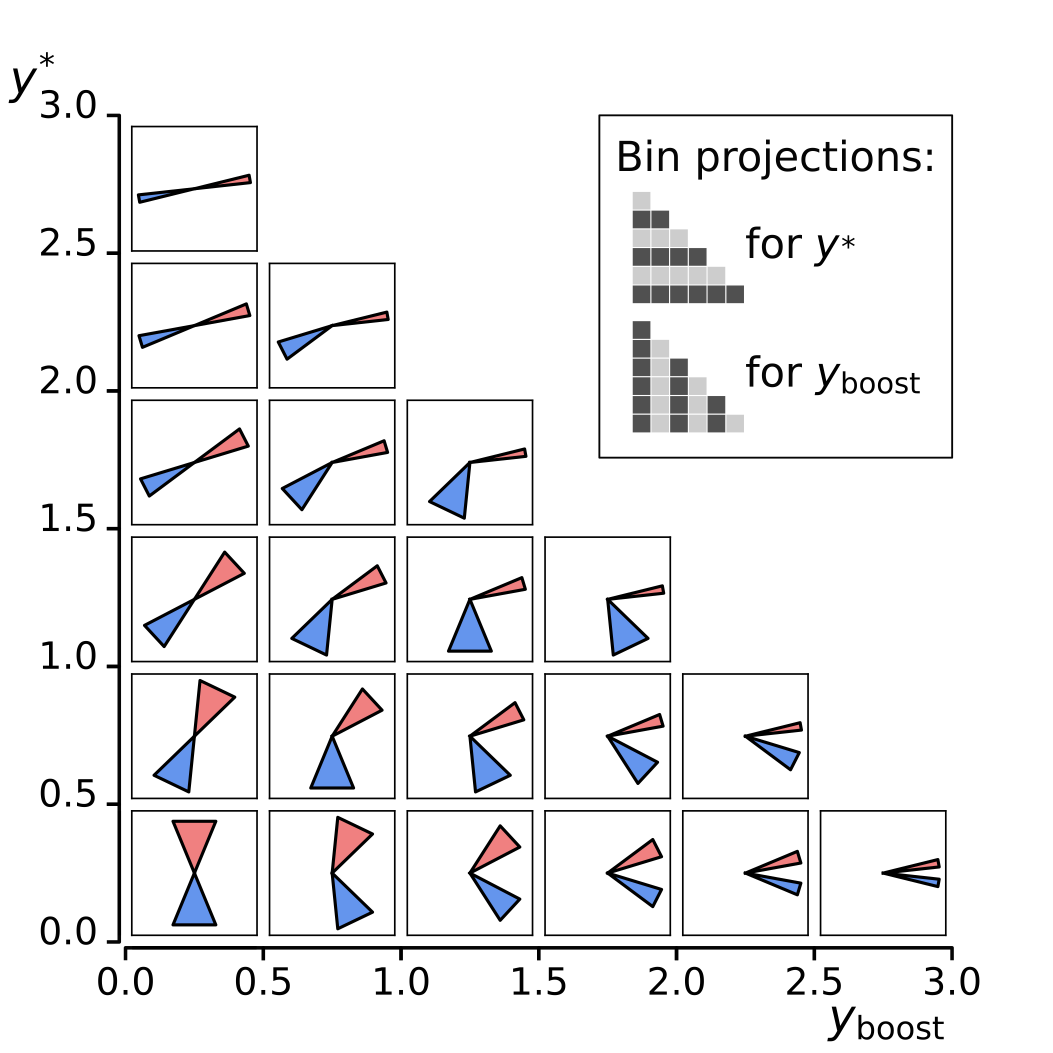}
\caption{An illustration of the dijet event topology in the laboratory frame as a function of the $\ystar{}=|y_1 - y_2|/2$ and $\yboost{}=|y_1 + y_2|/2$ observables, with $y_1$ and $y_2$ being the rapidities of the leading and subleading jets. Projections of individual rows (columns) to the \ystar (\yboost) axis correspond to the bin configurations of the measurement. In each subfigure, the horizontal axis corresponds to the beam direction, while the vertical axis represents the $x$--$y$ plane.}
\label{fig:crossSection:dijetTopology}
\end{figure}

The invariant dijet mass binning of \mjj is optimised to obtain a uniform purity\footnote{Purity represents the conditional probability of observing a particle-level dijet (T) for a given reconstructed-level dijet (R) entering the response matrix, expressed as $P(T|R) = P(T \cap R) / P(R)$.} of $50$\% and a corresponding stability\footnote{Stability represents the conditional probability of observing a reconstructed-level dijet (R) from a given particle-level dijet (T) entering the response matrix, expressed as $P(R|T) = P(T \cap R) / P(T)$.} above $50$\% derived from MC simulations. Additionally, the \mjj binning guarantees at least $10$ events in the reconstructed ATLAS data and maintains reasonable statistical uncertainty in ATLAS MC simulations.
\FloatBarrier


\FloatBarrier

\section{Data, simulation, and event selection}\label{sec:DatasetSelection}
This measurement uses the complete LHC Run-2 data sample collected between 2015 and 2018, satisfying the ATLAS data quality requirements, ensuring that all relevant detector subsystems were fully operational~\cite{DAPR-2018-01}. The integrated luminosity reached $140$~\ifb with a corresponding uncertainty of 0.83\%~\cite{DAPR-2021-01}, determined with the van der Meer beam separation scan method using the LUCID-2 sub-detector~\cite{LUCID2} and a set of complementary measurements using the inner detector and the calorimeter.

The data events are selected with a set of single-jet triggers~\cite{TRIG-2019-01,TRIG-2019-02,TRIG-2019-04}
with $\pt$ thresholds ranging from $60$~\GeV to $420$~\GeV and a detector acceptance of $|\eta| < 3.2$. The lower threshold single-jet triggers are prescaled to reduce high trigger rates and record only a fraction of events for online processing and data storage at an acceptable level. Only the highest \pt threshold single-jet triggers are unprescaled to record each observed event with high-\pT jets.
A leading-jet trigger strategy~\cite{Lendermann_2009}
is employed, where the trigger is accepted only if the leading jet fell within a predefined \pT range, achieving an efficiency of 99.5\%.  To obtain a smooth \mjj spectrum at the reconstructed level, events are weighted according to the trigger prescale.

The event reconstruction includes particle flow jets~\cite{PERF-2015-09}, which combine topological clusters~\cite{PERF-2014-07} in the calorimeter system with tracks from the inner detector to improve the jet energy resolution and the reconstruction efficiency at low \pT. The jet energy calibration begins by mitigating pile-up effects through a data-driven event-by-event pile-up \pt-density and jet–area subtraction, followed by residual pile-up MC-based corrections for the dependence on the number of interactions per bunch crossing and the number of primary vertices. An absolute MC-based calibration is then applied to restore the jet four-momentum to the particle level in bins of $\eta$ and $\pT$. The global sequential calibration is subsequently applied to reduce flavour- and topology-dependence. Finally, residual data-to-MC differences are corrected using in-situ techniques in $\gamma$+jet, $Z$+jet, dijet, and multijet events. The resulting jet energy scale reaches per-cent precision for central jets with $\pT>100$~\GeV, with uncertainties reduced by a factor of two to three compared with early Run~2~\cite{JETM-2018-05,JETM-2022-06}.

The simulated events are generated using \PYTHIA[8.230]~\cite{Sjostrand:2014zea} with leading-order (\LO) matrix elements for $2 \rightarrow 2$ dijet production, including the default \PYTHIA Lund string hadronisation model~\cite{Andersson:1983ia,Sjostrand:1984ic}, transverse momentum ordering~\cite{Sj_strand_2005} parton showers, and the A14~\cite{ATL-PHYS-PUB-2014-021} set of tuned parameters. The renormalisation and factorisation scales are set to the geometric mean of the squared transverse masses of the two outgoing particles, $m_\text{T,1}, m_\text{T,2}$, in the matrix element, $\sqrt{m^2_\text{T,1} \cdot m^2_\text{T,2}} = \sqrt{(\pTX[2][1] + m_1^2) (\pTX[2][2] + m_2^2)}$. The \NNPDF[2.3lo] PDF set~\cite{Ball:2012cx} is used in the matrix element generation, the parton shower, and in the simulation of the multi-parton interactions. For the modelling of heavy hadron decays, \EVTGEN~\cite{Lange:2001uf} is used.

The simulated events are propagated through the full detector simulation~\cite{SOFT-2010-01} of the ATLAS detector using \GEANT~\cite{Agostinelli:2002hh} and overlaid with inelastic minimum-bias interactions generated using \PYTHIA[8.1]~\cite{Sjostrand:2007gs} to simulate pile-up interactions. The simulated events are re-weighted according to the normalised distribution of an actual number of pile-up interactions to match the measured data precisely. The simulated particles are reconstructed with the same software~\cite{SOFT-2022-02} as the ATLAS data.

The jet cleaning requirements are applied to both data and simulated events~\cite{ATLAS-CONF-2015-029}, which suppress background from coherent noise, pathological calorimeter cells, and beam-induced effects. Additionally, events must contain at least one well-reconstructed primary vertex built from at least two associated inner detector tracks with $\pT > 500$~\MeV.

\FloatBarrier


%
%

\FloatBarrier

\section{Unfolding of detector effects}
\label{sec:unfolding}

To obtain the final dijet cross-sections at particle level, the reconstructed data distributions are corrected for detector inefficiencies and resolution effects using a modified Bayesian technique, the iterative, dynamically stabilised (IDS) method of data unfolding~\cite{malaescu2009}.

Migrations of dijet events across analysis bins between the particle and reconstructed levels are described with a response matrix built using MC-simulated dijet events.\footnote{Migrations in both (\mjj{}, \ystar{}) are simultaneously accounted for in the unfolding, as the response matrix is constructed to include all \mjj{} bins across the six \ystar{} intervals. The same approach is applied for (\mjj{}, \yboost{}).}
The particle-level and reconstructed-level jets are geometrically matched according to their angular distance.
Two jets are considered to be matched if they are closer to each other than to any other jet and lie within a radius of $\Delta{}R=0.3$.
The leading and subleading reconstructed jets must be matched to the leading and subleading particle jets.\footnote{The order of the two jets can be swapped between the reconstructed and particle levels due to the resolution effects, allowing for leading-to-subleading and subleading-to-leading matches.}
The response matrix is filled with dijet masses calculated using particle-level and reconstructed-level dijet events that simultaneously satisfy the analysis selection criteria described in~\Cref{sec:DatasetSelection} and that are geometrically matched between the two levels.
Events passing the selection on the reconstructed (particle) level that either fail to satisfy the selection on the particle (reconstructed) level or do not pass the matching procedure are used for impurity (inefficiency) corrections during the unfolding.

Reconstructed data are unfolded to the particle level using a three-step procedure, which corrects for the matching impurity at the reconstructed level, migrations of dijet events between analysis bins, and matching inefficiency at the particle level.
The number of dijet events, $N^\text{part}_i$, at the particle level in bin~$i$ is calculated from the number of events, $N^\text{reco}_k$, at the reconstructed level in bin~$k$, according to
\begin{equation*}
N^\text{part}_{i} = \sum_{k}^{}N^\text{reco}_{k} \cdot{} \mathcal{P}_{k} \cdot{} \mathcal{U}_{ik} / \mathcal{E}_{i},
\end{equation*}
where $\mathcal{P}_{k}$ and $\mathcal{E}_{i}$ denote matching purity and matching efficiency, respectively, in corresponding bins.
The matching purity, $\mathcal{P}_k$, is defined using the previously described matching procedure as the fraction of reconstructed-level dijet events matched to corresponding particle-level events in the reconstructed bin~$k$.
Similarly, the matching efficiency, $\mathcal{E}_i$, is defined as a fraction of particle-level dijet events matched to corresponding reconstructed-level events in the particle bin~$i$.
Elements of the unfolding matrix, $\mathcal{U}_{ik}$, are calculated from the response matrix and describe the probability of a dijet event reconstructed in a bin~$k$ to originate from a particle-level bin~$i$.

The particle-level MC-simulated spectrum and, therefore, the unfolding matrix, $\mathcal{U}$, are iteratively improved in the IDS unfolding method.
The residual shape uncertainty of the unfolded spectrum is evaluated using a data-driven closure test~\cite{malaescu2009}.
The particle-level MC-simulated spectrum is reweighted so that the agreement at the reconstructed level between the data and MC simulation is improved.
The relative difference between the reweighted reconstructed-level MC spectrum unfolded by the nominal response matrix and the reweighted particle-level MC spectrum is interpreted as a bias of the unfolding procedure.
The number of iterations of the IDS method is chosen such that the residual bias in each analysis bin is smaller than the respective statistical uncertainty, considering both data and MC contributions.
One iteration is chosen for \ystar{}-binned cross-section and two iterations are chosen in the case of \yboost{}.

\FloatBarrier


\FloatBarrier

\section{Experimental uncertainties}
\label{sec:experimentalUncertainties}

The total experimental uncertainty in the cross-sections, combining effects of statistical fluctuations of data and MC simulation and various sources of systematic uncertainties, is shown in~\Cref{fig:uncertainties:totalUncertainty}.
\begin{figure}[htbp]
\centering
\subfloat[]{\includegraphics[width=0.49\textwidth]{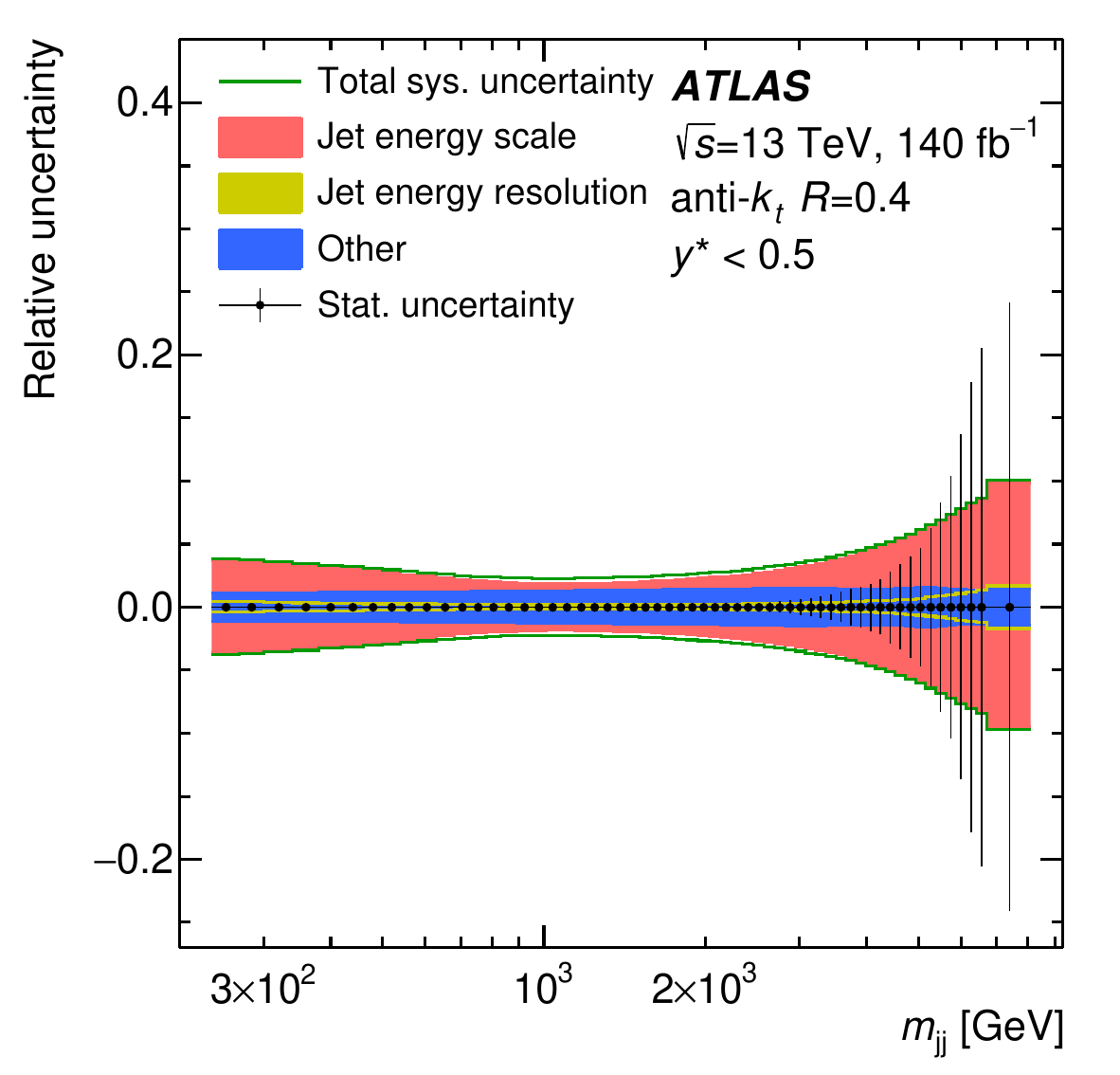}}
\subfloat[]{\includegraphics[width=0.49\textwidth]{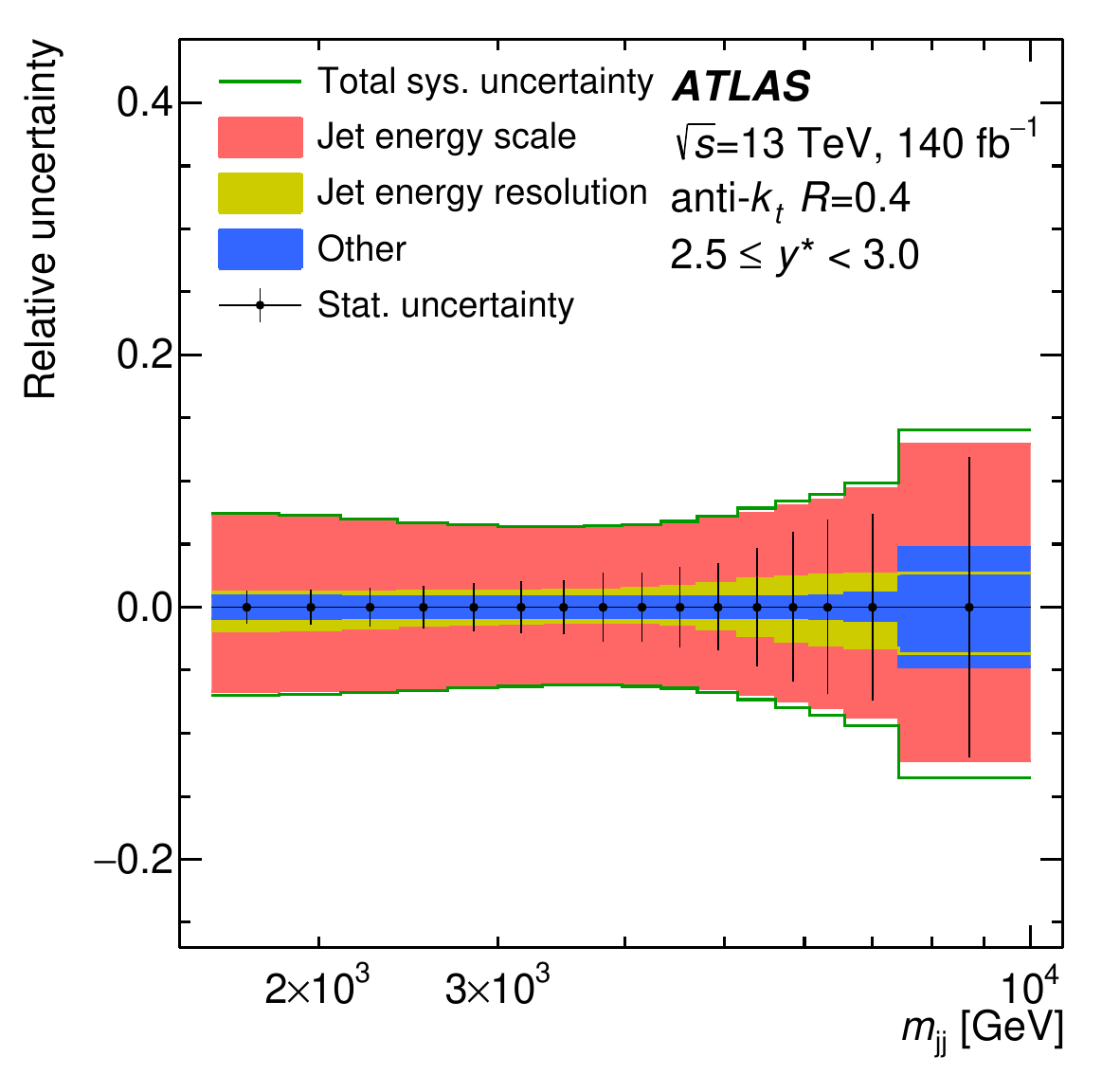}}
\\
\subfloat[]{\includegraphics[width=0.49\textwidth]{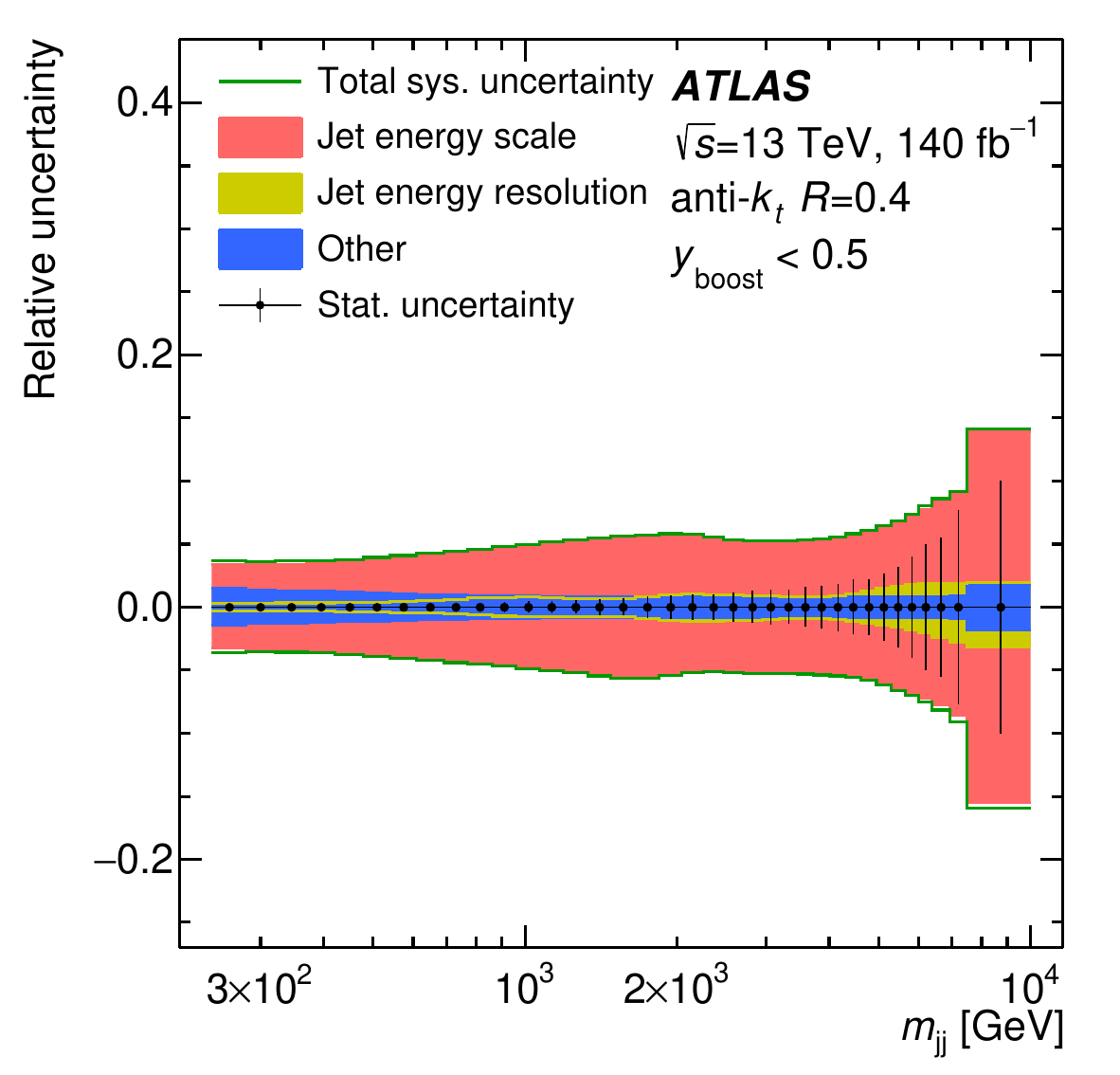}}
\subfloat[]{\includegraphics[width=0.49\textwidth]{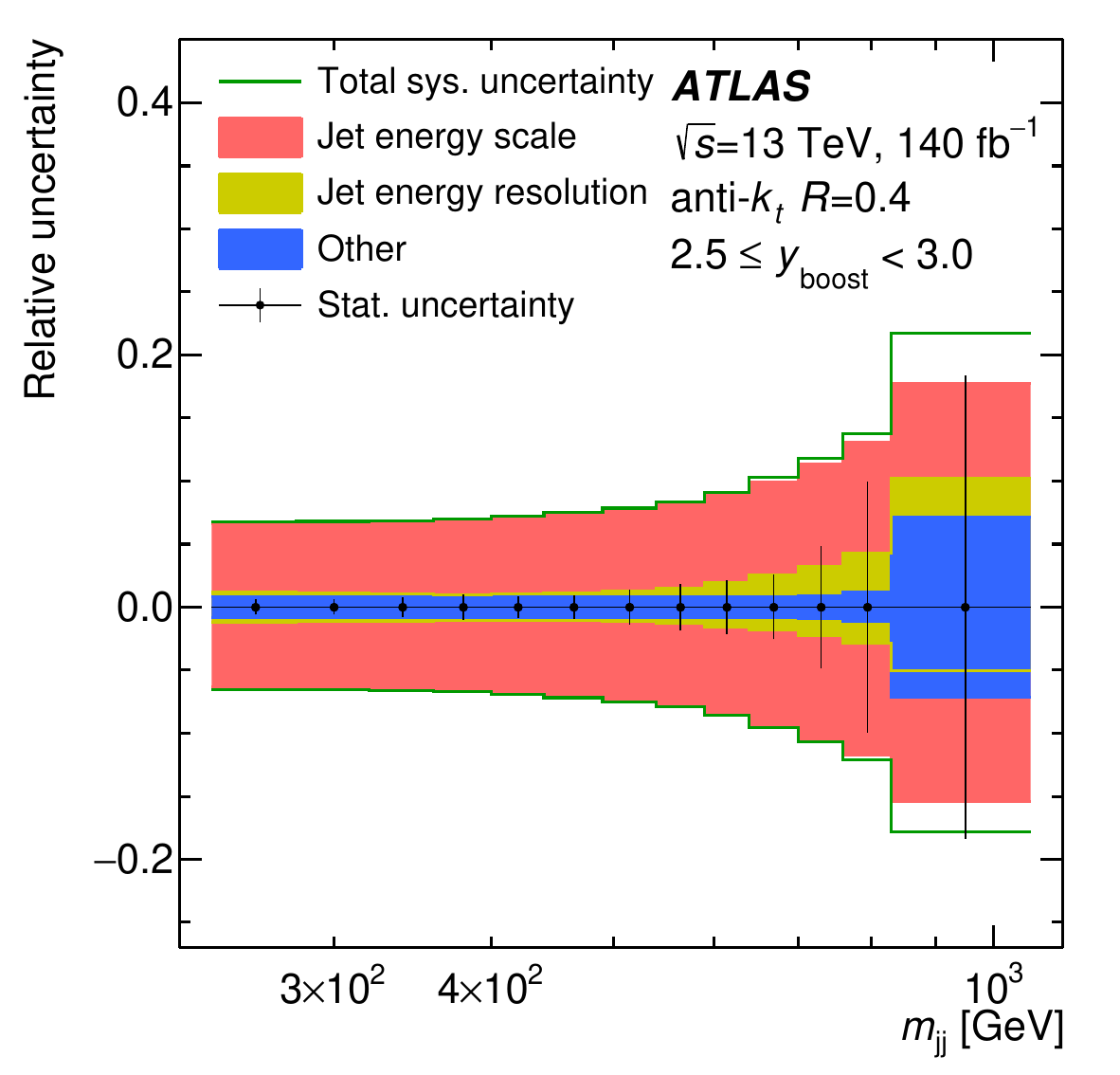}}
\caption{Relative systematic uncertainty in the dijet cross-section as a function of \mjj{} in the first and last \ystar{} (a,~b) and \yboost{} (c,~d) analysis bins. The components of the uncertainty are displayed as colour bands---Jet energy scale, Jet energy resolution, and Other (combining the residual unfolding bias, the effect of disabled Tile calorimeter modules, and the luminosity uncertainty). The total systematic uncertainty is obtained as a quadrature sum of the components and is compared with the statistical component.}
\label{fig:uncertainties:totalUncertainty}
\end{figure}

The statistical uncertainty in the measurement consists of data and MC simulation contributions, which are propagated through the unfolding procedure using a set of 1000 pseudo experiments.
Each pseudo experiment is created by fluctuating weights of the dijet events according to a Poisson distribution with a mean equal to one, producing varied data spectra and MC-simulated response matrices.
The data contribution to the total statistical uncertainty is evaluated by unfolding the fluctuated data spectra using the nominal response matrix, whereas the MC contribution is propagated by unfolding the nominal data spectrum using the fluctuated response matrices.
The corresponding statistical uncertainties are calculated as the root mean square error from the resulting distributions.
Bootstrap replicas are also used to produce a covariance matrix describing correlations between the analysis bins.

The total statistical uncertainty of the measurement is calculated as a quadrature sum of the data and MC contributions.
In the first \ystar{} analysis bin, the uncertainty is about 0.1\% at lower \mjj{} values and increases up to 24\% at higher \mjj{}.
In the last \ystar{} bin, the uncertainty is about 1\% at lower \mjj{} values and increases up to 12\% at higher \mjj{}.
Similarly, in the first \yboost{} bin, the uncertainty is about 0.2\% at lower \mjj{} values and increases up to 10\% at higher \mjj{}.
In the last \yboost{} bin, the uncertainty is about 0.5\% at lower \mjj{} values and increases up to 18\% at higher \mjj{}.

The jet energy scale (JES) uncertainty is the dominant systematic uncertainty of this measurement.
It is evaluated using various simulation-based techniques and in-situ measurements described in Refs.~\cite{JETM-2018-05,JETM-2022-06}.
The full JES uncertainty set of 1172 individual components, including the latest improvements\footnote{Jet energy scale uncertainty improvements come mainly from reduced jet flavour response and single-particle deconvolution uncertainties~\cite{JETM-2022-06}.} discussed in Ref.~\cite{JETM-2022-06}, is used.
All components of the JES uncertainty are propagated through the unfolding procedure using an MC simulation.
Each reconstructed jet \pT{} is scaled up and down by one standard deviation of a given component.
The \mjj{} spectra produced using the varied jets are unfolded with the nominal response matrix.
The relative difference to the nominal unfolded MC spectrum is taken as a corresponding systematic uncertainty, which is applied to the unfolded data.

At higher \mjj{} values, the $\eta$~intercalibration method uncertainty (described in Ref.~\cite{JETM-2018-05}) is the dominant JES uncertainty component in all \yboost{} and high-\ystar{} bins.
In low-\ystar{} bins, the uncertainty in the energy response to isolated hadrons (described in Refs.~\cite{JETM-2022-06,JETM-2018-08}) is the most significant component.
At lower \mjj{} values, the pile-up uncertainty is the dominant component in low-\ystar{} and low-\yboost{} bins, being overtaken by the $\eta$~intercalibration method uncertainty for higher \ystar{} and \yboost{} values.

The jet energy resolution (JER) is measured using various techniques described in Ref.~\cite{JETM-2018-05}.
The JER systematic uncertainty is split into 34 individual components.
Each component is propagated through the unfolding procedure using an MC simulation by smearing the jet \pT{} according to a Gaussian function with a width $\sigma_\text{smear}$,
\begin{equation*}
\sigma^2_\text{smear} =  \left( \sigma_\text{nom} + \left| \sigma_\text{NP} \right| \right)^2 - \sigma^2_\text{nom},
\end{equation*}
where $\sigma_\text{nom}$ is the nominal resolution in the MC sample and $\sigma_\text{NP}$ is a one-standard-deviation variation of a given uncertainty component~\cite{JETM-2018-05}.
Components of the JER uncertainty can involve JER improvement in part of the phase space, but JER degradation in another, relative to the nominal resolution $\sigma_\text{nom}$.
To preserve anti-correlations in such cases, the JER improvement and degradation are treated differently.
The effect of JER degradation is propagated by smearing the \pT{} of MC-simulated jets filled into the reconstructed spectrum.
In contrast, the effective JER improvement is realised by smearing the \pT{} of MC jets used for the response matrix construction.
The relative difference between the varied reconstructed spectrum unfolded by the varied response matrix and the nominal reconstructed spectrum unfolded by the nominal response matrix is taken as the corresponding systematic uncertainty.

The relative residual unfolding bias described in~\Cref{sec:unfolding} is applied to the unfolded data as a source of systematic uncertainty.
It stays mostly below one per mil and reaches up to a few per cent for large values of \mjj{}.

The effect of certain Tile calorimeter modules being disabled due to technical problems during the Run~2 data taking~\cite{TCAL-2021-01} is evaluated by comparing the nominal cross-section and one produced while removing dijet events with jets pointing to the disabled modules in both the data and MC simulations.
The difference, reaching up to approximately one per cent in some \mjj{} bins, is taken as a systematic uncertainty.

The relative luminosity uncertainty of 0.83\%~\cite{DAPR-2021-01} is applied across all \mjj{} bins.

No additional sources of the experimental systematic uncertainty are considered.
For particle flow jets, the effect of the jet angular resolution is negligible relative to the other sources of the uncertainty.
There is no uncertainty associated with the jet quality selection.

To mitigate the effects of statistical fluctuations of the systematic uncertainty estimates, a smoothing procedure is applied to the JES and JER components, the effect of disabled Tile modules, and to the unfolding bias.
First, statistical uncertainties in the uncertainty components are estimated by using pseudo experiments (100 for each component).
Second, neighbouring \mjj{} bins are combined until the statistical significance in each bin is larger than two standard deviations.
Last, a Gaussian kernel smoothing~\cite{PERF-2011-03} is applied, restoring the original, fine binning.

The total systematic uncertainty on the dijet cross-section measurement is calculated as a quadrature sum of the components.
In the first \ystar{} analysis bin, the uncertainty is about 2\% at medium \mjj, increasing to 4\% at low \mjj and up to 10\% at high \mjj. In the last \ystar{} bin, the uncertainty is higher: about 6\% at medium \mjj, increasing to 7\% at low \mjj and up to 14\% at high \mjj.
Similarly, in the first \yboost{} bin, the uncertainty is below 4\% at low \mjj and rises to 16\% at high \mjj. In the last \yboost{} bin, it increases from below 7\% at low \mjj to 22\% at high \mjj.

\FloatBarrier


\FloatBarrier

\section{Theoretical prediction}
\label{sec:theoreticalPrediction}
Theoretical predictions of the dijet cross-sections are obtained using pQCD, supplemented by corrections for non-perturbative and electroweak effects.
While the previous ATLAS dijet publication~\cite{STDM-2016-03} relied on next-to-leading-order (NLO) leading colour pQCD predictions from \NLOjet~\cite{PhysRevD.68.094002,PhysRevLett.88.122003},
the theoretical precision is now improved by employing \NNLO predictions with full colour information, obtained using \NNLOJET~\cite{NNLOJET:2025rno,PhysRevLett.119.152001,Czakon_2019,Chen_2022} interfaced to \PINEAPPL~\cite{Carrazza:2020gss} and \APPLGRID~\cite{Carli:2010rw} for fast convolutions with varied input parameters (PDFs, renormalisation and factorisation scales and the strong coupling constant).
The predictions are calculated using several recent PDF sets provided by the LHAPDF6~\cite{Buckley:2014ana} library: \CT{18}~\cite{Hou:2019efy}, \MSHT{}~\cite{Bailey:2020ooq}, \NNPDF{3.1}~\cite{NNPDF:2017mvq}, \NNPDF{4.0}~\cite{NNPDF:2021njg} and \ATLASPDF{21}~\cite{STDM-2020-32}.
The corresponding NNLO sets are used with the value of the strong coupling constant taken from the corresponding PDF set. The applied renormalisation and factorisation scale of dijet invariant mass, $\muRmuF = \mjj$, is the one proposed in Ref.~\cite{PhysRevLett.119.152001}.

The uncertainties in the theoretical predictions come from the uncertainties on the PDFs, the renormalisation and factorisation scales, and the uncertainty in the value of the strong coupling constant. The PDF uncertainties are defined at 68\% confidence level (CL) and propagated through the calculation following the recommendation of each PDF set. The scale uncertainty is estimated as the envelope obtained by varying the renormalisation and factorisation scales up and down by a factor of $2$ from the nominal values, excluding the opposite variations in the two scales. The uncertainty in the value of the strong coupling constant is estimated by varying its value by $0.001$ from the nominal value $\alphas(m_\text{Z}) = 0.118$ in the matrix element and applying the corresponding shifted PDF sets.

The representative uncertainties in the theoretical predictions from the \NNLO calculation using the \CT{18NNLO} (\NNPDF{4.0}) PDF sets are shown in~\Cref{fig:Theory:pQCDSys_CT18NNLO_MJJ_SCALE} (~\Cref{fig:Theory:pQCDSys_NNPDF40_MJJ_SCALE}). At the highest \mjj, \CT{18NNLO} has PDF uncertainties of up to 15--22\%, whereas \NNPDF{4.0} reaches about 2\% (except in the highest \mjj region of the highest \yboost bin, where it can reach about 6\%), consistent with the 1–2\% precision\footnote{Each PDF set employs different input data samples and methodologies. The 1–2\% level of precision achieved by \NNPDF{4.0} is widely recognised within the PDF community and is confirmed in several benchmarking studies~\cite{NNPDF:2021njg}. Despite this, \CT{18NNLO} is adopted as the default PDF set in this measurement because its PDF uncertainty is comparable to that of other modern global fits (e.g. \MSHT, \ATLASPDF{21}). The choice is consistent with the most recent dijet CMS measurement~\cite{CMS-SMP-21-008}.} quoted by the \NNPDF{} Collaboration~\cite{NNPDF:2021njg}. For \NNPDF{4.0}, the PDF uncertainty is comparable to the $\alphas$ and statistical components, while the scale uncertainty becomes the dominant contribution.

\begin{figure}[htbp]
\centering
\subfloat[]{\includegraphics[width=0.49\textwidth]{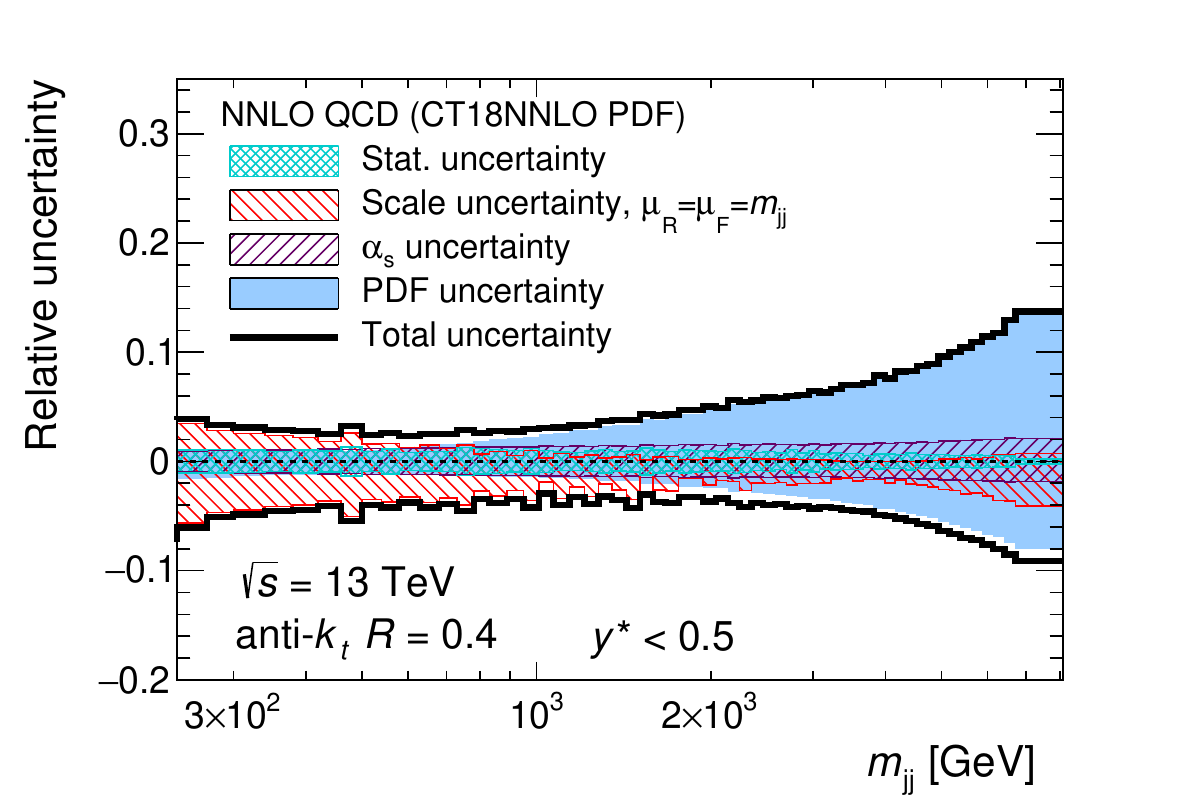}}
\subfloat[]{\includegraphics[width=0.49\textwidth]{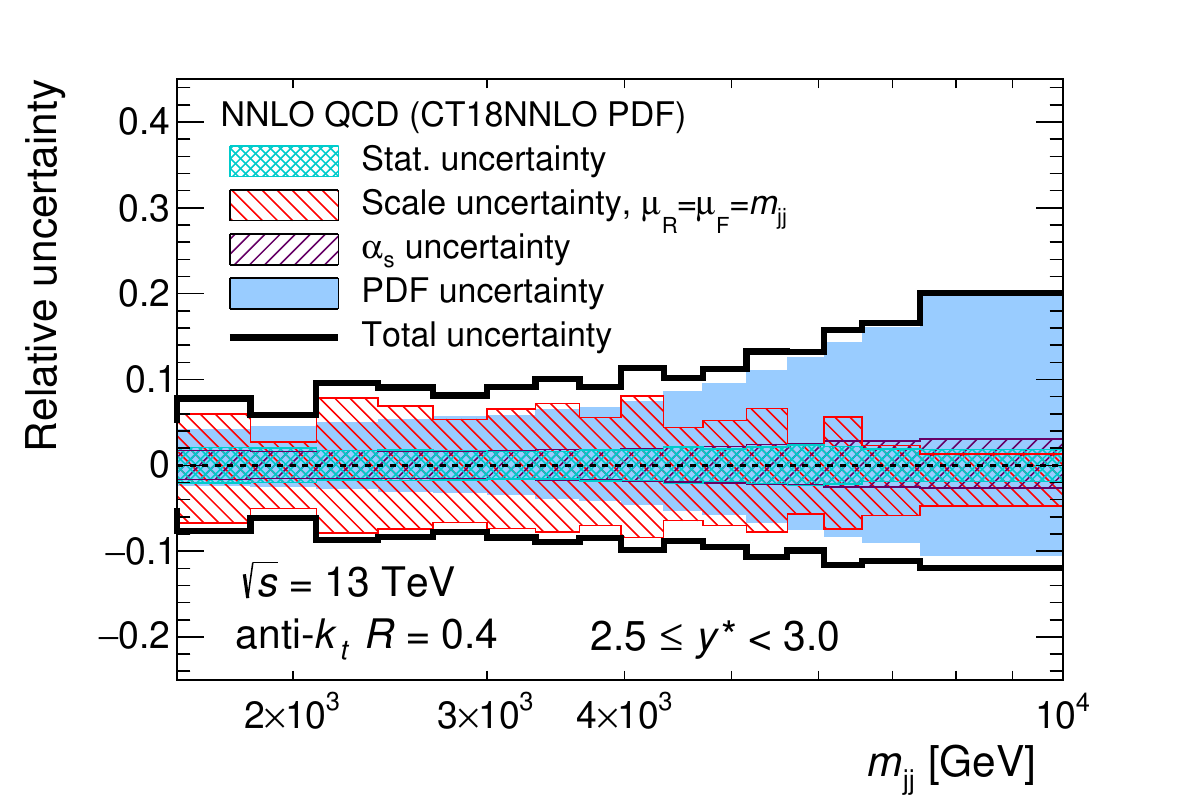}} \\
\subfloat[]{\includegraphics[width=0.49\textwidth]{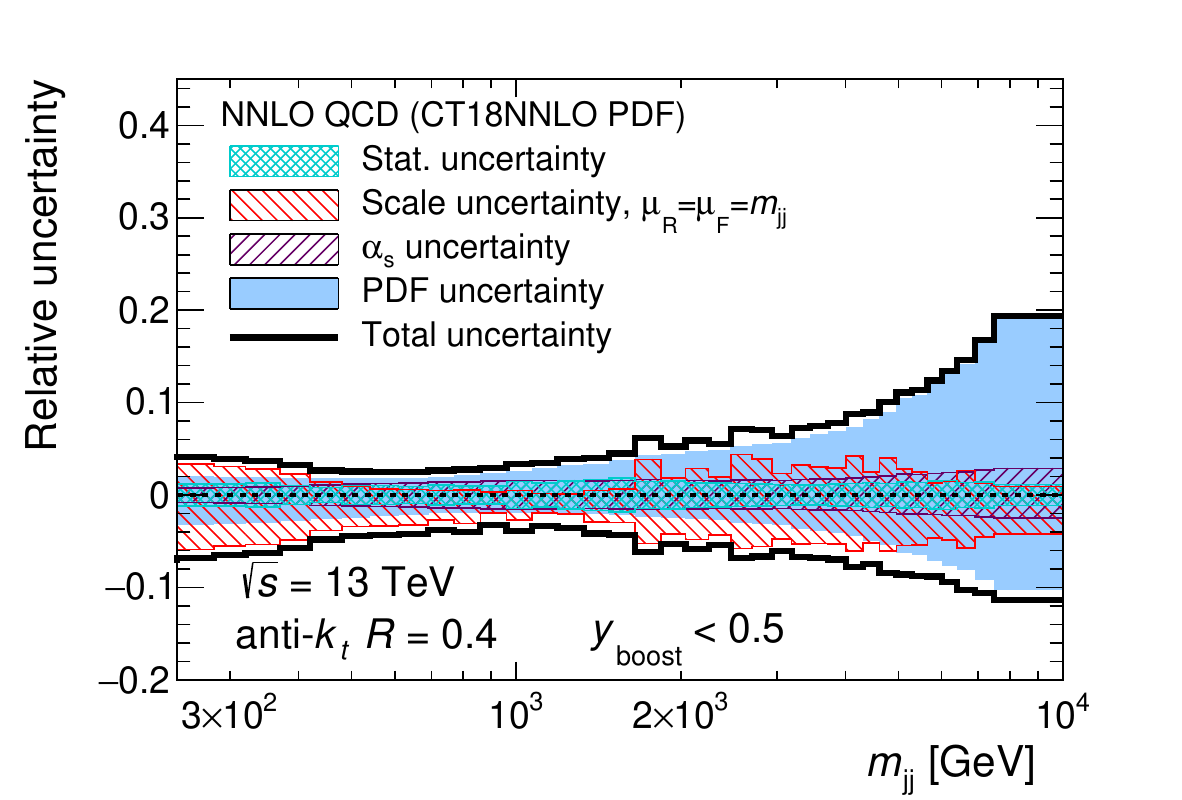}}
\subfloat[]{\includegraphics[width=0.49\textwidth]{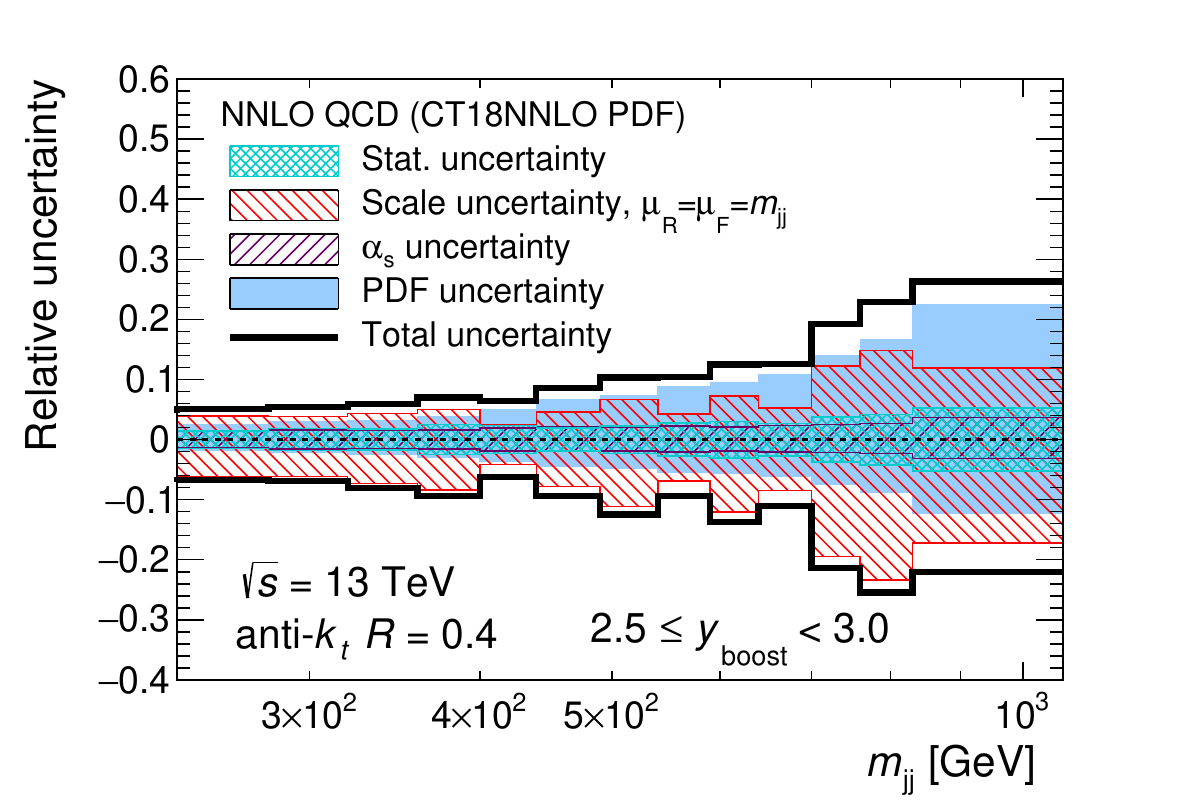}}
\caption{Relative \NNLO pQCD theoretical uncertainties for dijet cross-section predictions for the \CT{18NNLO} PDF set. Panels a,~b (c,~d) correspond to the first and last \ystar{} (\yboost{}) analysis bins. The uncertainties due to the renormalisation and factorisation scales, the value of \alphas{}, the choice of particular PDF, the statistical uncertainty on the calculation and the total uncertainty are shown. The total uncertainty is calculated by adding the individual uncertainties in quadrature.}
\label{fig:Theory:pQCDSys_CT18NNLO_MJJ_SCALE}
\end{figure}

\begin{figure}[htbp]
\centering
\subfloat[]{\includegraphics[width=0.49\textwidth]{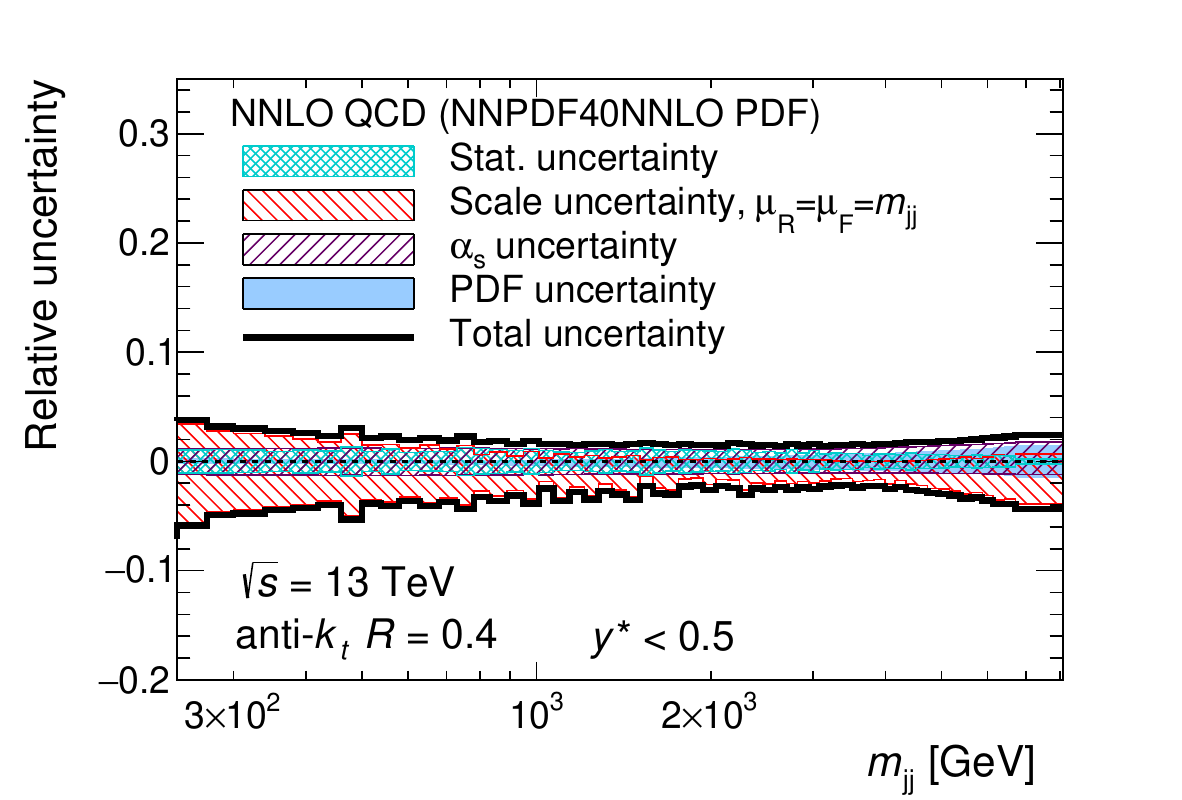}}
\subfloat[]{\includegraphics[width=0.49\textwidth]{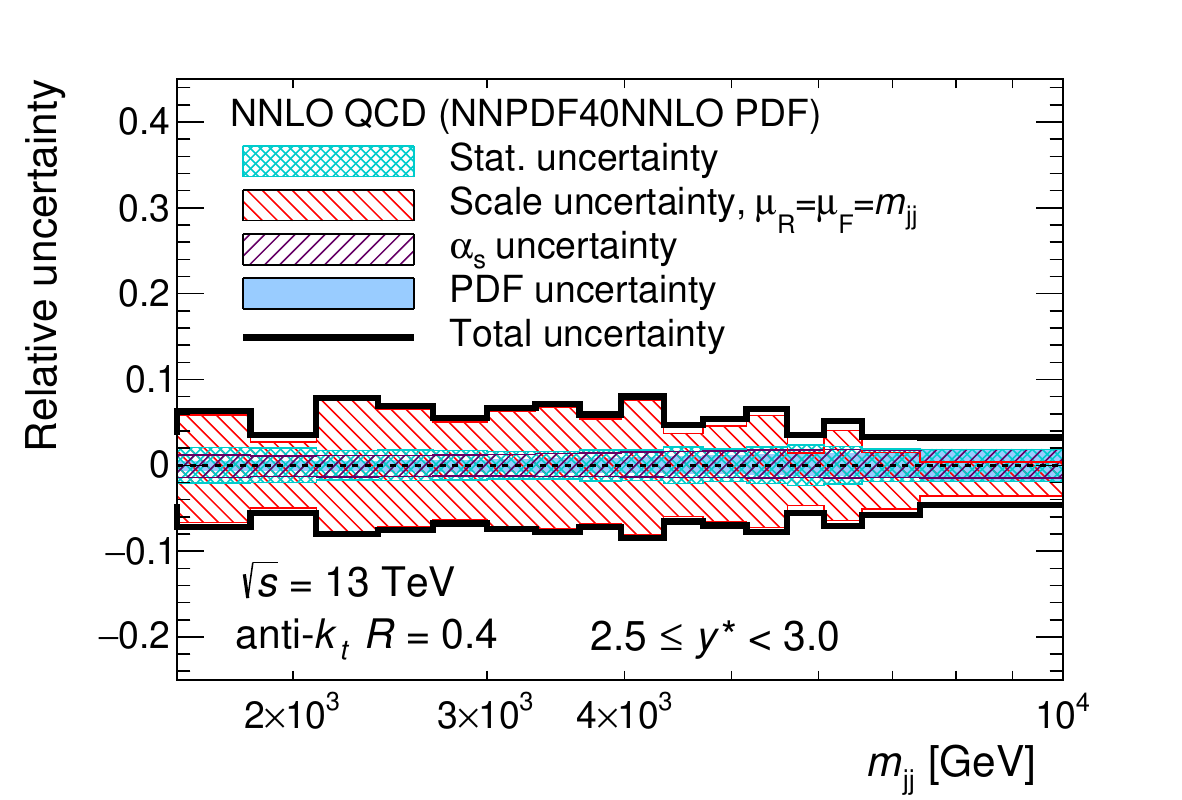}} \\
\subfloat[]{\includegraphics[width=0.49\textwidth]{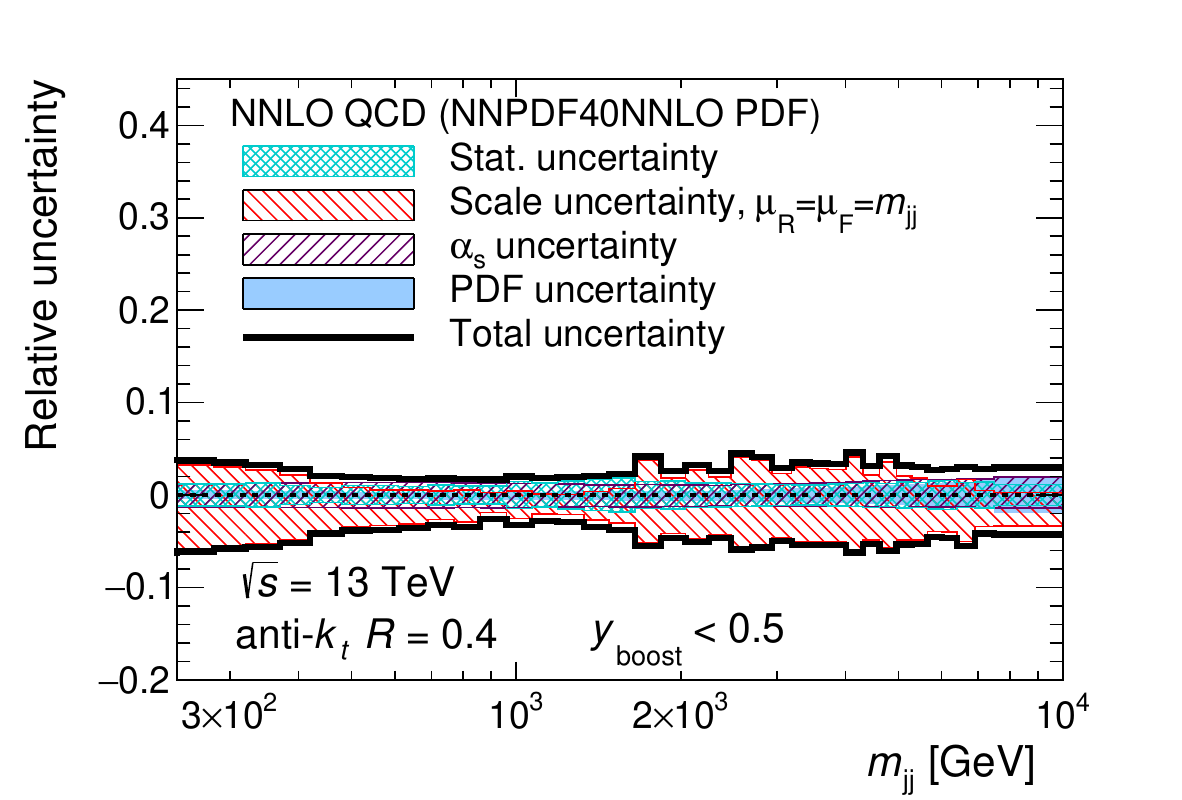}}
\subfloat[]{\includegraphics[width=0.49\textwidth]{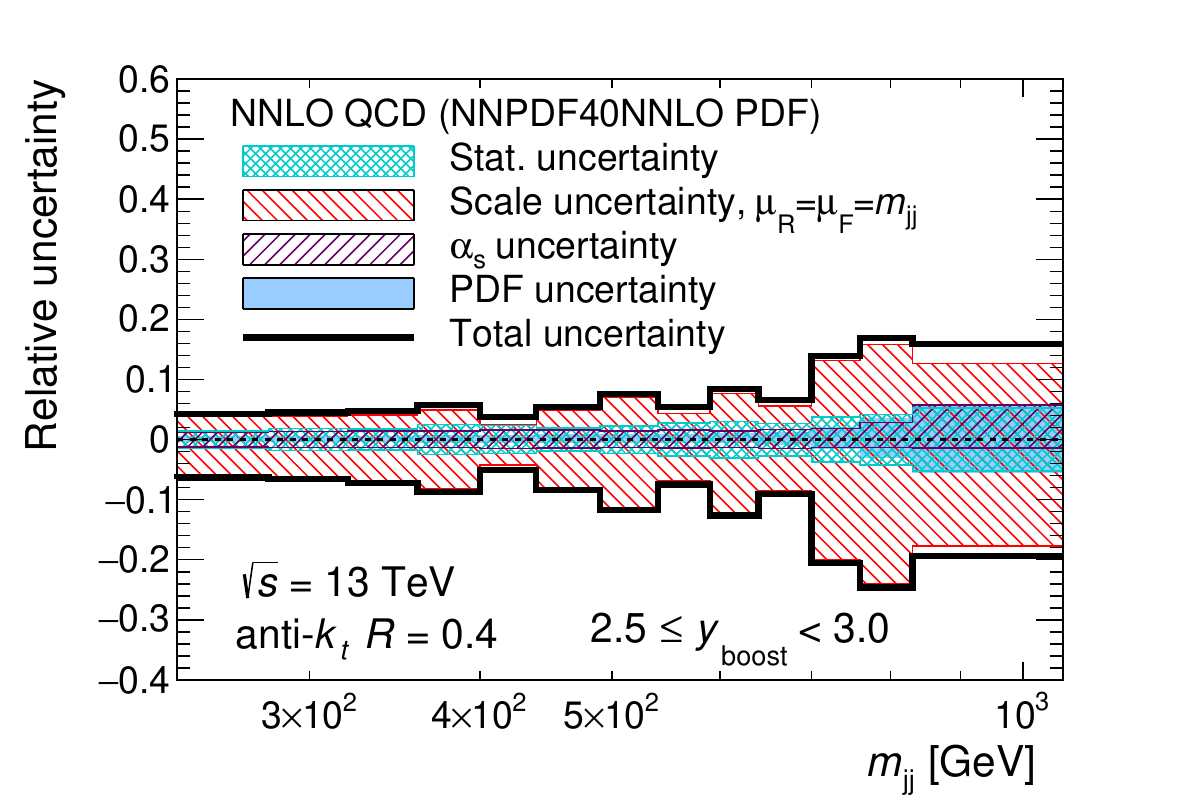}}
\caption{Relative \NNLO pQCD theoretical uncertainties for dijet cross-section predictions for the \NNPDF{4.0} PDF set. Panels a,~b (c,~d) correspond to the first and last \ystar{} (\yboost{}) analysis bins. The uncertainties due to the renormalisation and factorisation scales, the value of \alphas{}, the choice of particular PDF, the statistical uncertainty on the calculation and the total uncertainty are shown. The total uncertainty is calculated by adding the individual uncertainties in quadrature.}
\label{fig:Theory:pQCDSys_NNPDF40_MJJ_SCALE}
\end{figure}

To obtain the predicted cross-section at the particle level, non-perturbative (NP) corrections are applied to the parton-level pQCD predictions. These corrections are obtained using LO MC generators, as the bin-by-bin ratios of the nominal particle-level MC prediction to the prediction from the partons remaining after showering (PS), but with hadronisation (HAD) and multi-parton interactions (MPI) modelling switched off, according to
\begin{equation*}
\text{NP corrections} = \frac{\sigma^\text{Particle level}}{\sigma^\text{Parton level}} = \frac{\sigma^\text{PS+HAD+MPI}}{\sigma^\text{PS}}.
\end{equation*}
In order to account for various HAD and MPI models, several MC generators with different hadronisation models, PDFs and sets of tuned parameters due to the underlying events\footnote{The \Herwig{7.1} tuned parameters are fitted to LHC data at 900~\GeV, 7~\TeV, and 13~\TeV, with Tevatron measurements used only for validation. The \Pythia AU2 tuned parameters are primarily based on ATLAS minimum-bias and underlying-event data at 7~\TeV and 900~\GeV, with Tevatron (CDF) underlying-event observables included only with small weights to guide the energy extrapolation.
The later A14 tuned parameters are derived solely from LHC Run-1 data. For the non-perturbative correction, the A14 tuned parameters are used with the \NNPDF{2.3} PDF as nominal, and the others are used to evaluate the uncertainty.}
are considered as shown in~\Cref{tab:NPC_TUNES}.

\begin{table}[htbp] %
\small
\centering
\caption{List of MC generators, hadronisation models, PDF sets, and tuned parameters due to underlying event used for the determination of the NP corrections.}

\begin{tabular}{l|lll}
\hline
MC generator & Hadronisation & PDF & Tuned parameters\\
\hline
\PYTHIA[8.210]~\cite{Sjostrand:2014zea} & Lund String model~\cite{Sjostrand:1984ic,Andersson:1983ia} & \CTEQ{6L1}~\cite{Pumplin:2002vw} & C4~\cite{Corke2011}  \\
\PYTHIA[8.210] & Lund String model & \CTEQ{6L1} & AU2~\cite{ATL-PHYS-PUB-2012-003} \\
\PYTHIA[8.210] & Lund String model & \CT{10}~\cite{Lai:2010vv}        & AU2  \\
\PYTHIA[8.210] & Lund String model & \NNPDF{2.3}~\cite{Ball:2012cx}   & A14~\cite{ATL-PHYS-PUB-2014-021}  \\
\HERWIG[7.2.3p2]~\cite{Bellm:2019zci} & Cluster model~\cite{Webber:1983if} & \NNPDF{2.3} & H7.1-Default~\cite{herwig7tunes} \\
\HERWIG[7.2.3p2] & Cluster model & \NNPDF{2.3} & H7.1-BaryonicReconnection~\cite{Gieseke:2012ft}\\
\HERWIG[7.2.3p2] & Cluster model & \NNPDF{2.3} & H7.1-Soft \\

\hline
\end{tabular}
\label{tab:NPC_TUNES}
\end{table}
\begin{figure}[htbp]
\centering
\subfloat[]{\includegraphics[width=0.45\textwidth]{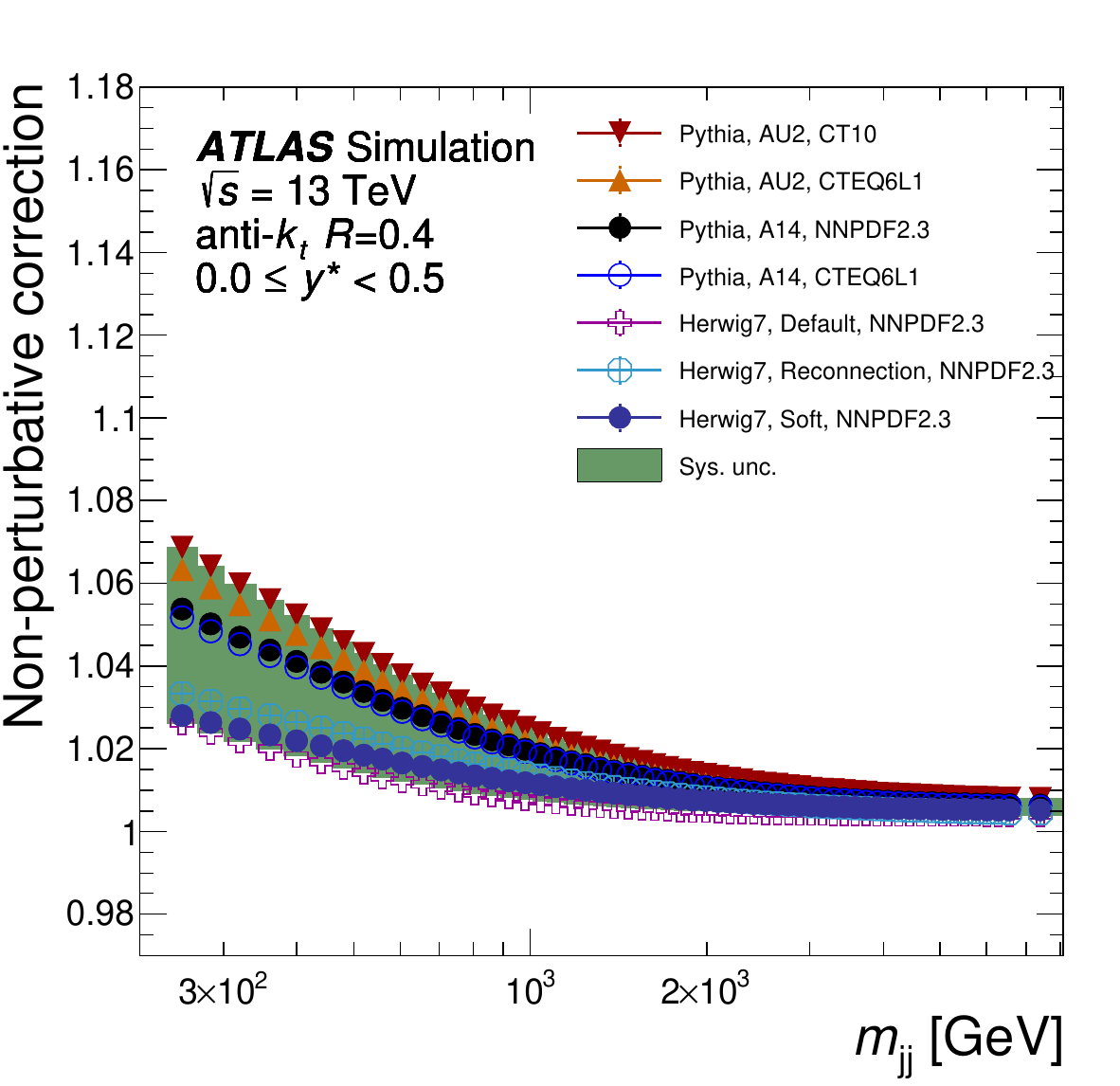}}
\subfloat[]{\includegraphics[width=0.45\textwidth]{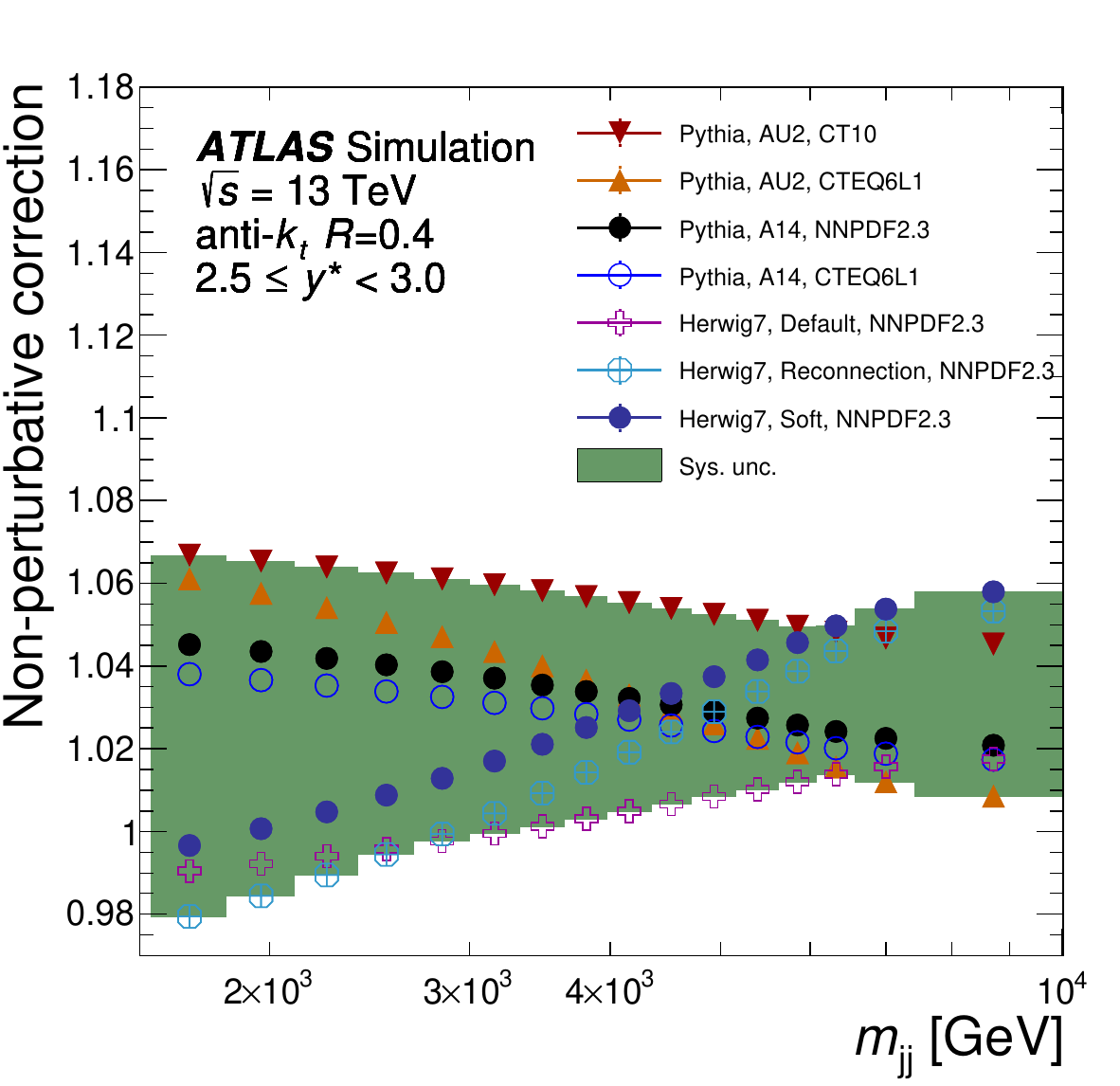}} \\
\subfloat[]{\includegraphics[width=0.45\textwidth]{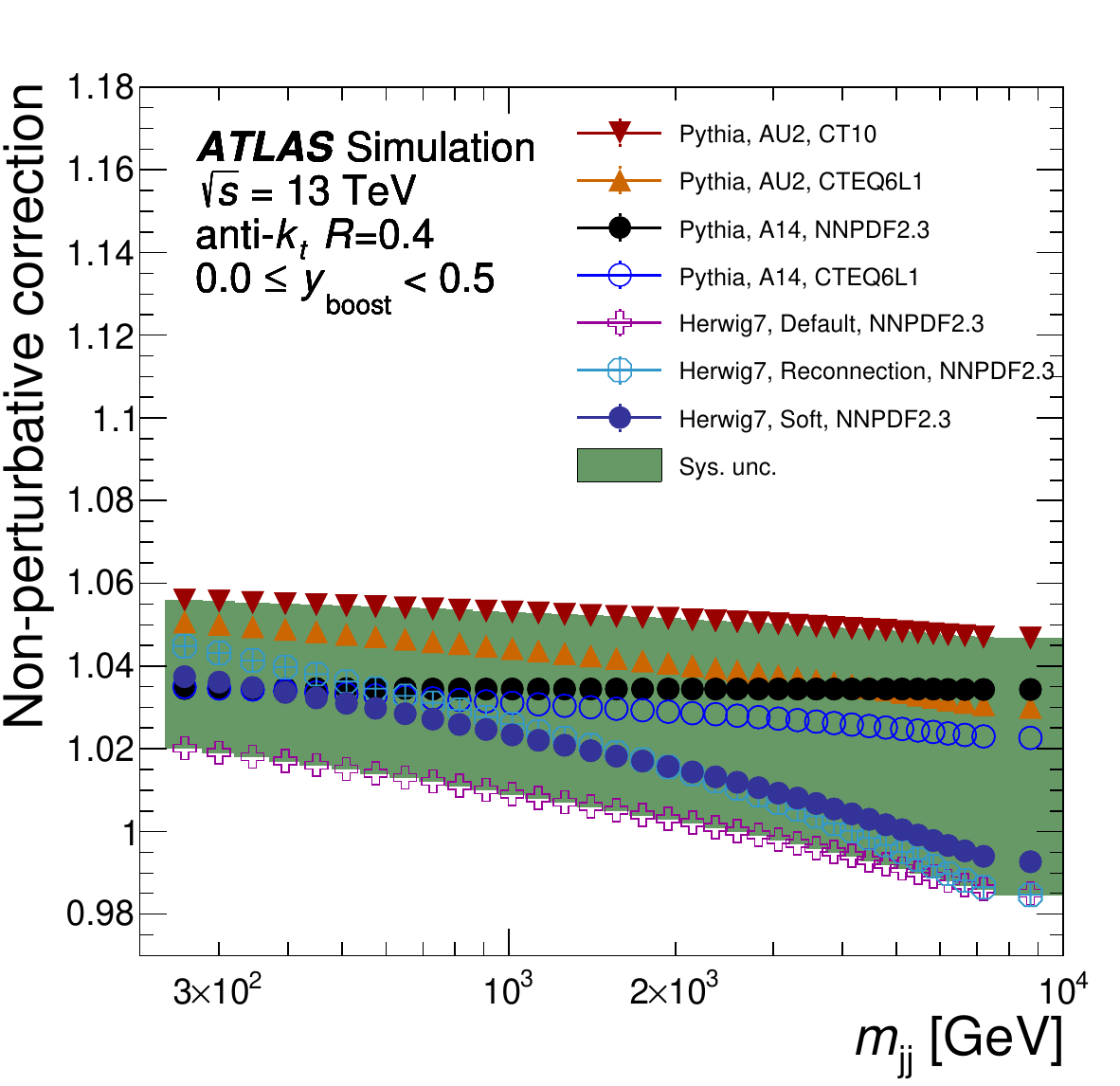}}
\subfloat[]{\includegraphics[width=0.45\textwidth]{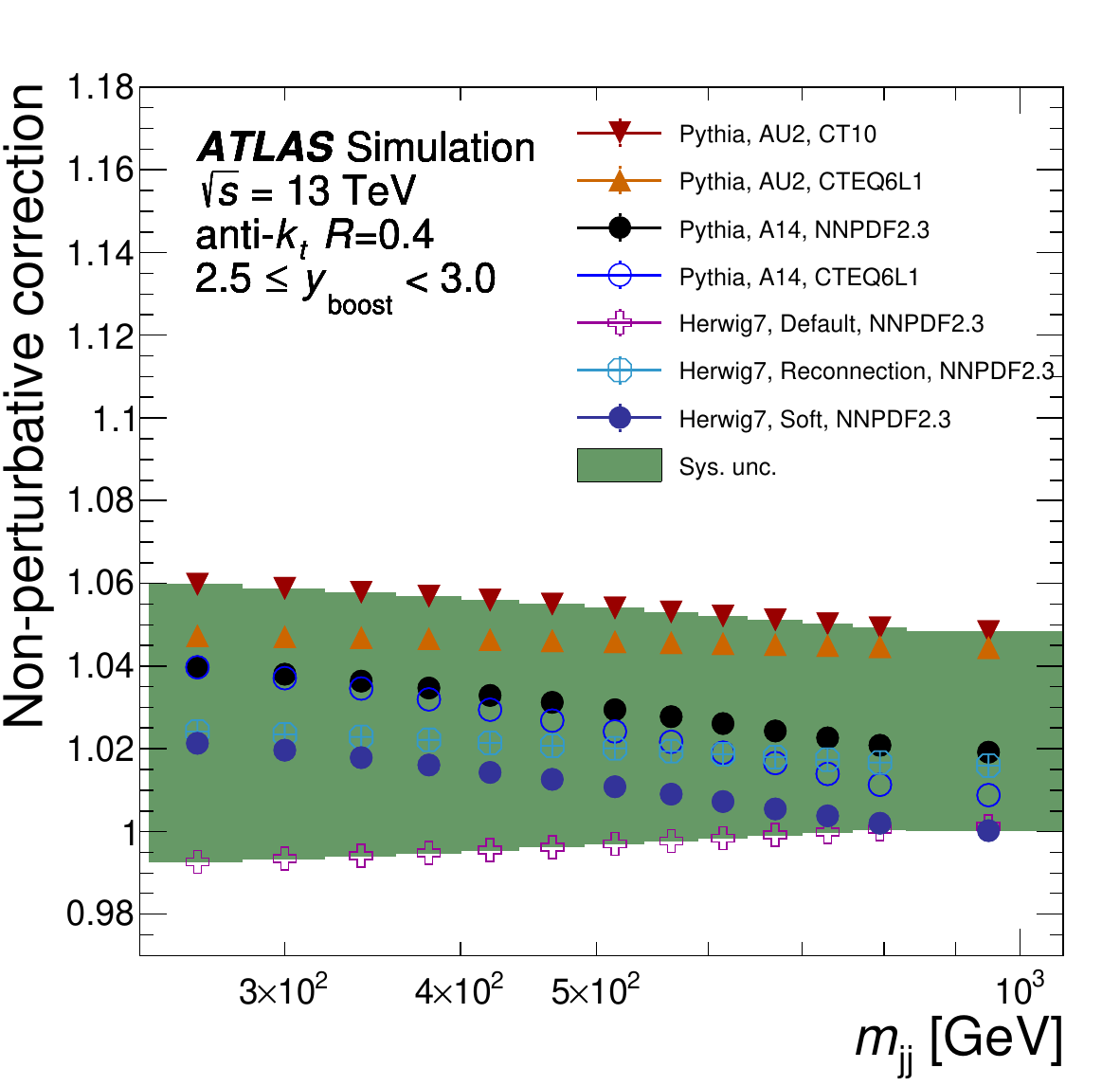}}
\caption{The non-perturbative correction factors in the first and last \ystar (a,~b) and \yboost bins (c,~d). The nominal non-perturbative correction factors are derived using \PYTHIA[8.210], \NNPDF{2.3} PDF set, and A14 tuned parameters, while the envelope defines the associated uncertainty applied to the theory.}
\label{fig:Theory_NPC}
\end{figure}

The final NP corrections (NPC) are evaluated as the envelope of all sets of the tuned parameters around the nominal prediction from \PYTHIA[8.210] with \NNPDF{2.3} and the A14 tuned parameters. The nominal central values are applied as multiplicative correction factors to the theory, while the envelope defines an additional systematic uncertainty. The smoothed out representative NPC factors are shown in~\Cref{fig:Theory_NPC}.
The NP corrections reach 5\% at low dijet mass region and decrease at high dijet mass in all \ystar bins. In contrast, the NPC factors stand approximately constant as a function of \mjj in the lower \yboost bins, whereas a decreasing trend is observed in high \yboost bins.

The pQCD predictions are subsequently corrected for electroweak (EW) effects calculated for $2\rightarrow 2$ processes up to NLO order using a tree and one-loop correction factors:
\begin{equation}
\text{EW corrections} = 1 + \delta^\text{tree} + \delta^\text{1-loop}.
\end{equation}
The individual terms $\delta^\text{tree}$ and $\delta^\text{1-loop}$ are defined as ratios relative to the LO pQCD cross-section. The contribution proportional to $\delta^\text{tree}$ accounts for tree-level effects of order $\alpha^2$ (pure EW), and $\alphas\alpha$ (QCD–EW interference).
The $\delta^\text{1-loop}$ term includes one-loop virtual corrections of order $\alphas^2\alpha$ at NLO\@. These correction factors, computed within the same phase space as the measurement, are provided by the authors of Ref.~\cite{Dittmaier_2012}. No uncertainty associated with these corrections is presently estimated.

The electroweak corrections are shown in \Cref{fig:EW:Theory_EW_Corrections}. Their impact is negligible at low \mjj but increases with dijet mass, reaching up to $+10\%$ in the lowest \ystar bins, then it suddenly drops up to $-5\%$ with increasing \ystar. In the \yboost bins, the corrections reach up to $-3\%$ in the high \mjj region.
\begin{figure}[htbp]
\centering
\subfloat[]{\includegraphics[width=0.49\textwidth]{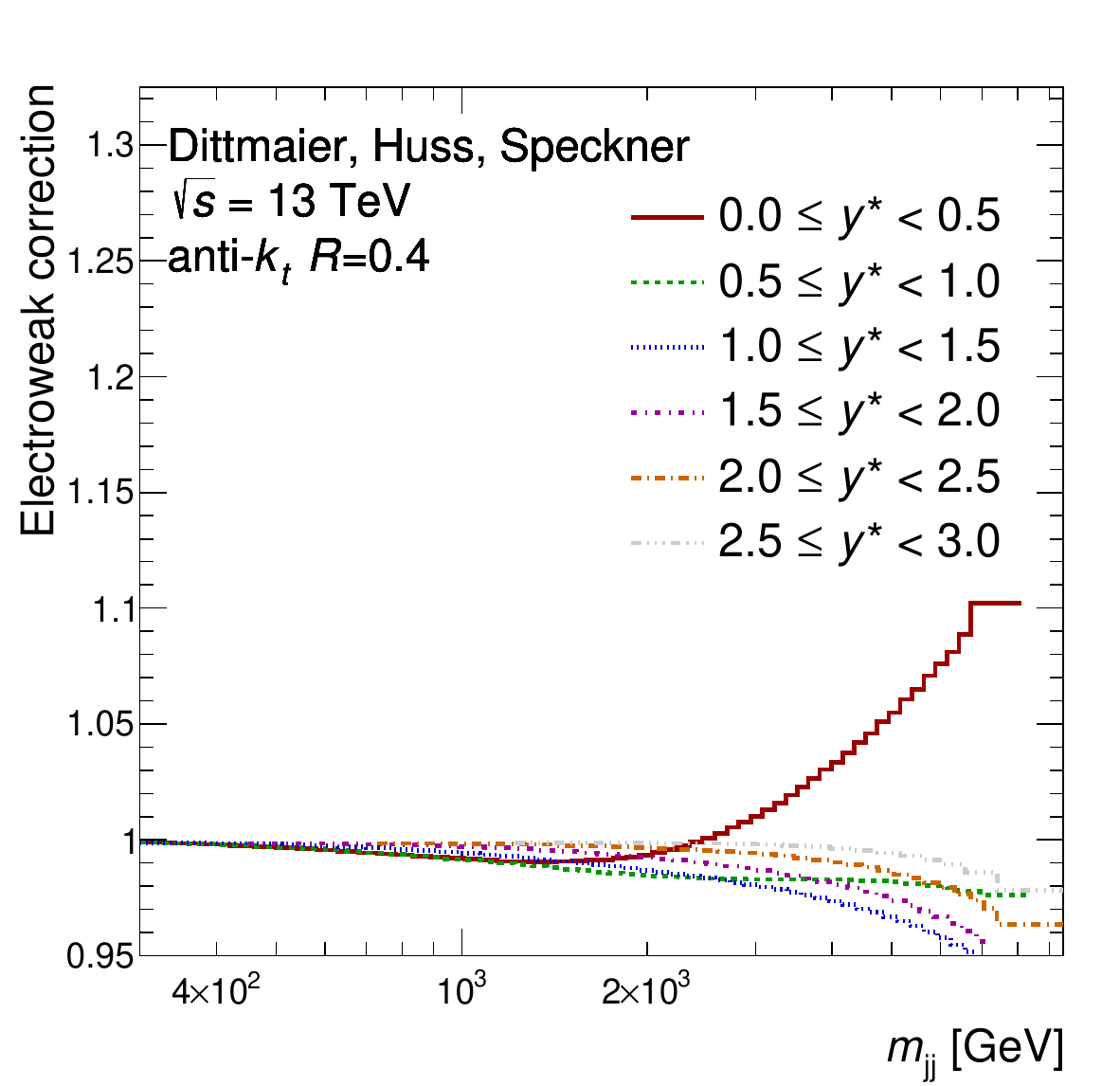}}
\subfloat[]{\includegraphics[width=0.49\textwidth]{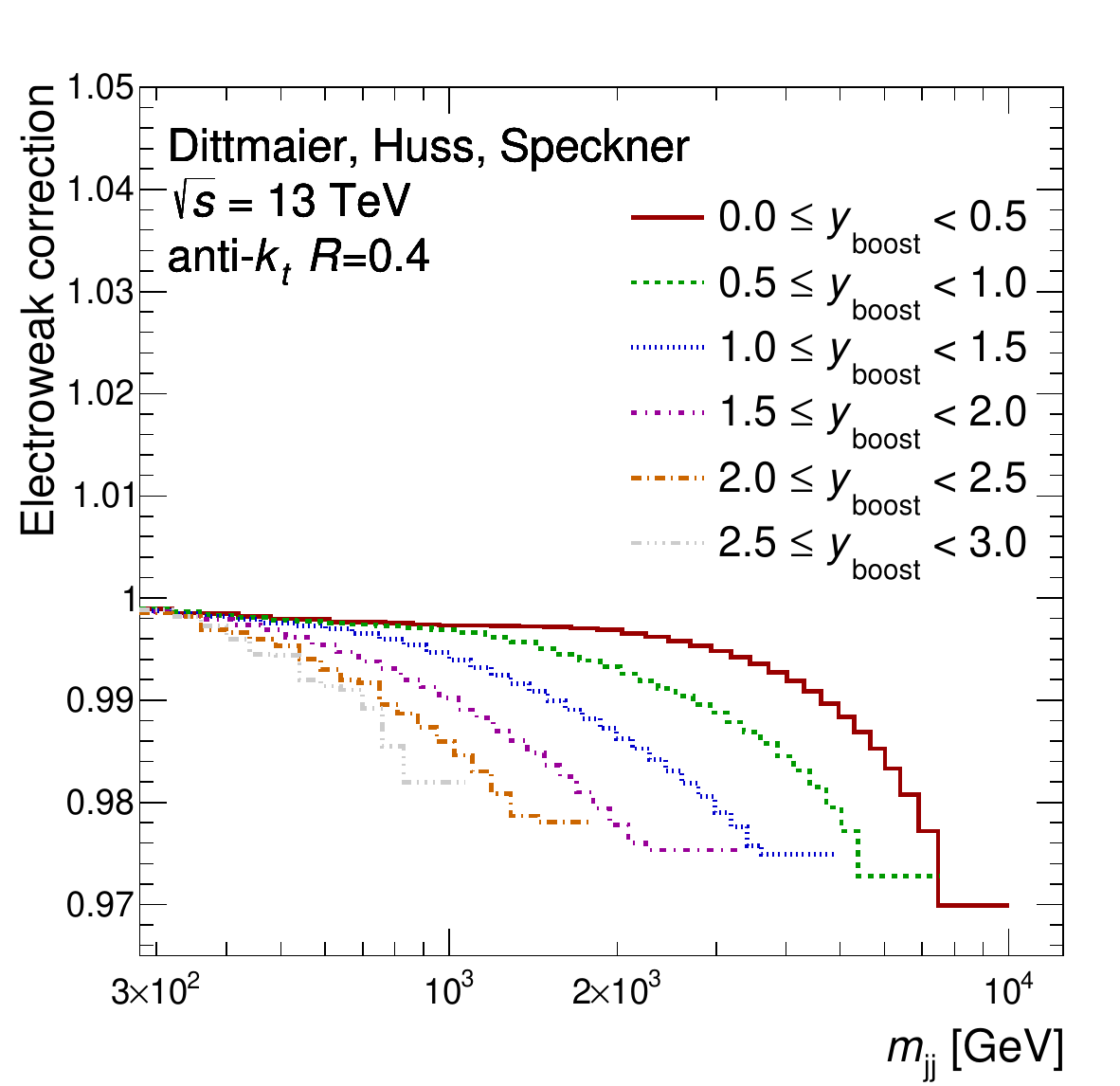}}
\caption{Electroweak correction factors for the dijet double-differential cross-section as a function of the invariant dijet mass \mjj in (a) six \ystar bins and (b) six \yboost bins. The calculation follows the methodology of Ref.~\cite{Dittmaier_2012}.}
\label{fig:EW:Theory_EW_Corrections}
\end{figure}

A strong dependence of the electroweak corrections is observed in the (\mjj, \ystar) phase space, particularly across the transition between the first and second \ystar bins. The competition between the two contributions $\delta^\text{tree}$ and $\delta^\text{1-loop}$ drives this behaviour: the positive and rapidly decreasing tree-level term, $\delta^\text{tree}$, and the negative and slowly decreasing one-loop level term, $\delta^\text{1-loop}$, which arises primarily from negative Sudakov logarithms. In the central region ($\ystar < 0.5$), the dominant $\delta^\text{tree}$ contributions originate from $qq$-initiated tree-level processes (most notably from partonic sub-processes such as $uu \rightarrow uu$ and $ud \rightarrow ud$) in the interference between QCD and EW at order $\mathcal{O}(\alpha_s \alpha)$,
where only the interference terms of the $t$-, $u$-, and $s$-channels contribute, while the squared topologies vanish due to the colour structure~\cite{Dittmaier_2012}. Notably, this interference is more central in rapidity than the dominant QCD contributions, which mainly originate from squared topologies of Mandelstam variables $u^2$, $t^2$, and $s^2$.

\FloatBarrier


\graphicspath{{figures/Results/}{fig/}}

\FloatBarrier

\section{Results}
\label{sec:Results}
The unfolded double-differential inclusive dijet cross-sections are presented as functions of $(\mjj, \ystar)$ and $(\mjj, \yboost)$ and compared with theoretical predictions based on pQCD, including electroweak and non-perturbative corrections. The measurements benefit from the full Run-2 data sample of $140~\ifb$, providing unprecedented statistics and substantially reduced statistical uncertainties with respect to the previous measurement. The improved jet energy calibration has also been implemented, reducing the jet energy scale uncertainty by up to a factor of three for central dijet configurations ($\ystar < 0.5$) with $\mjj > 1$~\TeV. These improvements lead to a smaller total experimental uncertainty and enable more precise comparisons with theoretical predictions. The measured kinematic region covers dijets with $\ystar, \yboost$ between $0.0$ and $3.0$ and $\mjj$ between $240$~\GeV\ and nearly $10$~\TeV, with the highest observed dijet mass reaching $9.54$~\TeV. Representative comparisons between data and theory are shown in \Cref{fig:MoneyPlot_mjj_yStar_NNLO_CT18__MJJ_SCALE,fig:MoneyPlot_mjj_yBoost_NNLO_CT18__MJJ_SCALE}, based on nominal \NNLO\ pQCD predictions using the \CT{18NNLO} PDF set.

The theory-to-data comparisons using \NLO and \NNLO matrix elements of pQCD calculations with the nominal \CT{18} PDF set are presented in \Cref{fig:DataToTheory_mjj_yStar_NNLO_CT18__MJJ_SCALE,fig:DataToTheory_mjj_yBoost_NNLO_CT18__MJJ_SCALE}. The \NLO matrix elements are coupled to the \NLO PDF set, while the \NNLO matrix elements are coupled to the \NNLO PDF\@.  Both \NLO and \NNLO predictions overestimate the data on average by 15--20\%.

At low \ystar ($\ystar < 1.5$), the theoretical predictions overestimate the data, with the theory to data ratio gradually decreasing as $\mjj$ increases. At higher $\ystar$ values, the differences become more pronounced, showing a clear rise in the theory to data ratio with increasing \mjj. A similar trend is observed across the \yboost bins, where the agreement progressively worsens toward larger \yboost and large \mjj.
For the (\mjj,~\ystar) measurement, the \NNLO predictions provide an improved description of the shape at low \ystar, which contains the majority of events in the phase space. At higher \ystar values, however, the \NNLO predictions remain consistently above the data, similar to the \NLO predictions. In the (\mjj, \yboost) measurement, a better agreement is also observed at low \mjj across all \yboost bins. Furthermore, the \NNLO predictions lead to an improved agreement with the data for all \mjj bins at higher \yboost values above $1.5$. This behaviour is expected since each \yboost bin is dominated by contributions from low-\ystar events, as illustrated in \Cref{fig:crossSection:dijetTopology}.

These observations might be attributed to the limitations of fixed-order pQCD calculations, including the residual dependence on the unphysical renormalisation and factorisation scales, which affect both the shape and normalisation of the predictions. The conventional prescriptions for estimating theory uncertainties—based on seven-point scale variations and PDF uncertainties—do not necessarily capture the differences between central values obtained with all possible scale choices or between different PDF sets~\cite{currie2018jetcrosssectionslhc,Currie:2018xkj}. Additional limitations may arise from the absence of resummation effects, which are still under development~\cite{generet2025smallradiusinclusivejet}.
Another difference may arise from the choice of jet radius $R$. Small-$R$ jets, including the ATLAS standard jets with $R=0.4$, might have an additional perturbative uncertainty\footnote{For small jet radii, perturbative emissions outside the jet lead to a logarithmic $R$-dependent shift in the jet momentum,
$\frac{\langle \delta p_T\rangle}{p_T} \sim \frac{\alpha_s}{\pi} L_i \ln(R)$, where $L_i$ encodes a dependence on the colour factor of the initiating parton. For $R<1.0$, the effect increases in magnitude with decreasing $R$~\cite{Mrinal:Dasgupta_2008}.}~\cite{Mrinal:Dasgupta_2008}. Larger radii might also be preferred at LHC energies, where gluon-initiated jets dominate.

The complementary comparison for the alternative PDF sets (\CT{18}, \NNPDF{3.1}, \NNPDF{4.0}, \MSHT, and \ATLASPDF{21T3} using fixed \NNLO matrix elements) are available in \Cref{fig:DataToTheory_pdf1_mjj_yStar__MJJ_SCALE,fig:DataToTheory_pdf1_mjj_yBoost__MJJ_SCALE,fig:DataToTheory_pdf2_mjj_yStar__MJJ_SCALE,fig:DataToTheory_pdf2_mjj_yBoost__MJJ_SCALE} showing an almost constant offset between each of them, all being above the unfolded data.
Focusing on the central values, the \NNPDF{4.0} PDF set shows the best overall agreement with the unfolded data—even in the high-\mjj region with a boosted topology (large \yboost), where most predictions tend to overestimate the data more steeply due to resummation effects. Since the \NNPDF{4.0} set provides significantly smaller PDF uncertainties—up to 2\% level precision—it may lead to higher $\chi^2$ values compared with other PDF sets, whose broader uncertainty bands can result in better quantitative agreement to data.

\begin{figure}[htbp]
\centering
\includegraphics[width=0.99\textwidth]{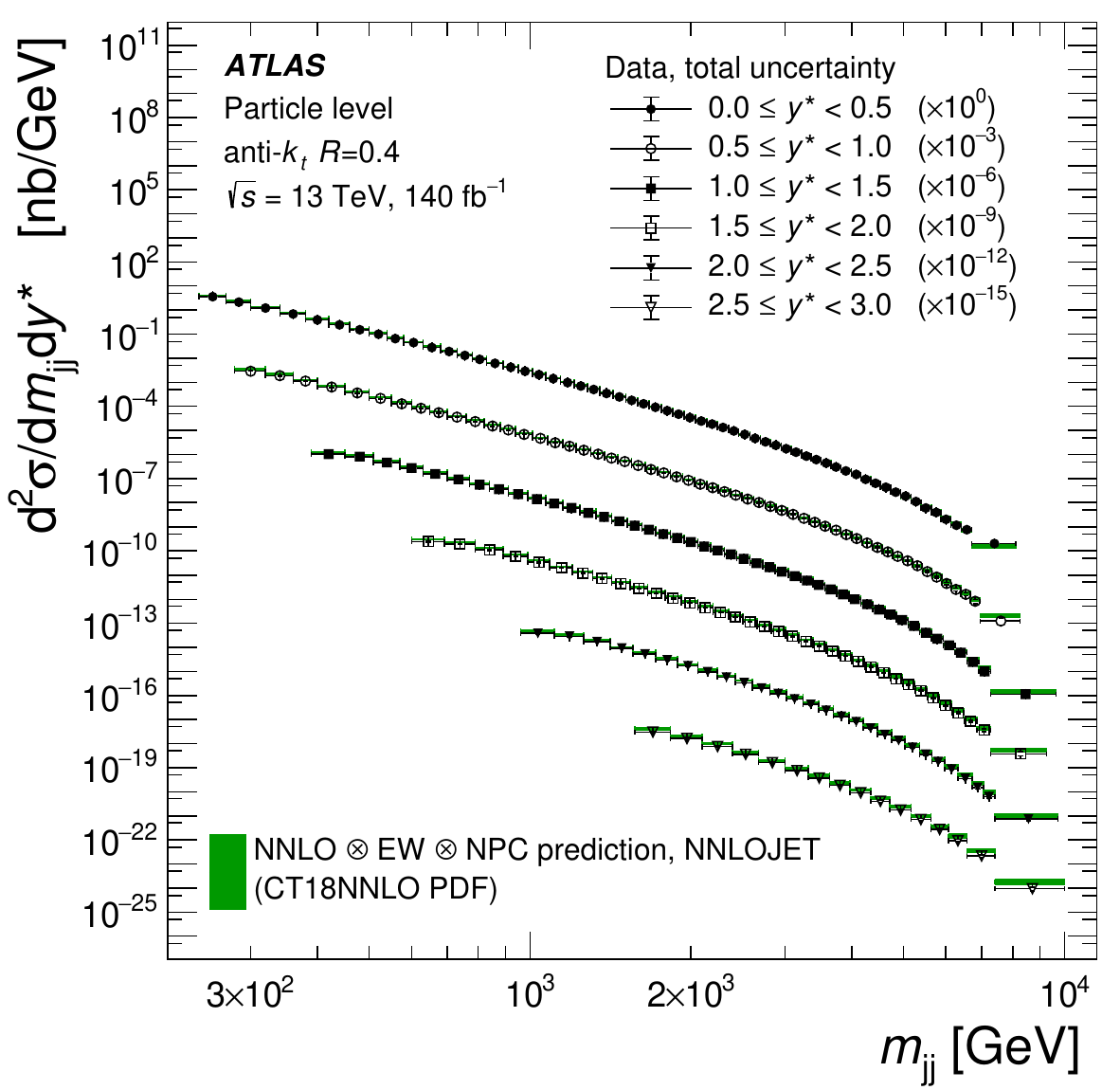}
\caption{Double-differential cross-section as a function of the invariant dijet mass \mjj in six equidistant \ystar bins. Each distribution is scaled by an additional factor $10^{-n}$, indicated in the legend, to improve visibility. The figure compares the ATLAS unfolded data with \NNLO predictions obtained using the \CT{18} PDF set, with electroweak and non-perturbative corrections applied. Coloured bands represent the total theoretical uncertainties, while vertical error bars show the experimental uncertainties.}
\label{fig:MoneyPlot_mjj_yStar_NNLO_CT18__MJJ_SCALE}
\end{figure}

\begin{figure}[htbp]
\centering
\includegraphics[width=0.99\textwidth]{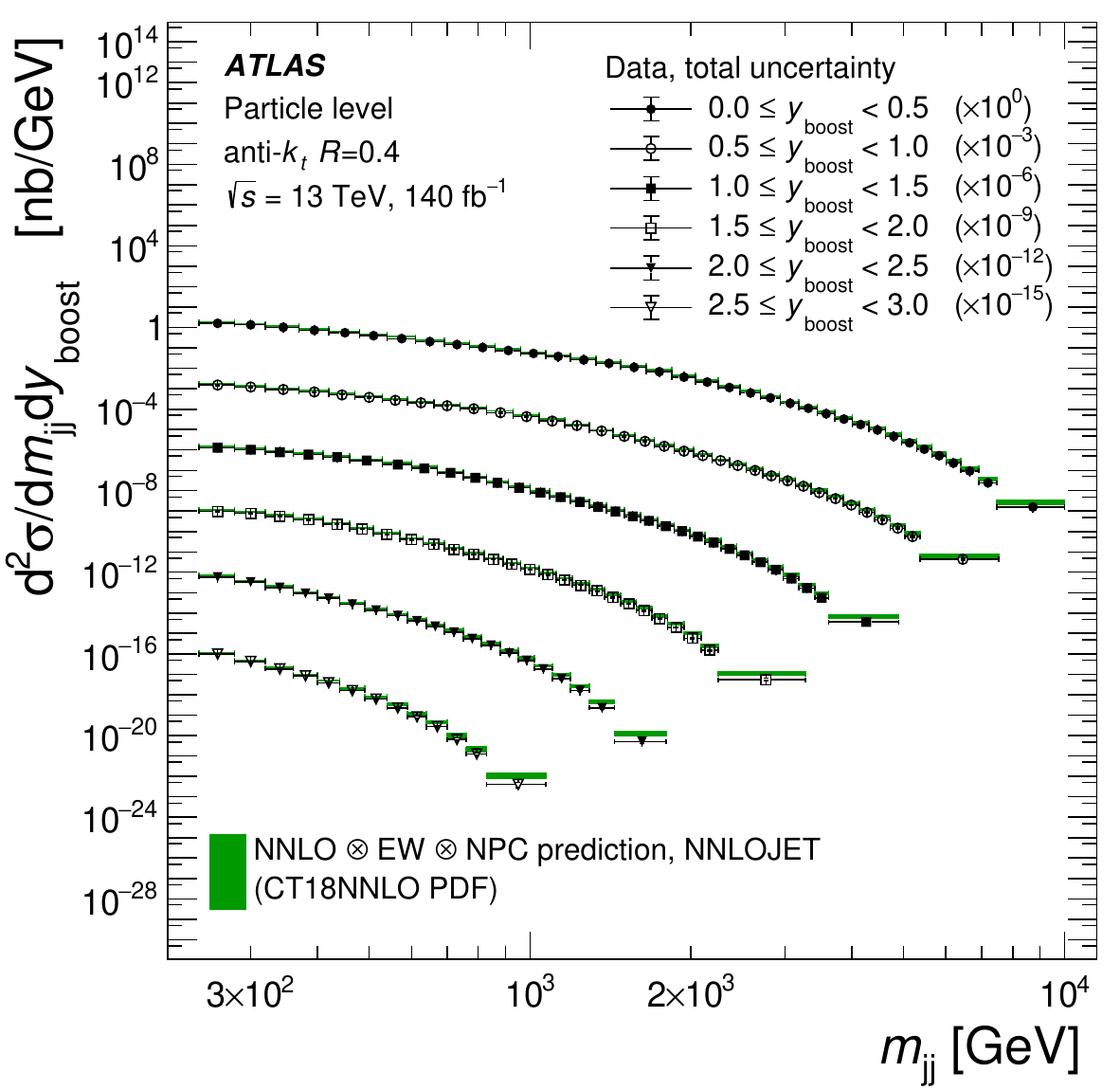}
\caption{Double-differential cross-section as a function of the invariant dijet mass \mjj in six equidistant \yboost bins. Each distribution is scaled by an additional factor $10^{-n}$, indicated in the legend, to improve visibility. The figure compares the ATLAS unfolded data with \NNLO predictions obtained using the \CT{18} PDF set, with electroweak and non-perturbative corrections applied. Coloured bands represent the total theoretical uncertainties, while vertical error bars show the experimental uncertainties.}
\label{fig:MoneyPlot_mjj_yBoost_NNLO_CT18__MJJ_SCALE}
\end{figure}

\FloatBarrier
\begin{figure}[htbp]
\centering
\begin{tabular}{lll}
\includegraphics[width=0.31\textwidth]{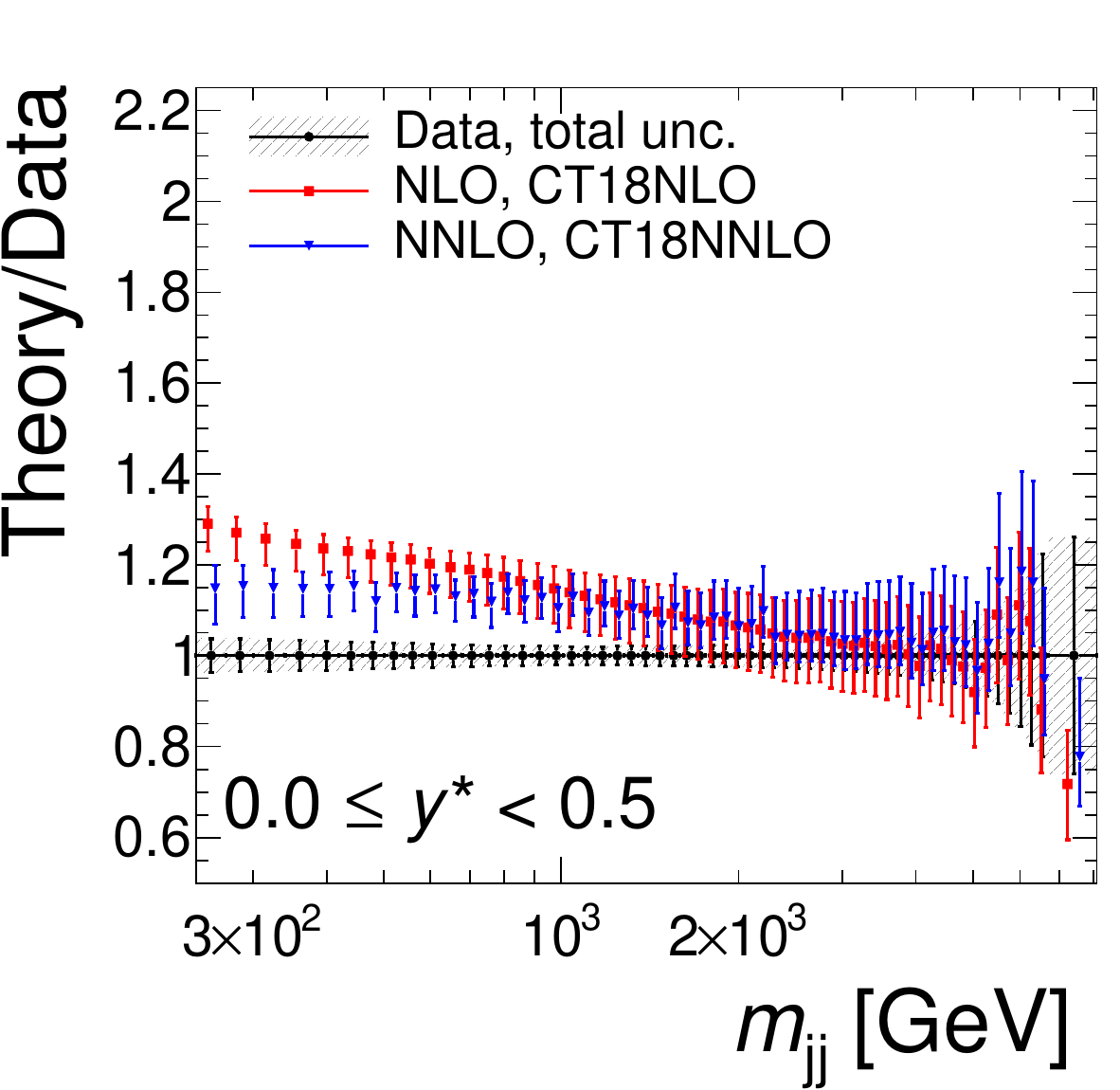} &
\includegraphics[width=0.31\textwidth]{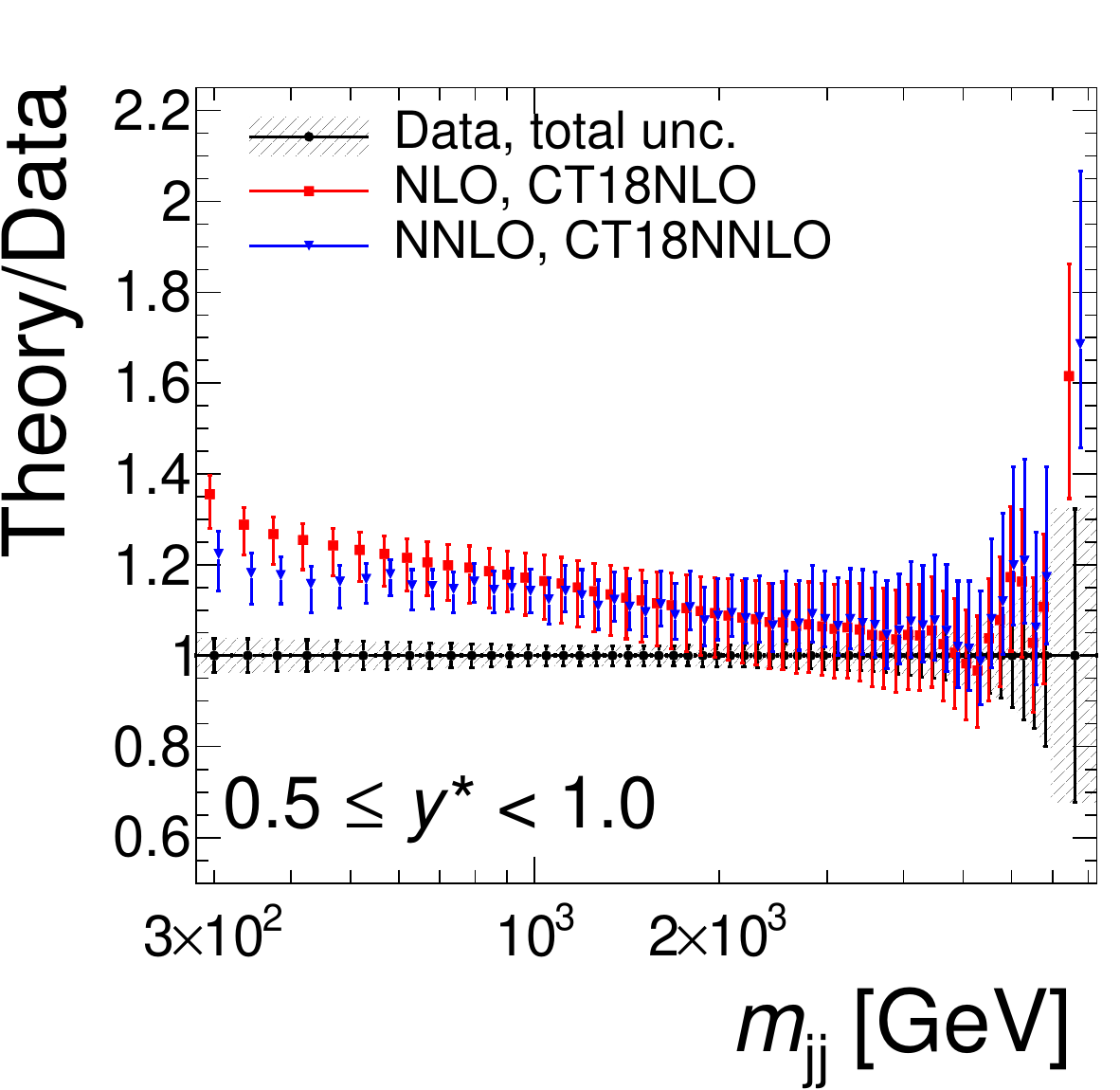} &
\includegraphics[width=0.31\textwidth]{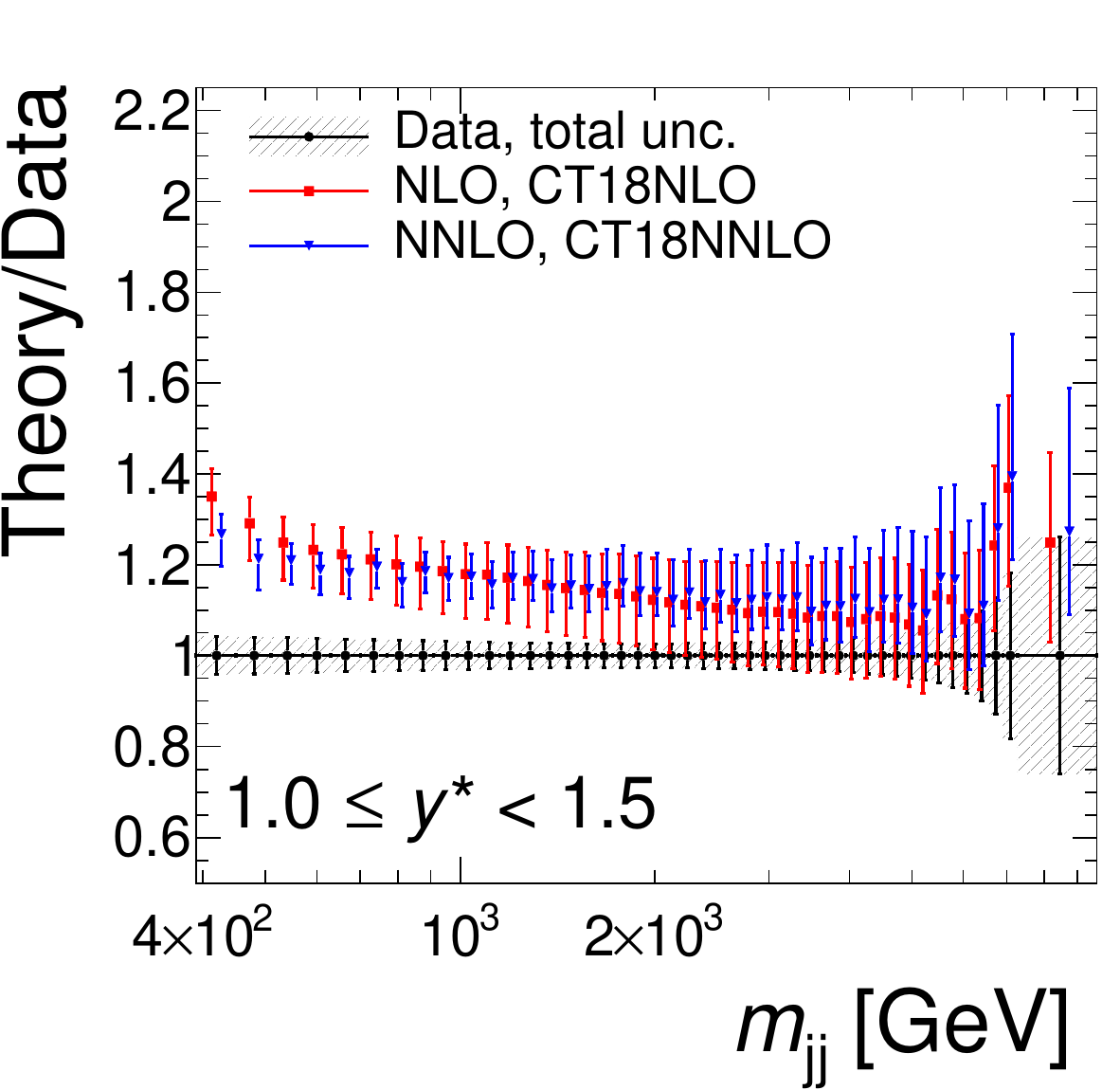} \\
\includegraphics[width=0.31\textwidth]{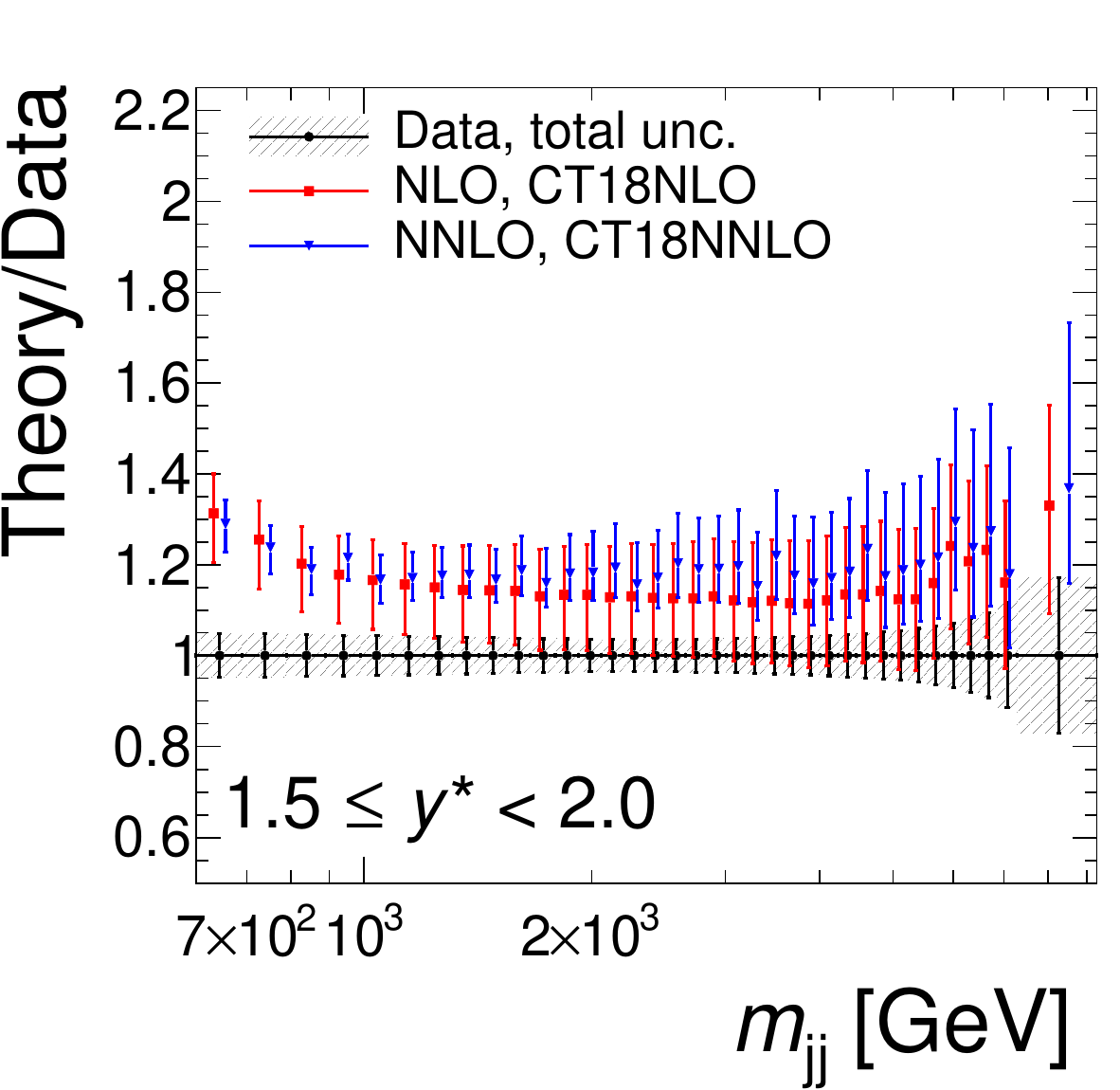} &
\includegraphics[width=0.31\textwidth]{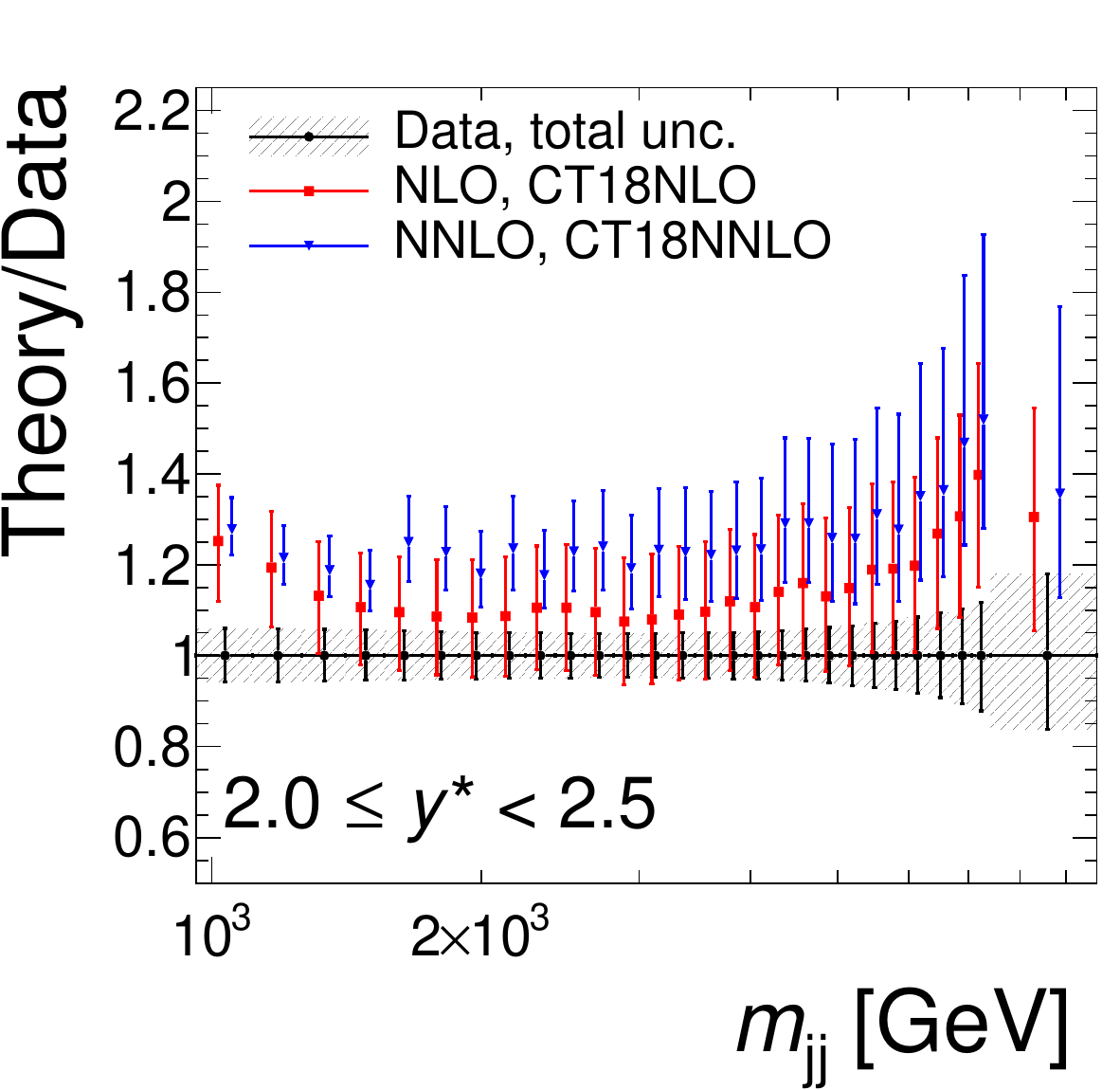} &
\includegraphics[width=0.31\textwidth]{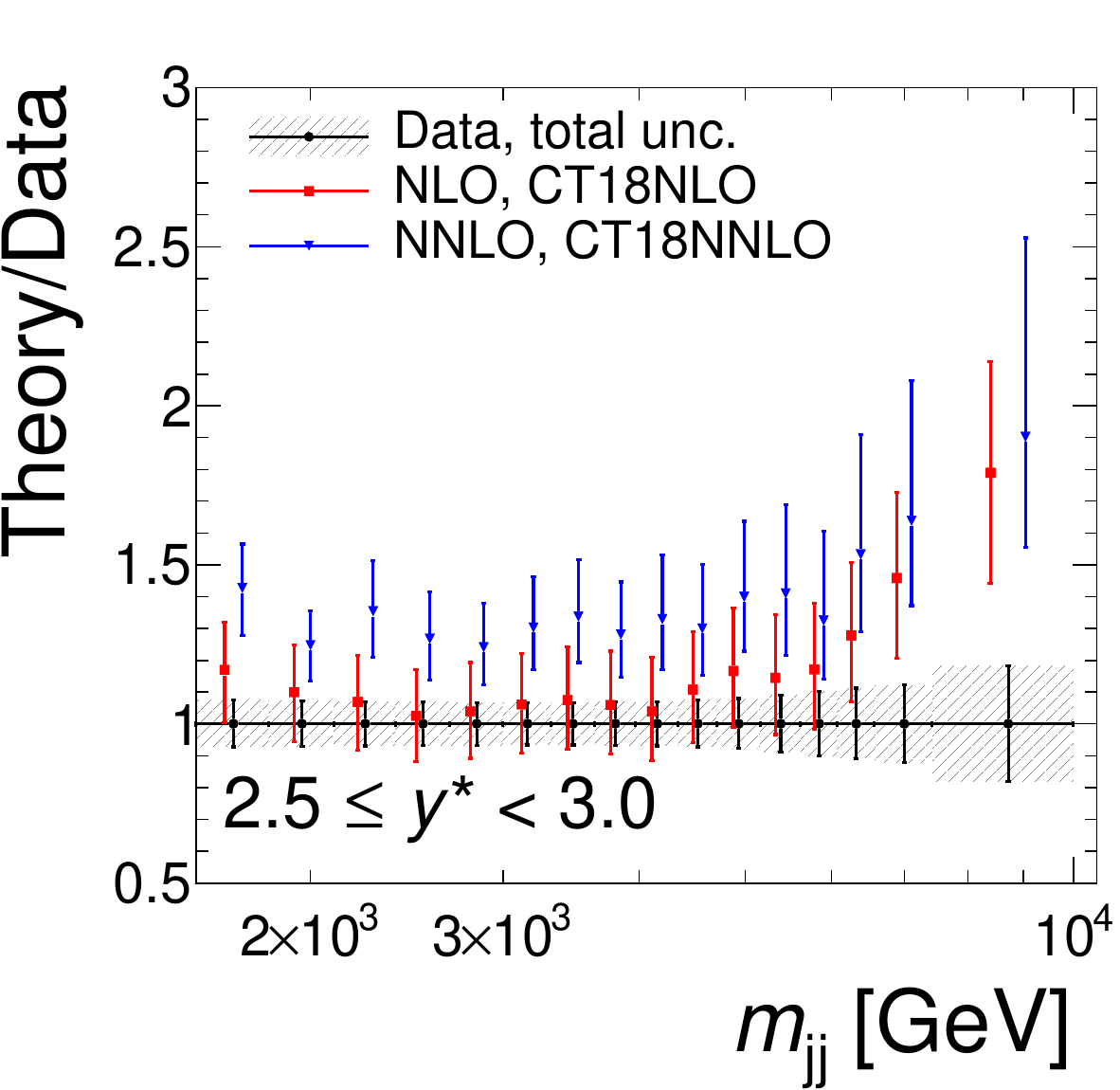} \\
\multicolumn{3}{c}{\includegraphics[width=1.0\textwidth]{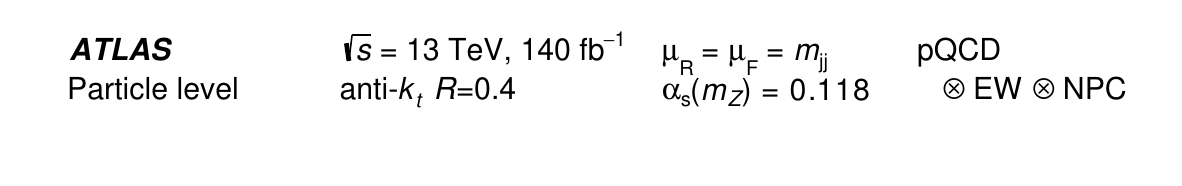}}  \\
\end{tabular}
\caption{   Theory-to-data ratio for the double-differential cross-section as a function of the invariant dijet mass \mjj in six equidistant \ystar bins for
\antikt $R=0.4$ jets at the particle level.
The sub-figures show different orders (\NLO and \NNLO) of pQCD full colour predictions via the NNLOJET programme at the parton level using the \CT{18} PDF set and renormalisation and factorisation scales of $\muRmuF = \mjj$.
The pQCD calculations are corrected by electroweak and non-perturbative corrections to obtain the complete predictions at the particle level.
The theory is compared with the data unfolded to the particle level using the IDS method with one iteration.
Coloured vertical line errors indicate the total systematic uncertainties in pQCD calculations, including the effect of non-perturbative corrections.
Vertical band errors around $1.0$ indicate total experimental uncertainty in the data.}
\label{fig:DataToTheory_mjj_yStar_NNLO_CT18__MJJ_SCALE}
\end{figure}

\begin{figure}[tbp]
\centering
\begin{tabular}{lll}
\includegraphics[width=0.31\textwidth]{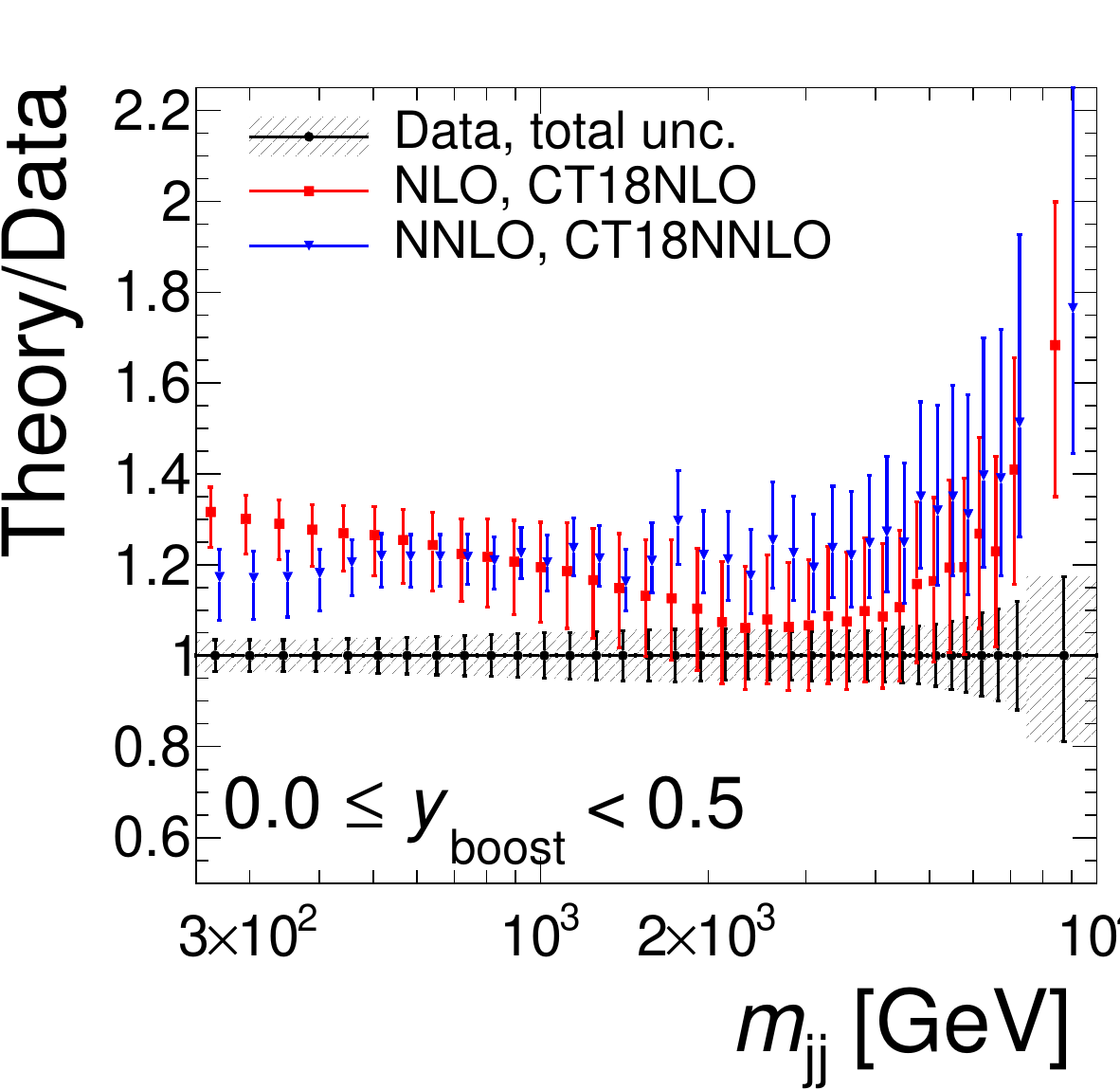} &
\includegraphics[width=0.31\textwidth]{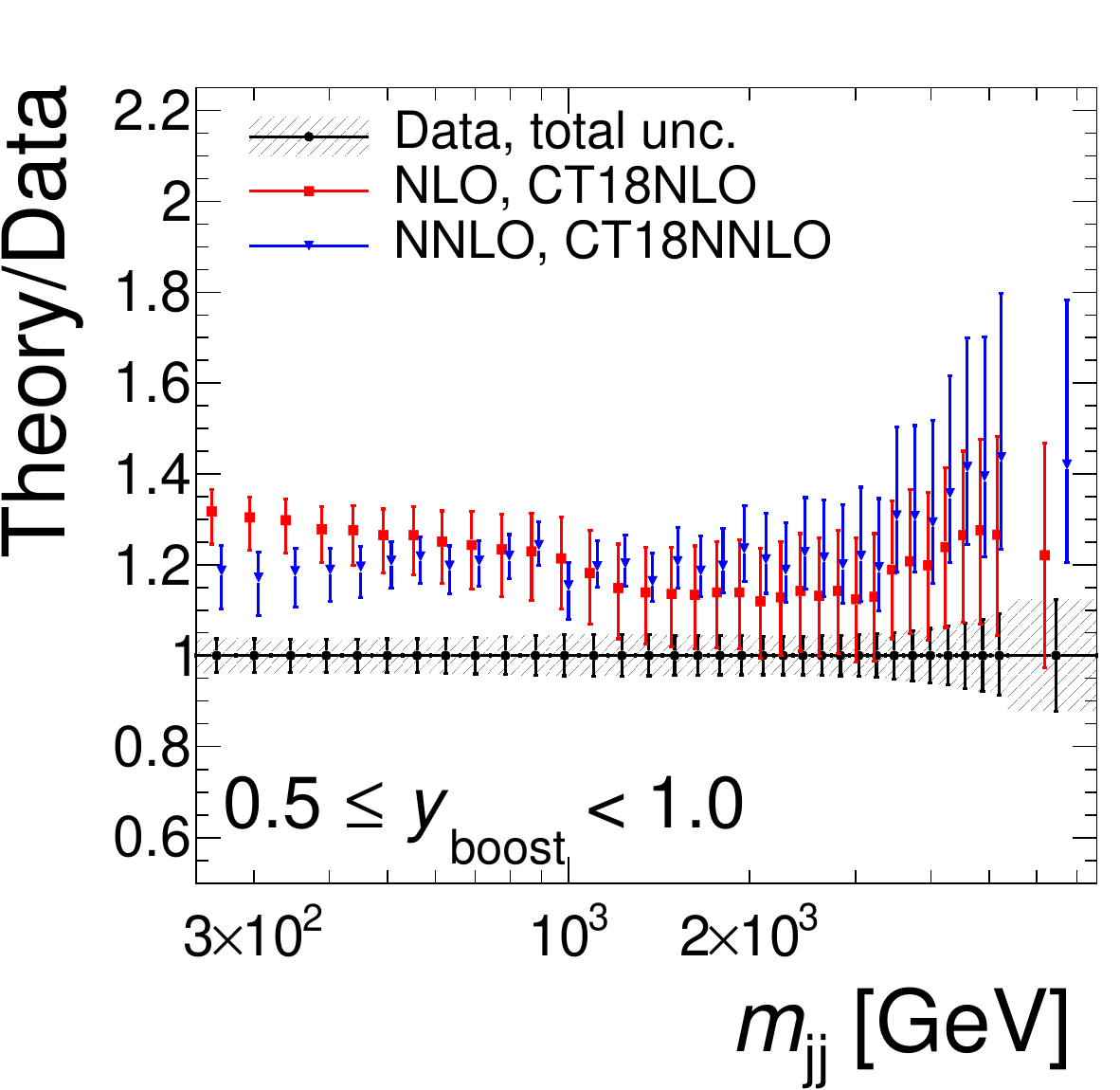} &
\includegraphics[width=0.31\textwidth]{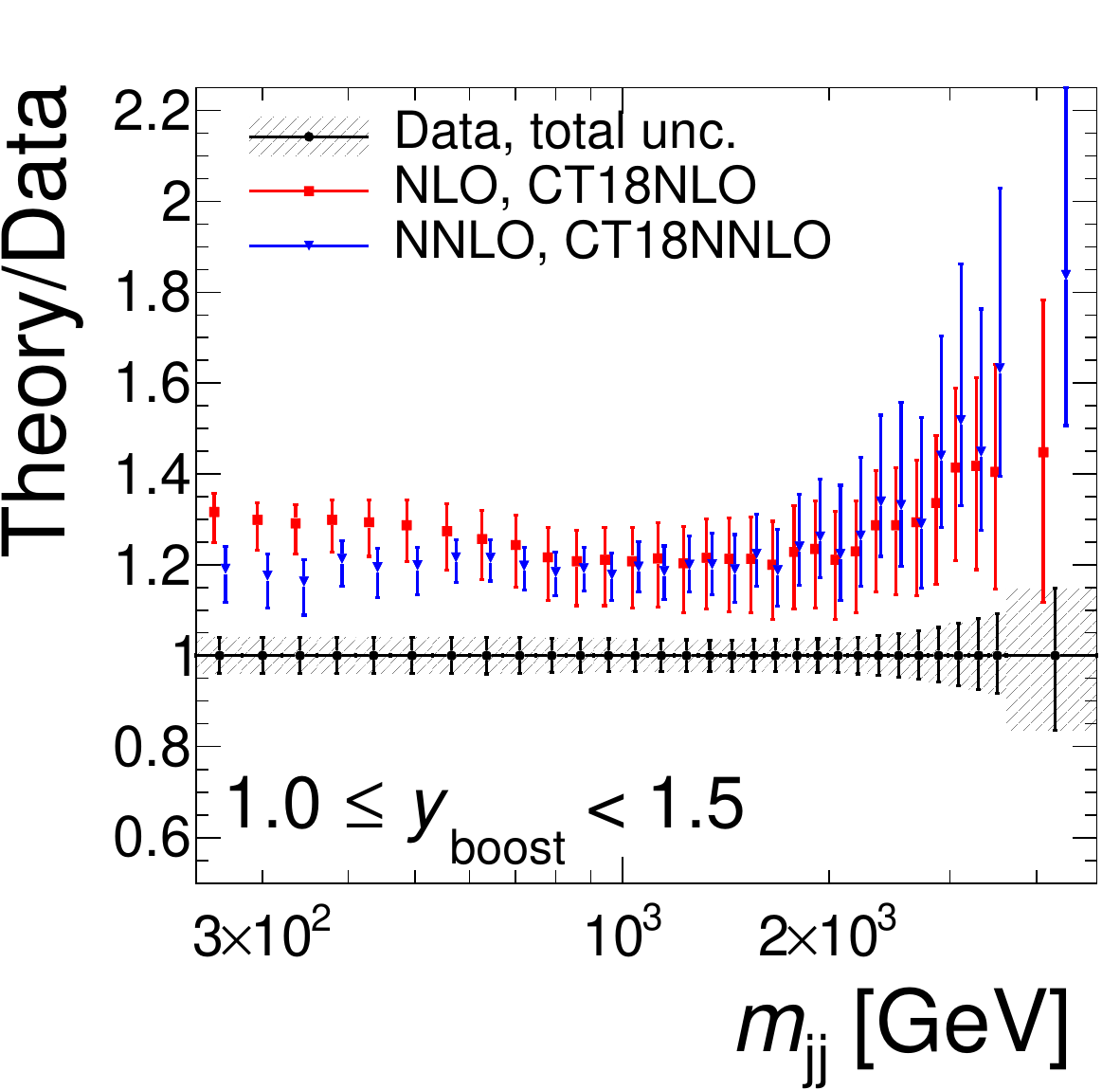} \\
\includegraphics[width=0.31\textwidth]{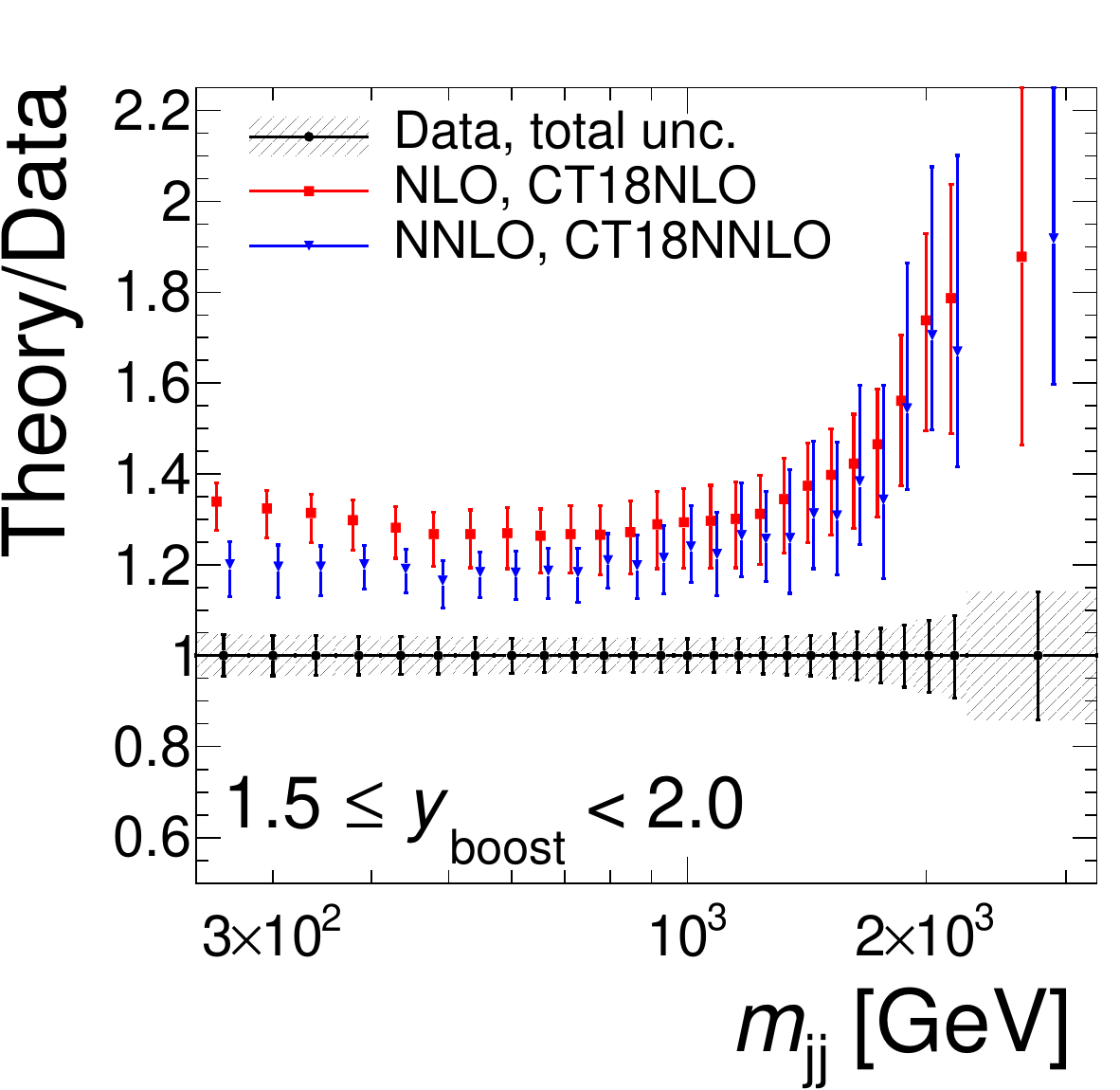} &
\includegraphics[width=0.31\textwidth]{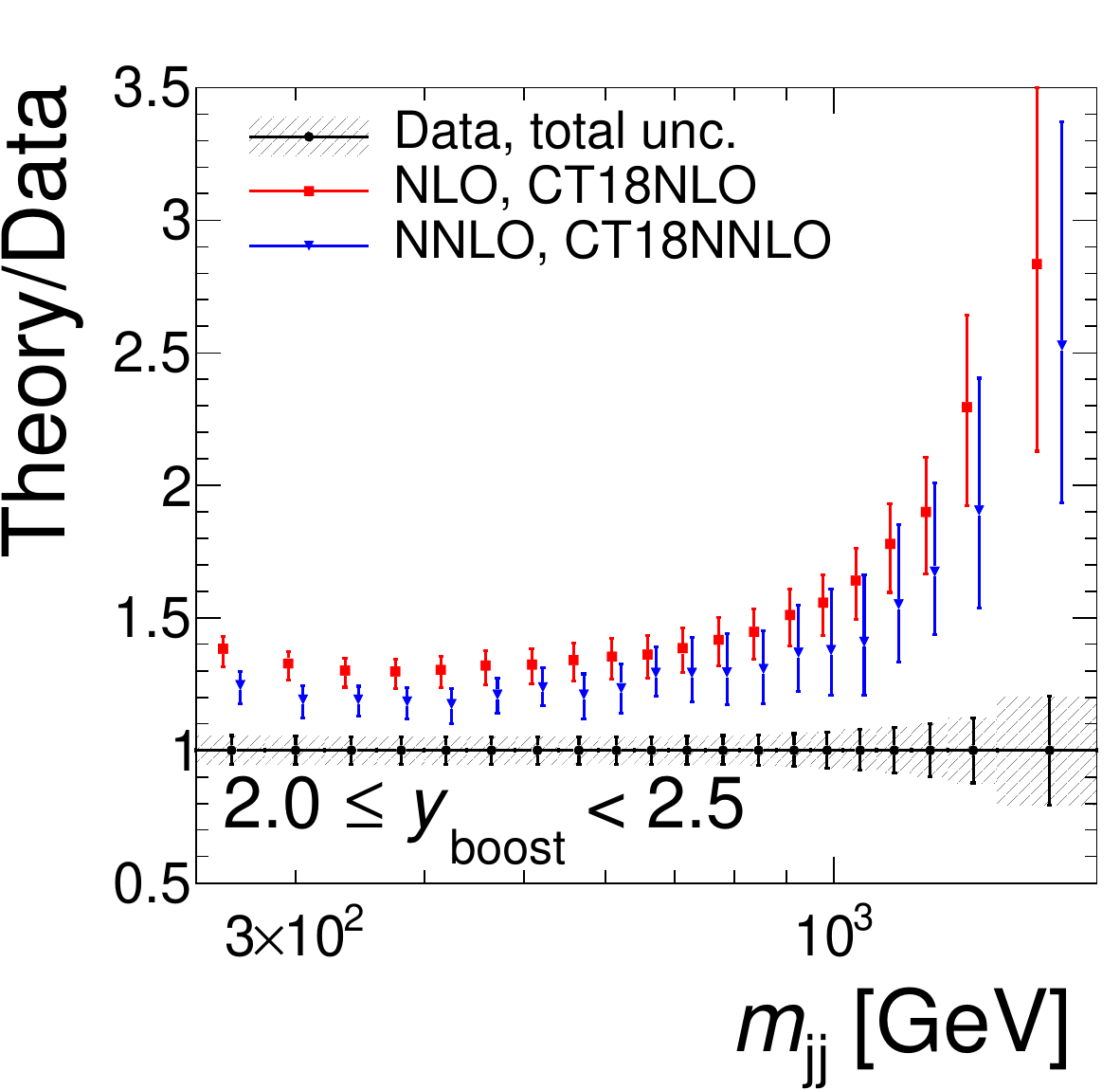} &
\includegraphics[width=0.31\textwidth]{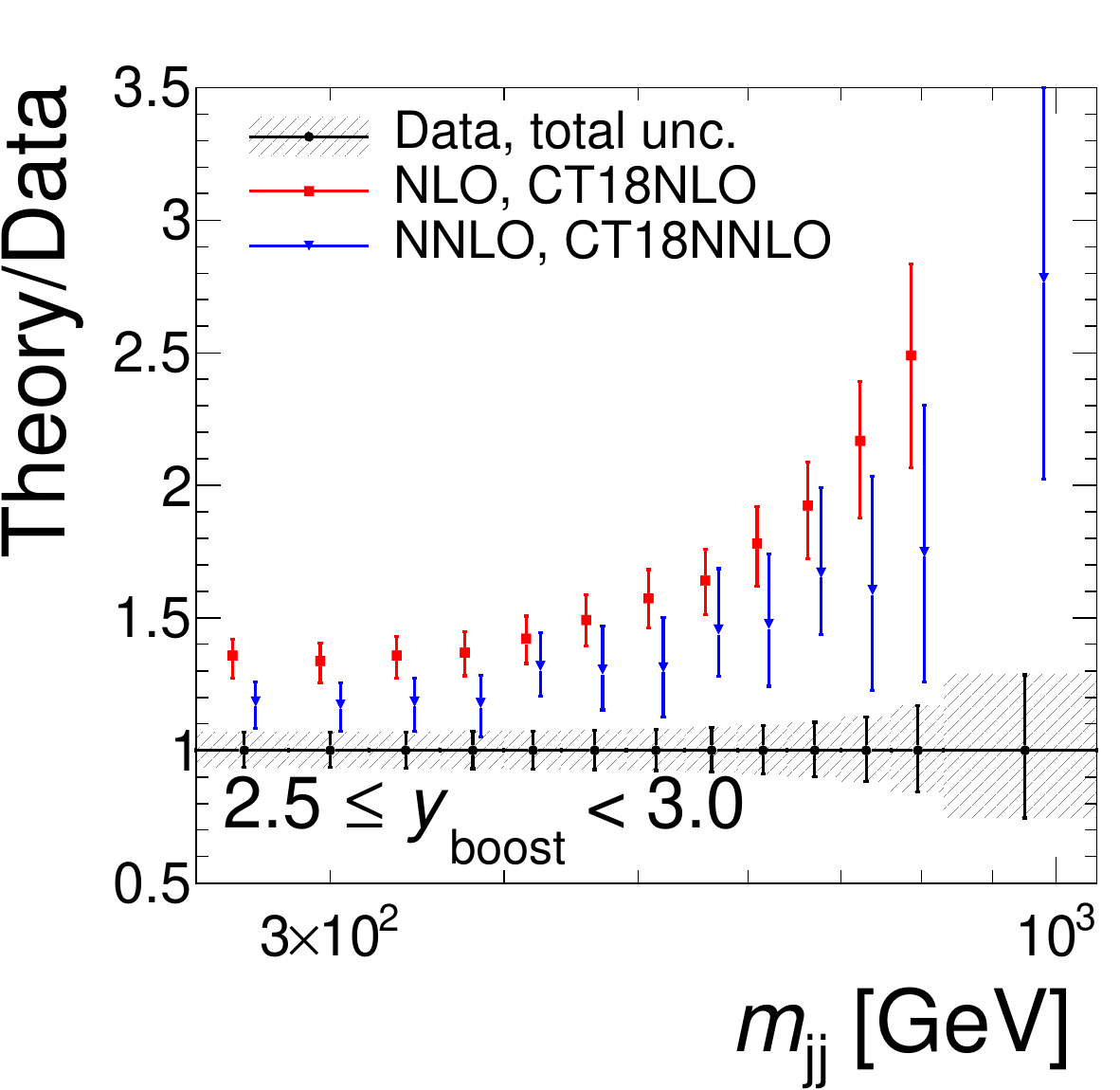} \\
\multicolumn{3}{c}{\includegraphics[width=1.0\textwidth]{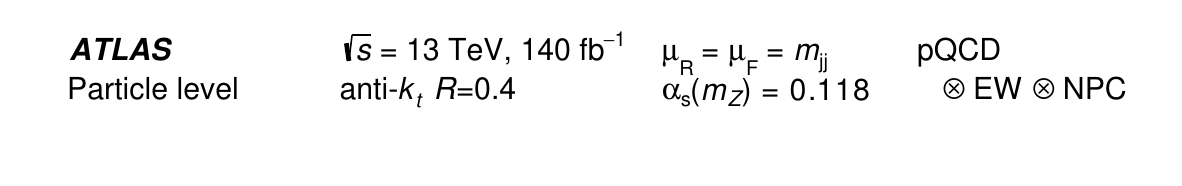}}  \\
\end{tabular}
\caption{   Theory-to-data ratio for the double-differential cross-section as a function of the invariant dijet mass \mjj in six equidistant \yboost bins for \antikt
$R=0.4$ jets at the particle level.
The sub-figures show different orders (\NLO and \NNLO) of pQCD full colour predictions via the NNLOJET programme at the parton level using the \CT{18} PDF set and renormalisation and factorisation scales of $\muRmuF = \mjj$.
The pQCD calculations are corrected by electroweak and non-perturbative corrections to obtain the complete predictions at the particle level.
The theory is compared with the data unfolded to the particle level using the IDS method with two iterations.
Coloured vertical line errors indicate the total systematic uncertainties in pQCD calculations, including the effect of non-perturbative corrections.
Vertical band errors around $1.0$ indicate total experimental uncertainty in the data.}
\label{fig:DataToTheory_mjj_yBoost_NNLO_CT18__MJJ_SCALE}
\end{figure}

\FloatBarrier
\begin{figure}[tbp]
\centering
\begin{tabular}{lll}
\includegraphics[width=0.31\textwidth]{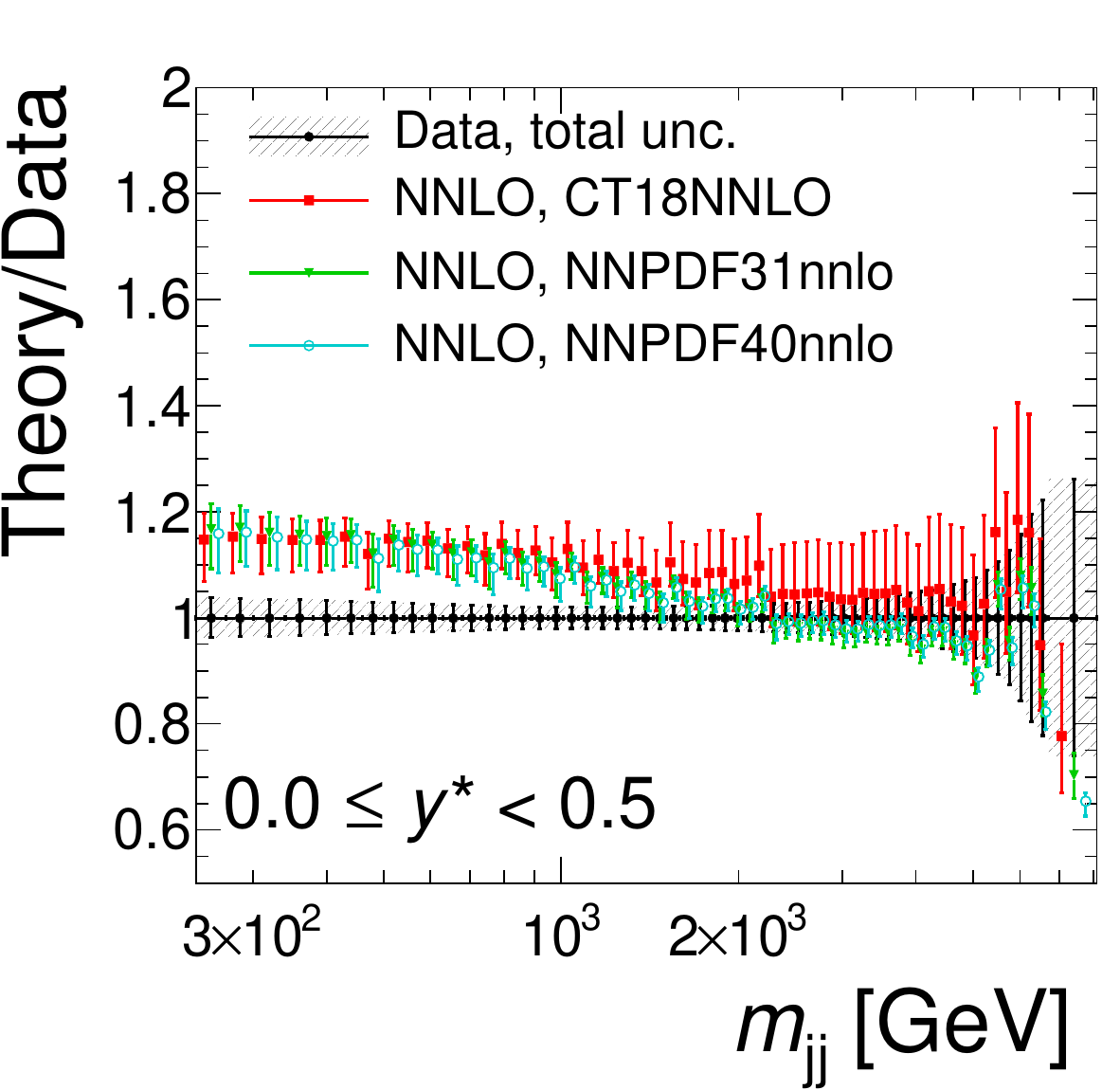} &
\includegraphics[width=0.31\textwidth]{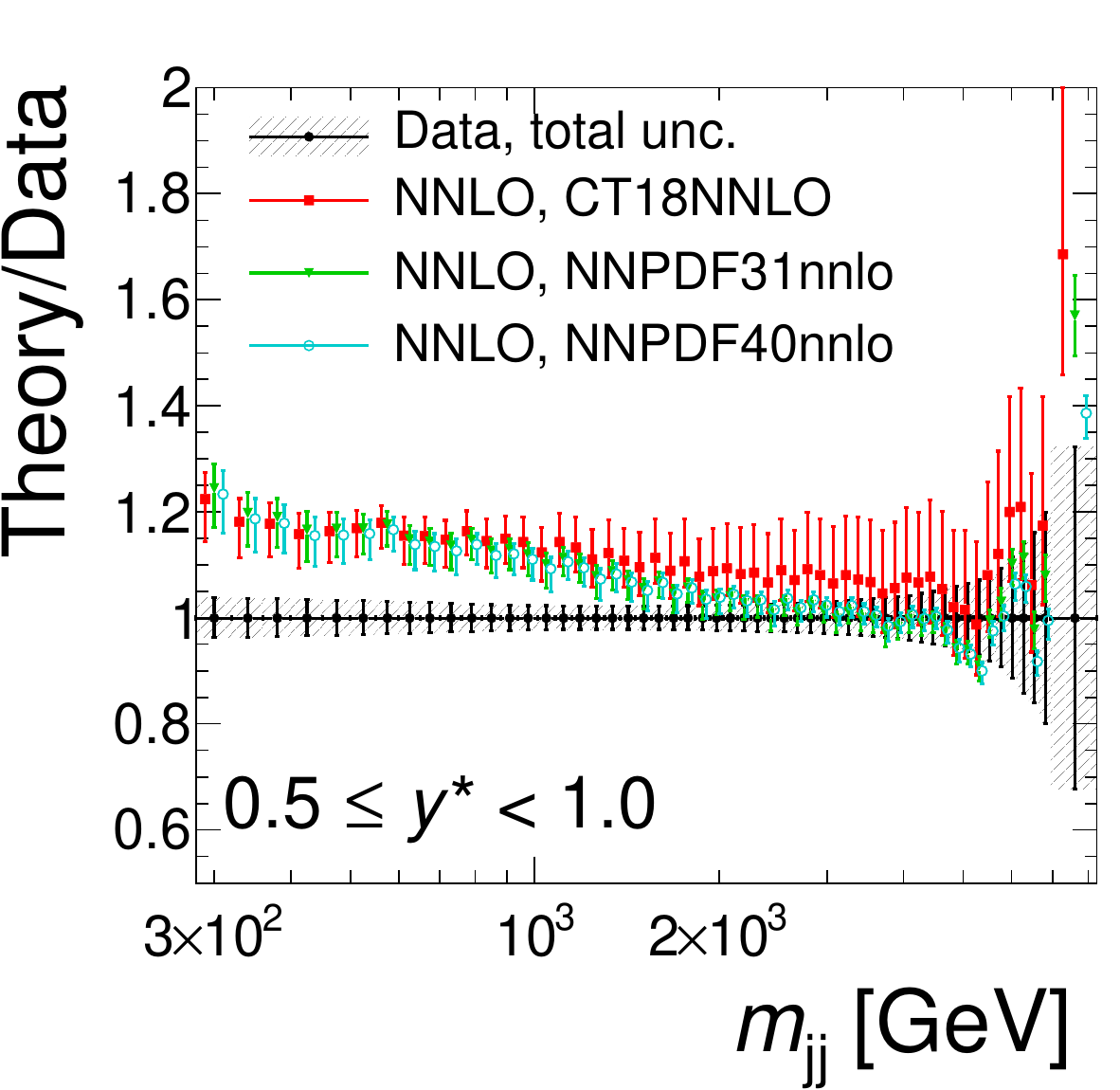} &
\includegraphics[width=0.31\textwidth]{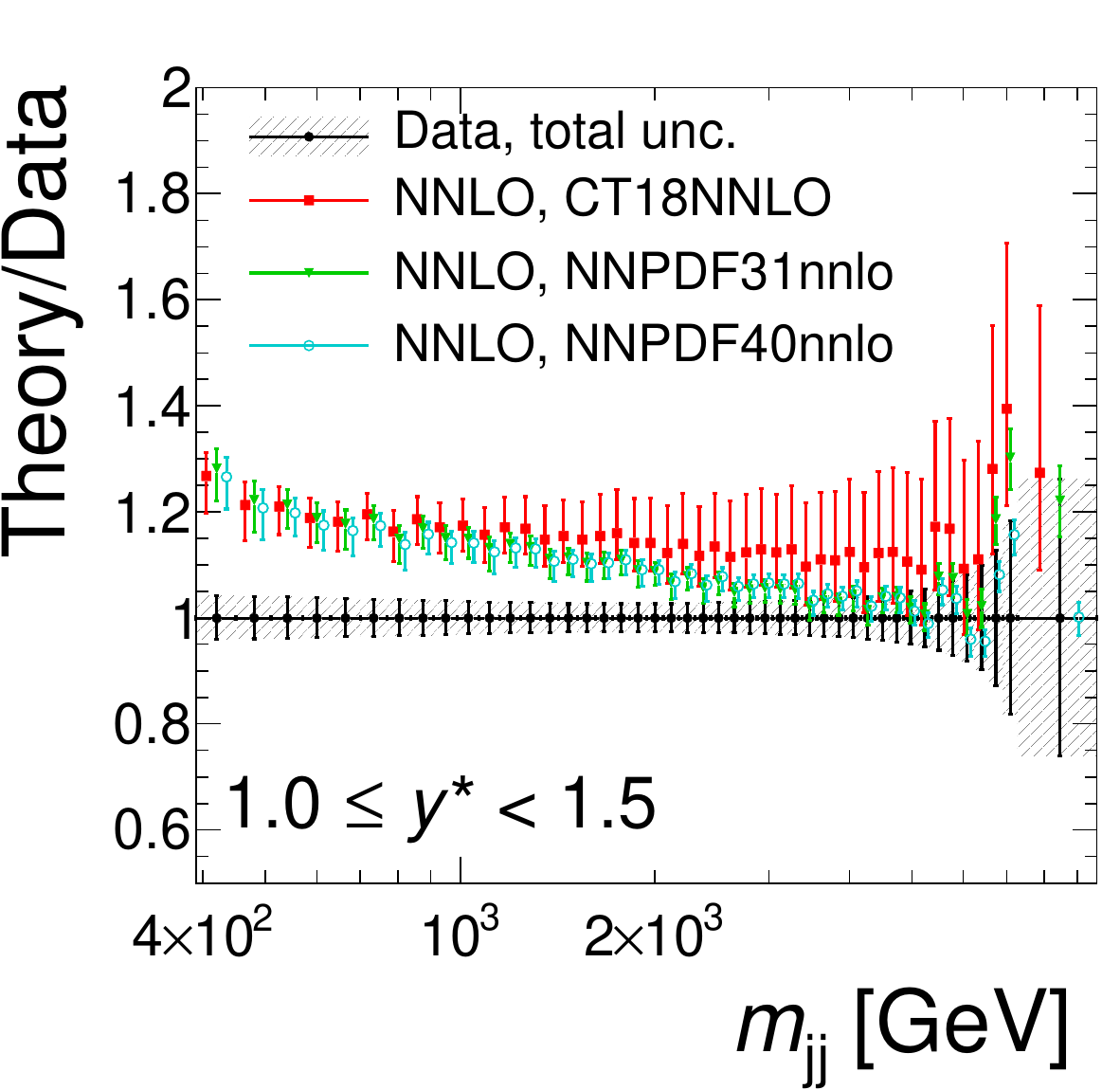} \\
\includegraphics[width=0.31\textwidth]{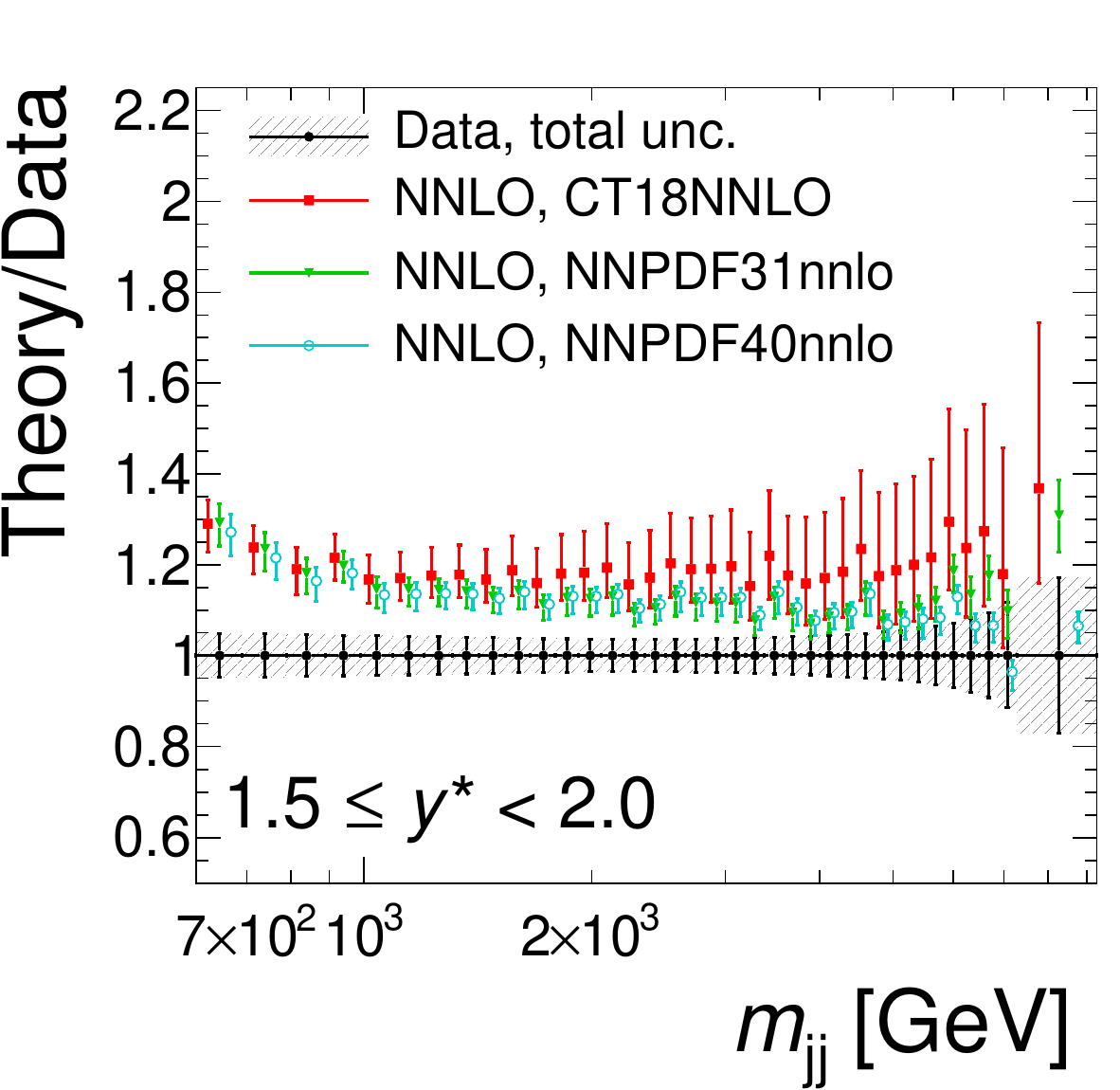} &
\includegraphics[width=0.31\textwidth]{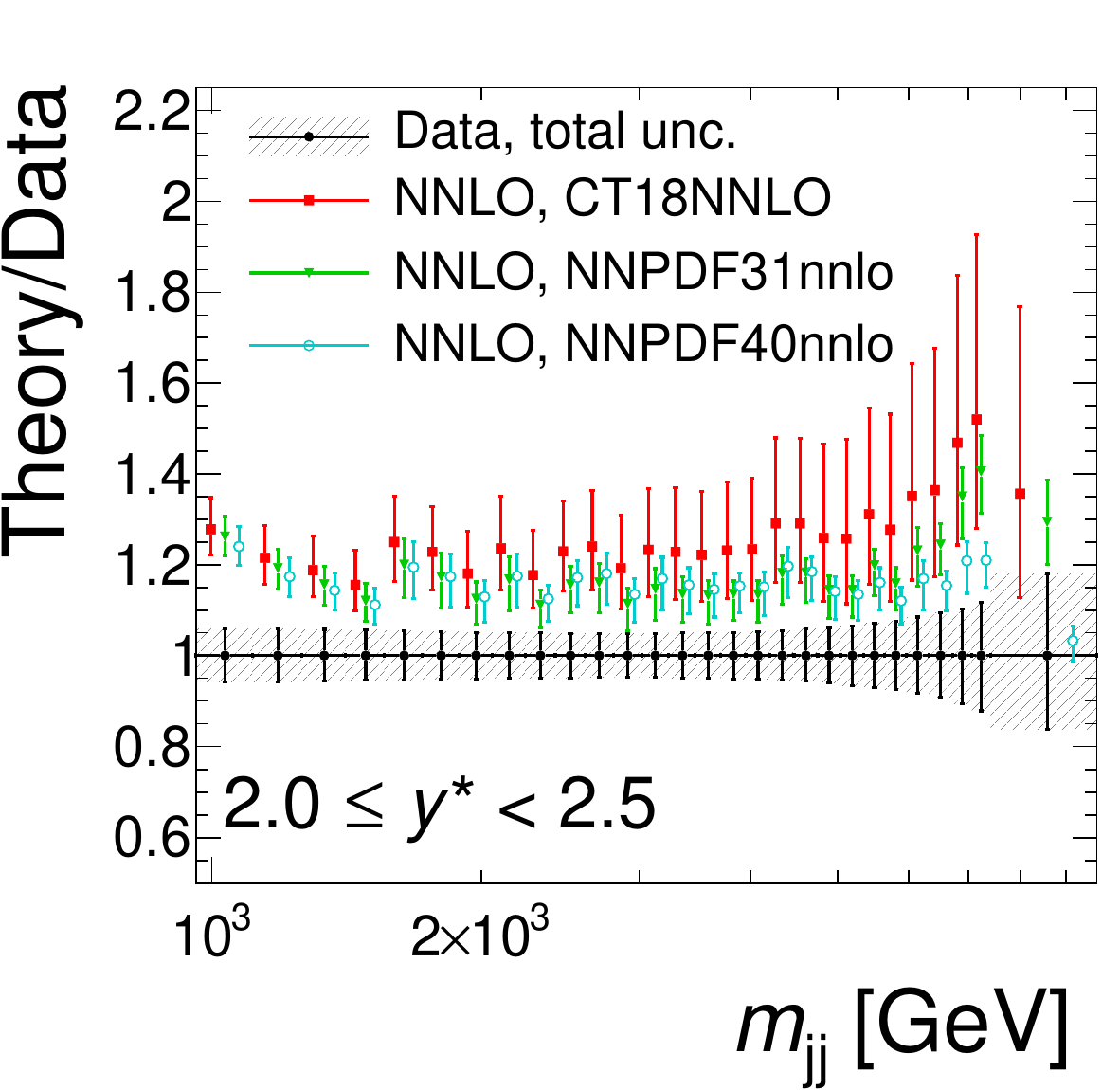} &
\includegraphics[width=0.31\textwidth]{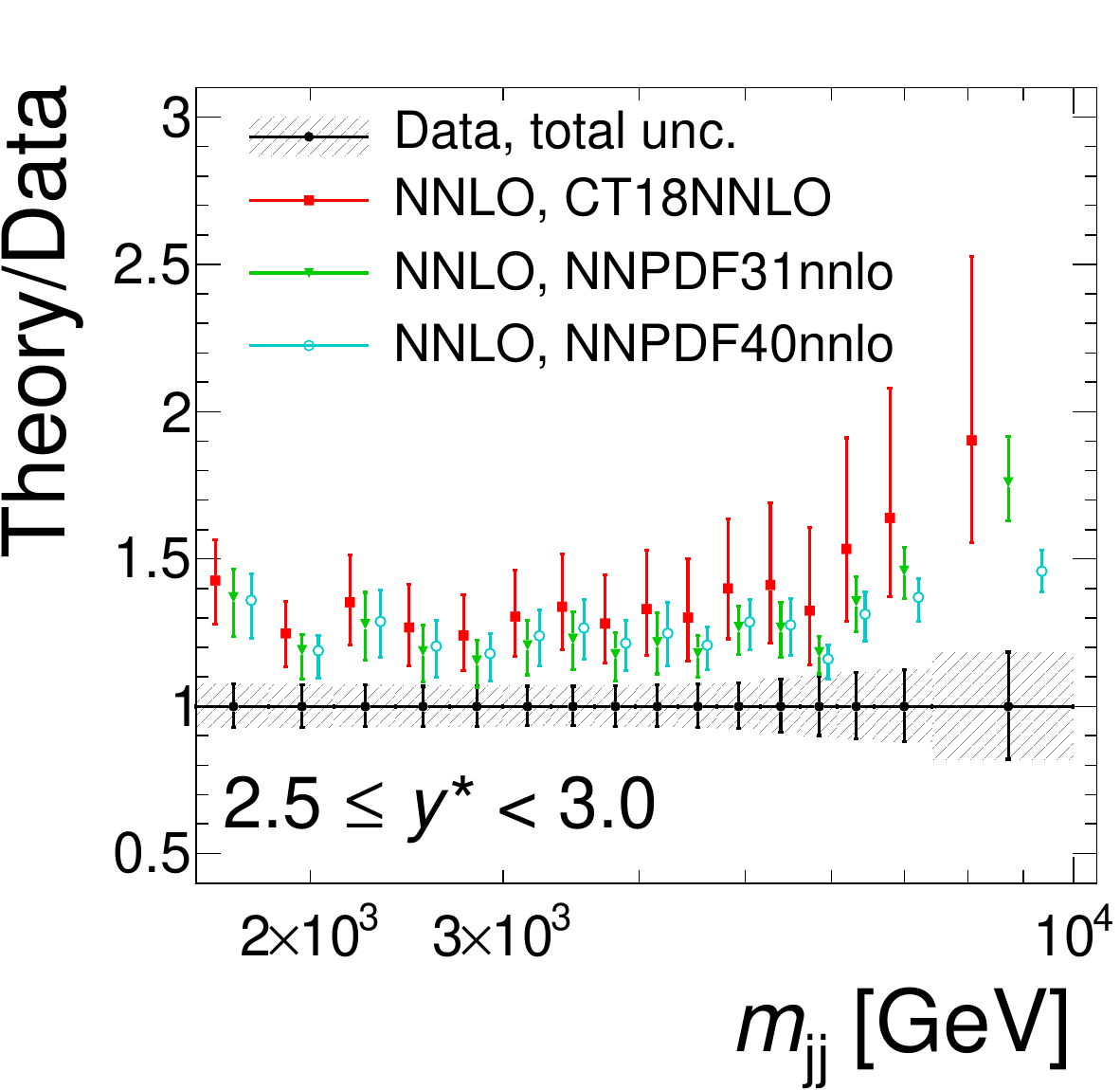} \\
\multicolumn{3}{c}{\includegraphics[width=1.0\textwidth]{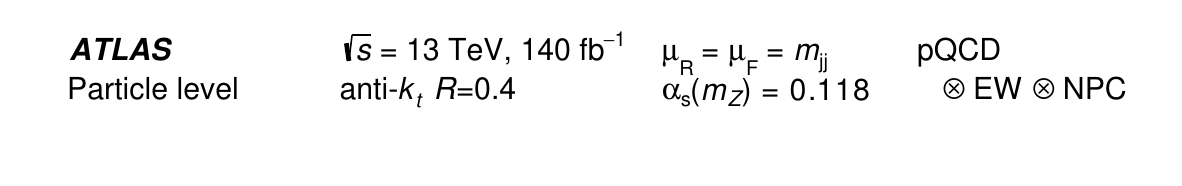}}  \\
\end{tabular}
\caption{   Theory-to-data ratio for the double-differential cross-section as a function of the invariant dijet mass \mjj in six equidistant \ystar bins for \antikt $R=0.4$ jets at the particle level.
The sub-figures compare \NNLO pQCD calculations with different PDF sets (\CT{18}, \NNPDF{3.1}, and \NNPDF{4.0}) to the ATLAS data.
The pQCD full colour predictions via the NNLOJET programme include renormalisation and factorisation scales of $\muRmuF = \mjj$.
The pQCD calculations are corrected by electroweak and non-perturbative corrections to obtain the complete predictions at the particle level.
The theory is compared with the data unfolded to the particle level using the IDS method with one iteration.
Coloured vertical line errors indicate the total systematic uncertainties in pQCD calculations, including the effect of non-perturbative corrections.
Vertical band errors around $1.0$ indicate total experimental uncertainty in the data.
}
\label{fig:DataToTheory_pdf1_mjj_yStar__MJJ_SCALE}
\end{figure}

\begin{figure}[tbp]
\centering
\begin{tabular}{lll}
\includegraphics[width=0.31\textwidth]{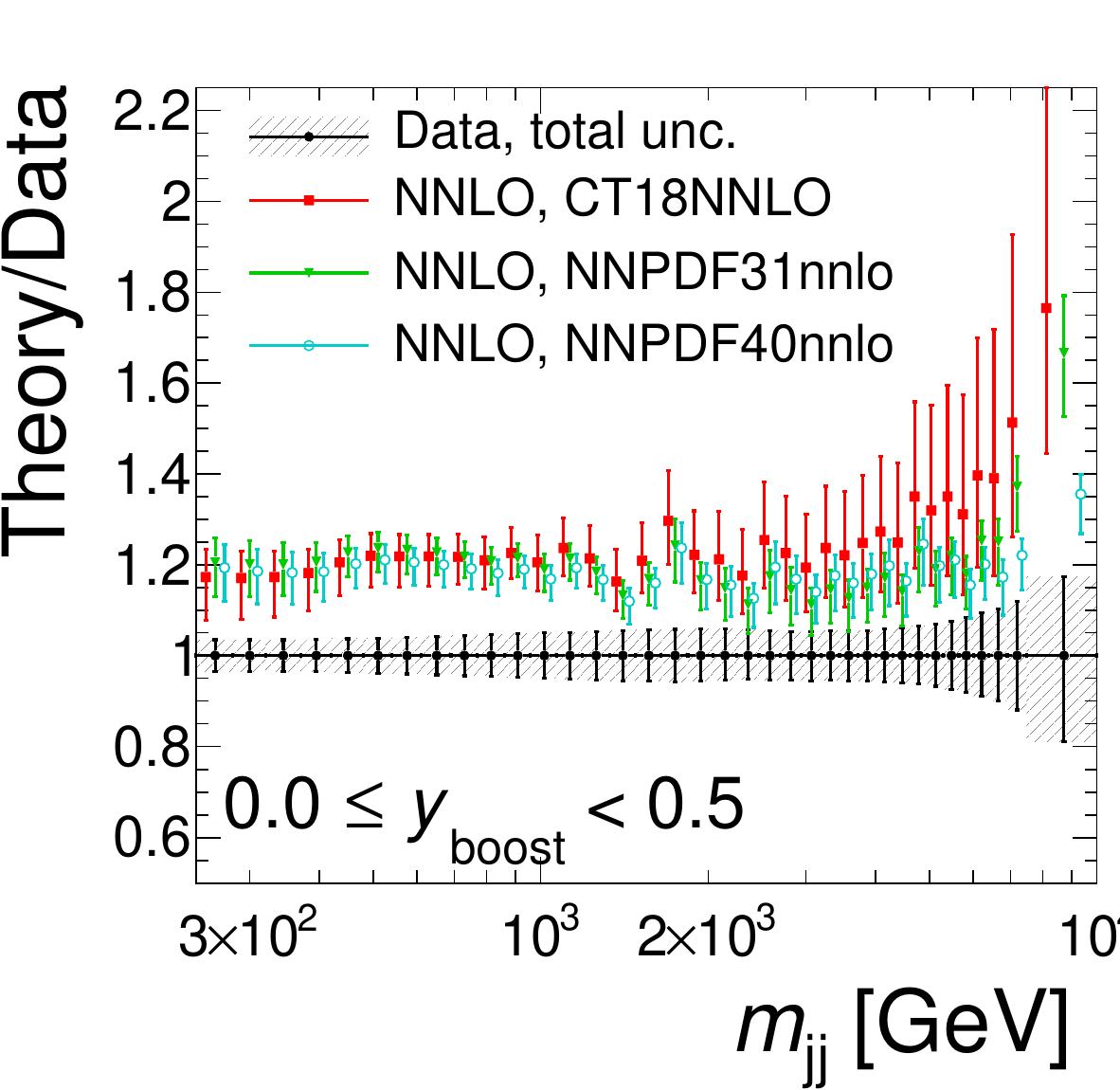} &
\includegraphics[width=0.31\textwidth]{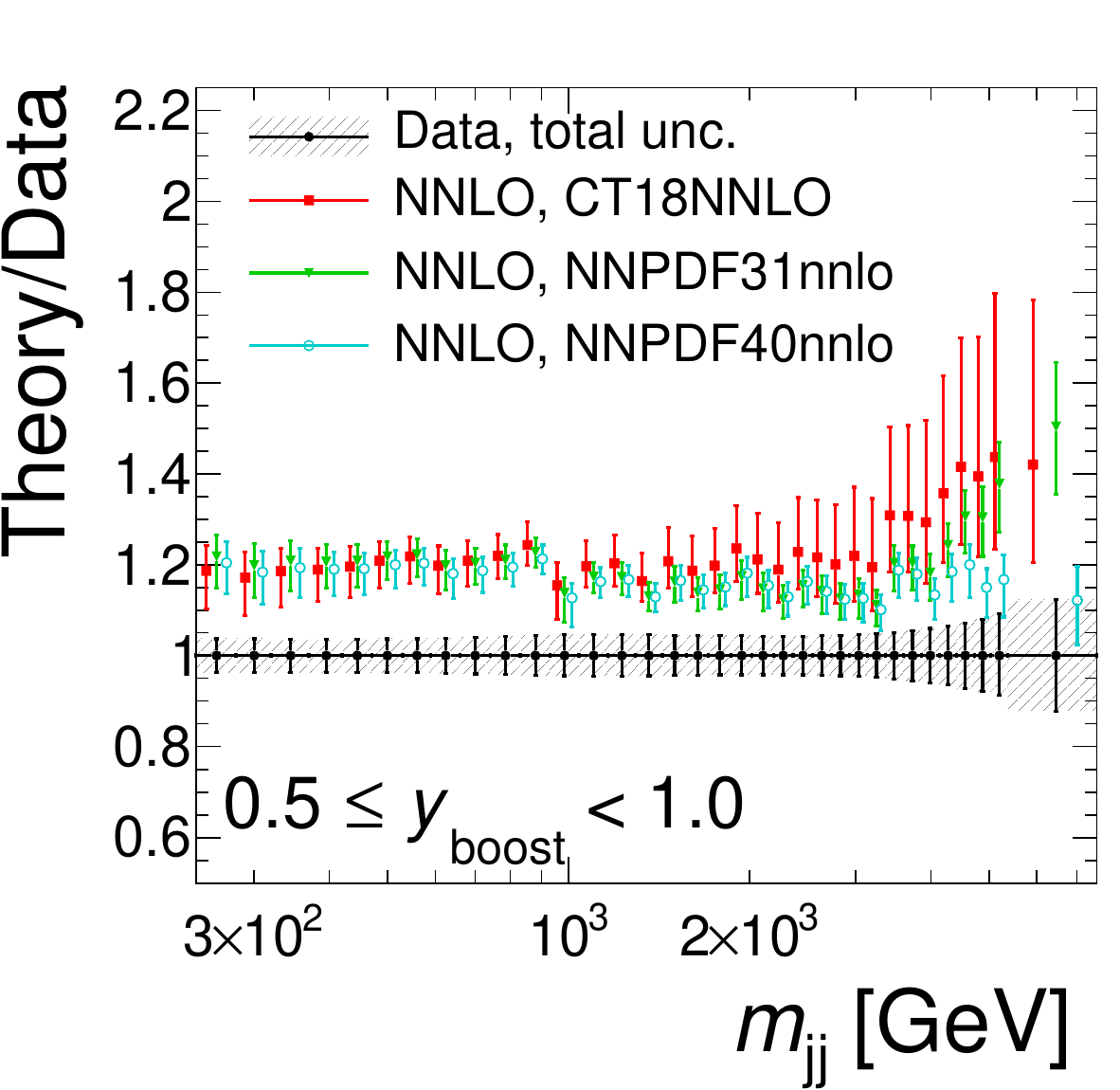} &
\includegraphics[width=0.31\textwidth]{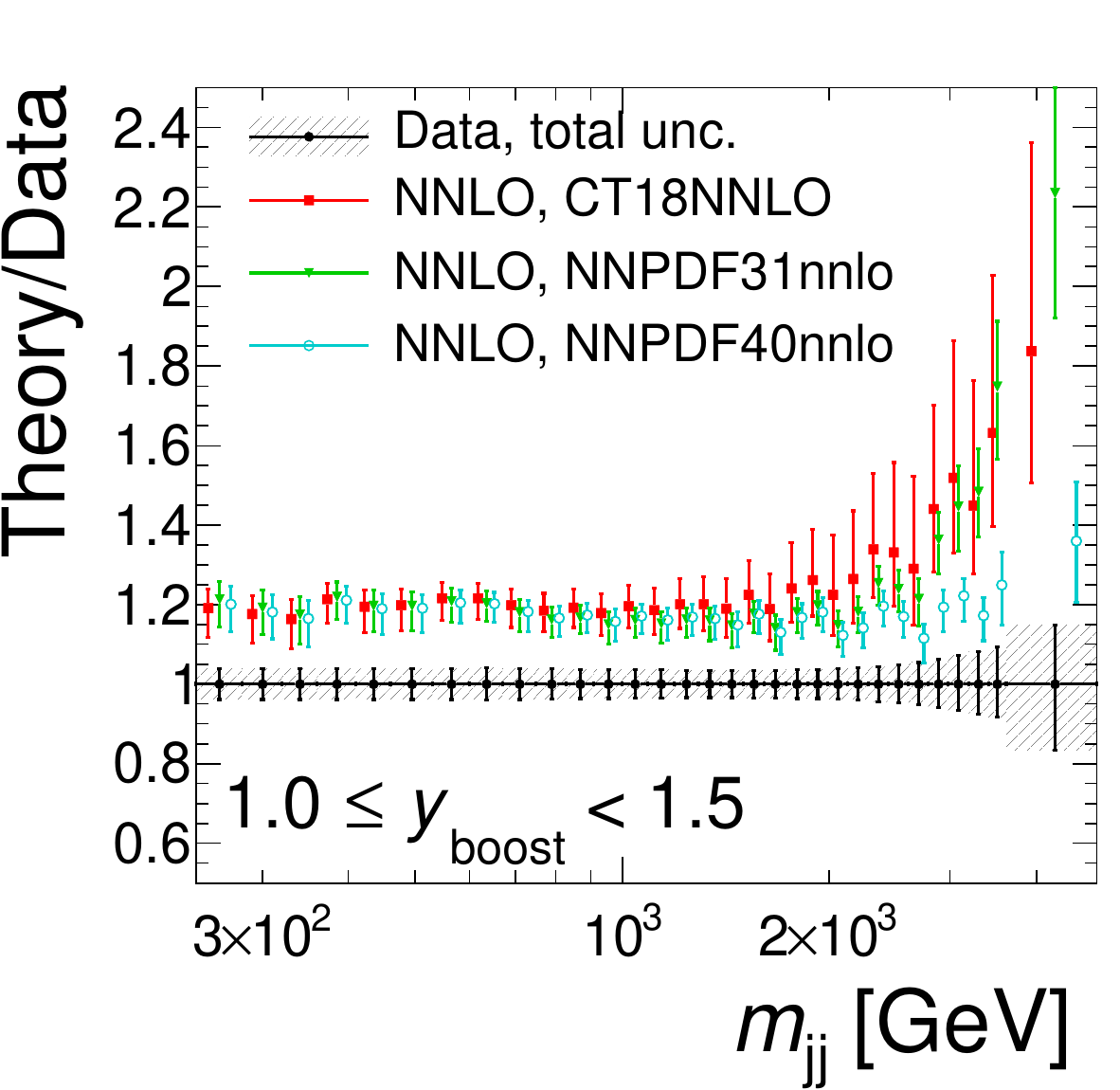} \\
\includegraphics[width=0.31\textwidth]{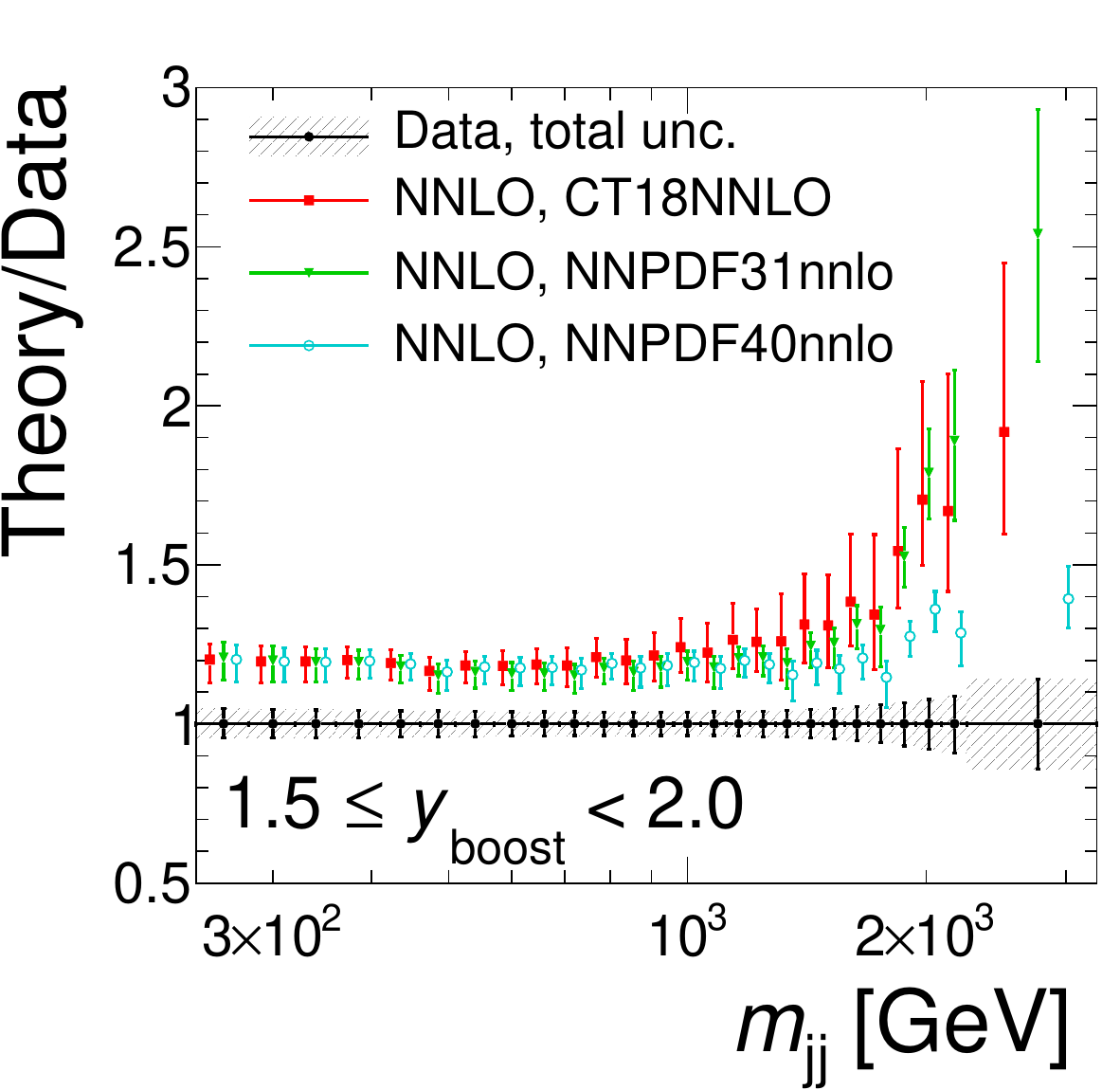} &
\includegraphics[width=0.31\textwidth]{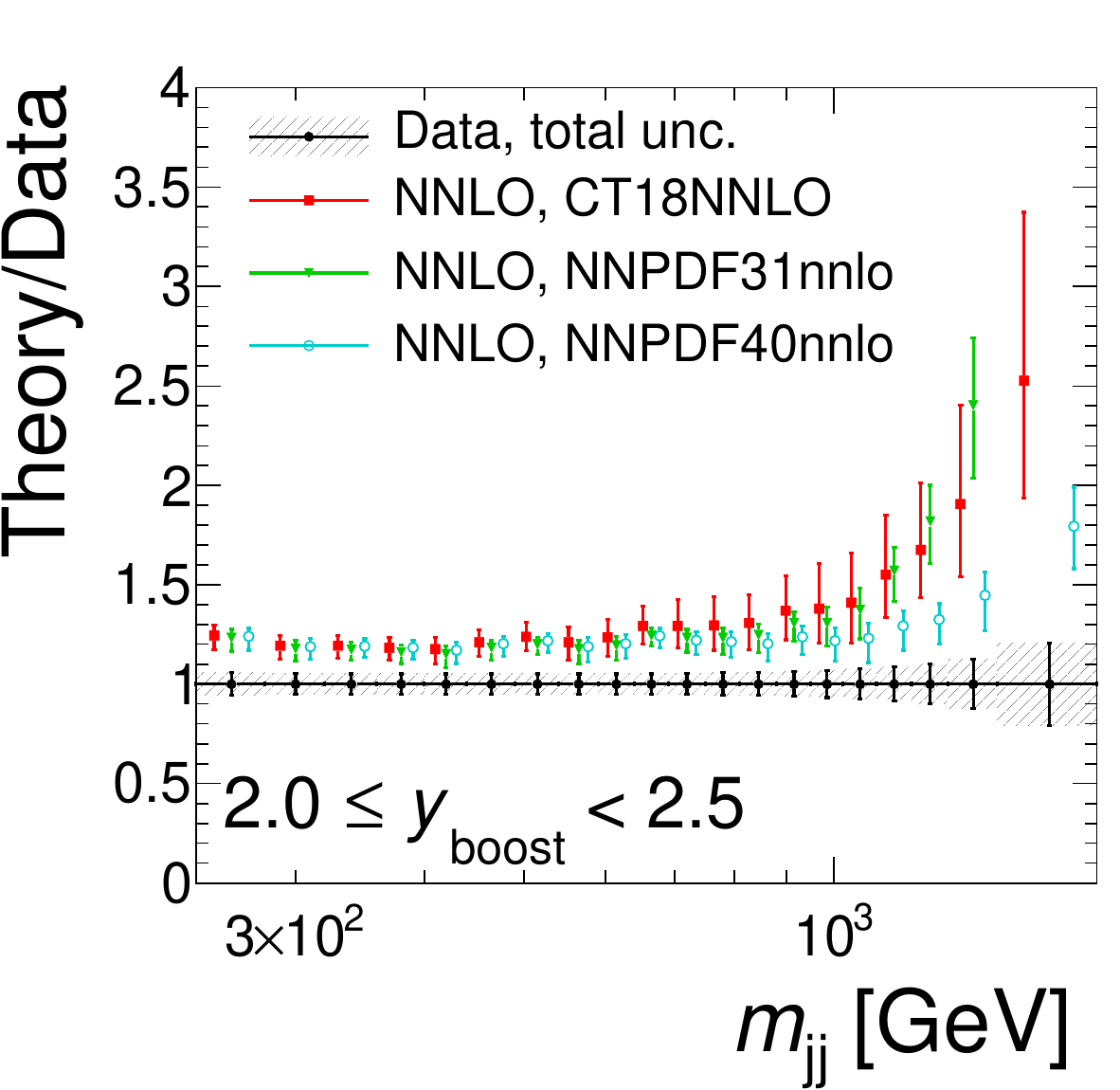} &
\includegraphics[width=0.31\textwidth]{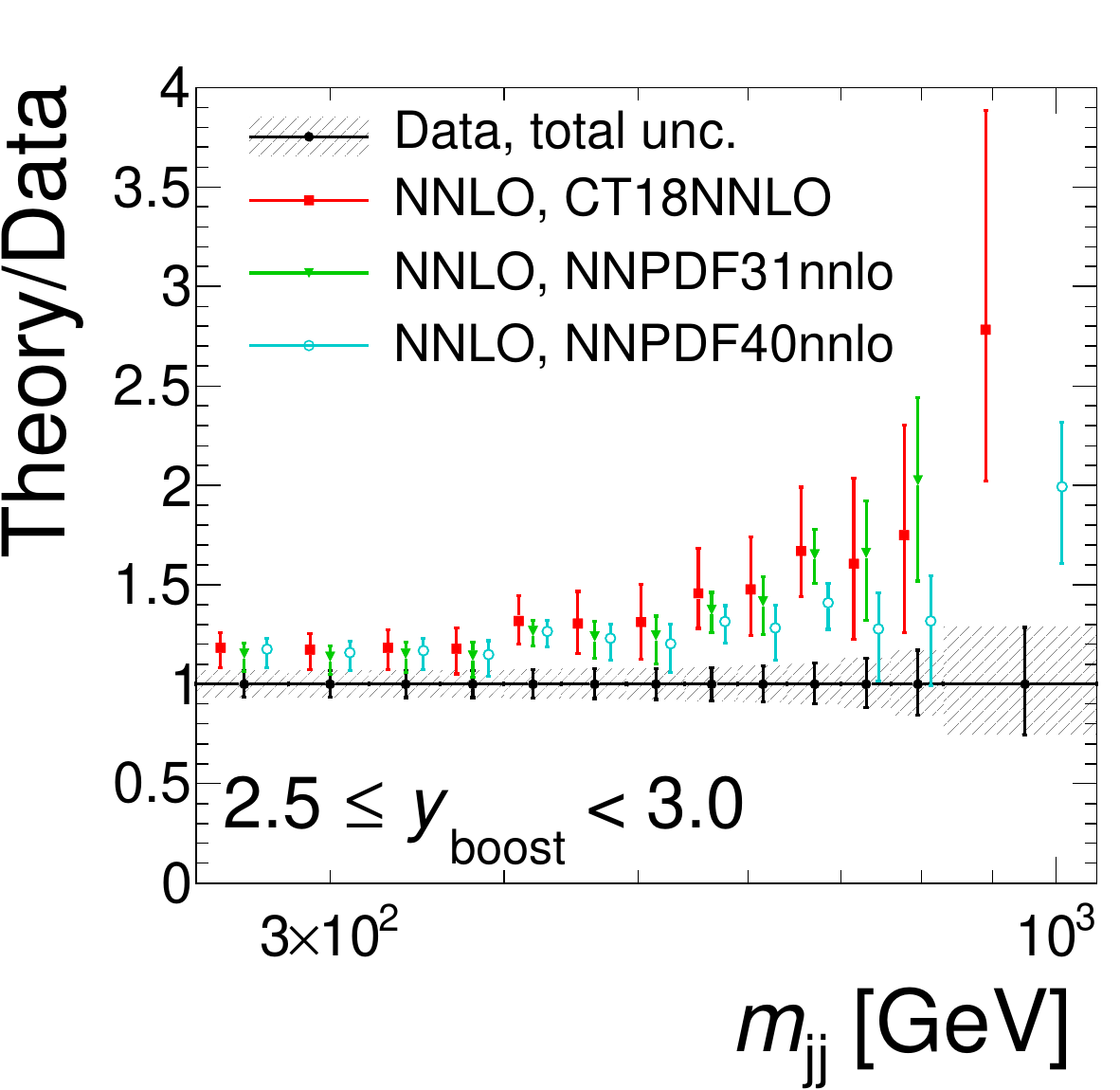} \\
\multicolumn{3}{c}{\includegraphics[width=1.0\textwidth]{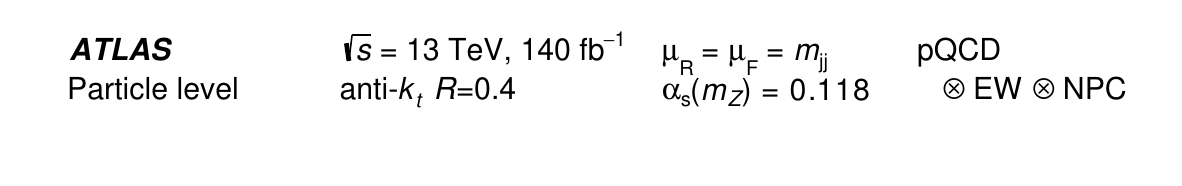}} \\

\end{tabular}
\caption{   Theory-to-data ratio for the double-differential cross-section as a function of the invariant dijet mass \mjj in six equidistant \yboost bins for \antikt $R=0.4$
jets at the particle level.
The sub-figures compare \NNLO pQCD calculations with different PDF sets (\CT{18}, \NNPDF{3.1}, and \NNPDF{4.0}) to the ATLAS data.
The pQCD full colour predictions via the NNLOJET programme include renormalisation and factorisation scales of $\muRmuF = \mjj$.
The pQCD calculations are corrected by electroweak and non-perturbative corrections to obtain the complete predictions at the particle level.
The theory is compared with the data unfolded to the particle level using the IDS method with two iterations.
Coloured vertical line errors indicate the total systematic uncertainties in pQCD calculations, including the effect of non-perturbative corrections.
Vertical band errors around $1.0$ indicate total experimental uncertainty in the data.
}
\label{fig:DataToTheory_pdf1_mjj_yBoost__MJJ_SCALE}
\end{figure}

\FloatBarrier
\begin{figure}[tbp]
\centering
\begin{tabular}{lll}
\includegraphics[width=0.31\textwidth]{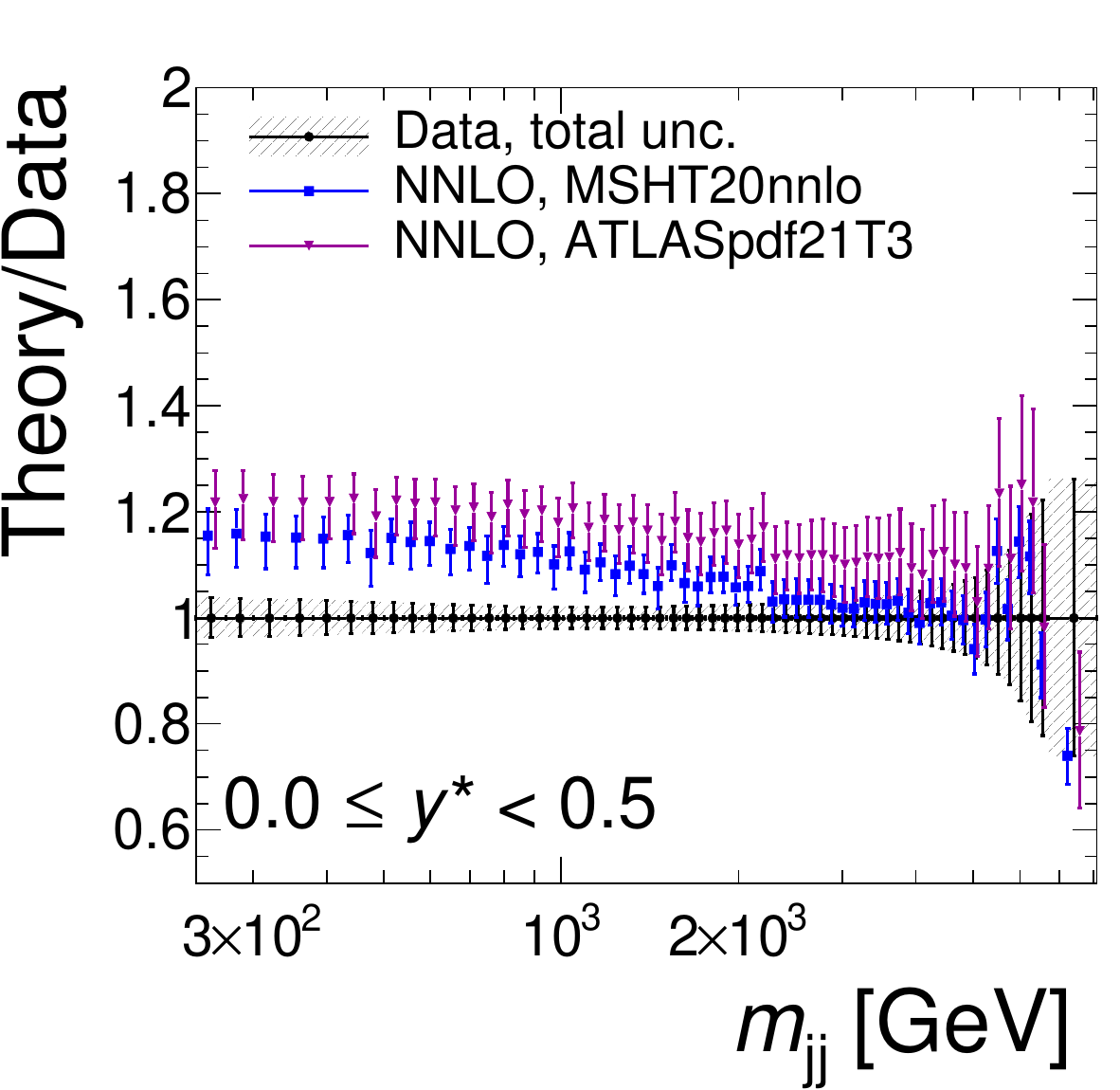} &
\includegraphics[width=0.31\textwidth]{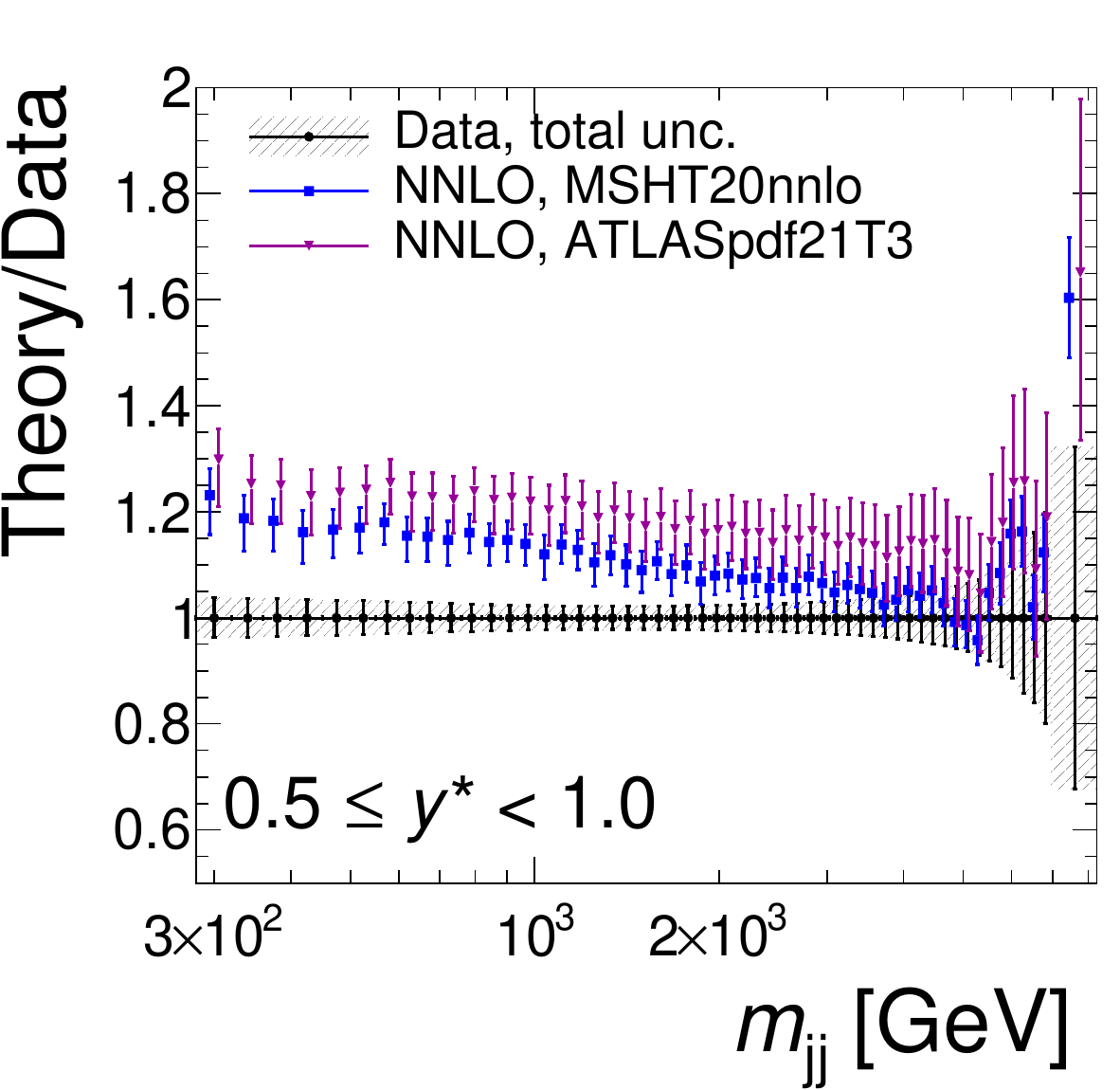} &
\includegraphics[width=0.31\textwidth]{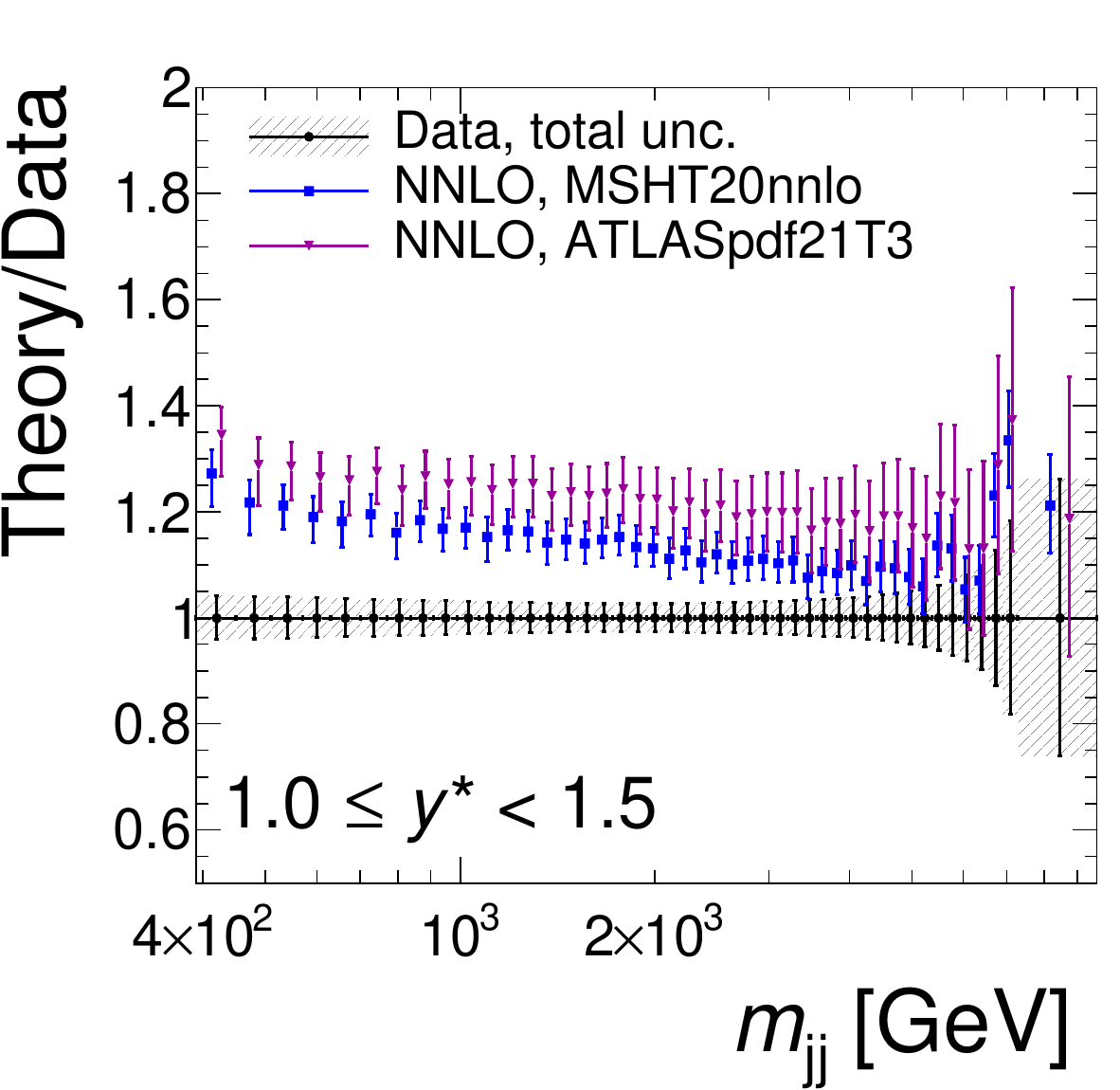} \\
\includegraphics[width=0.31\textwidth]{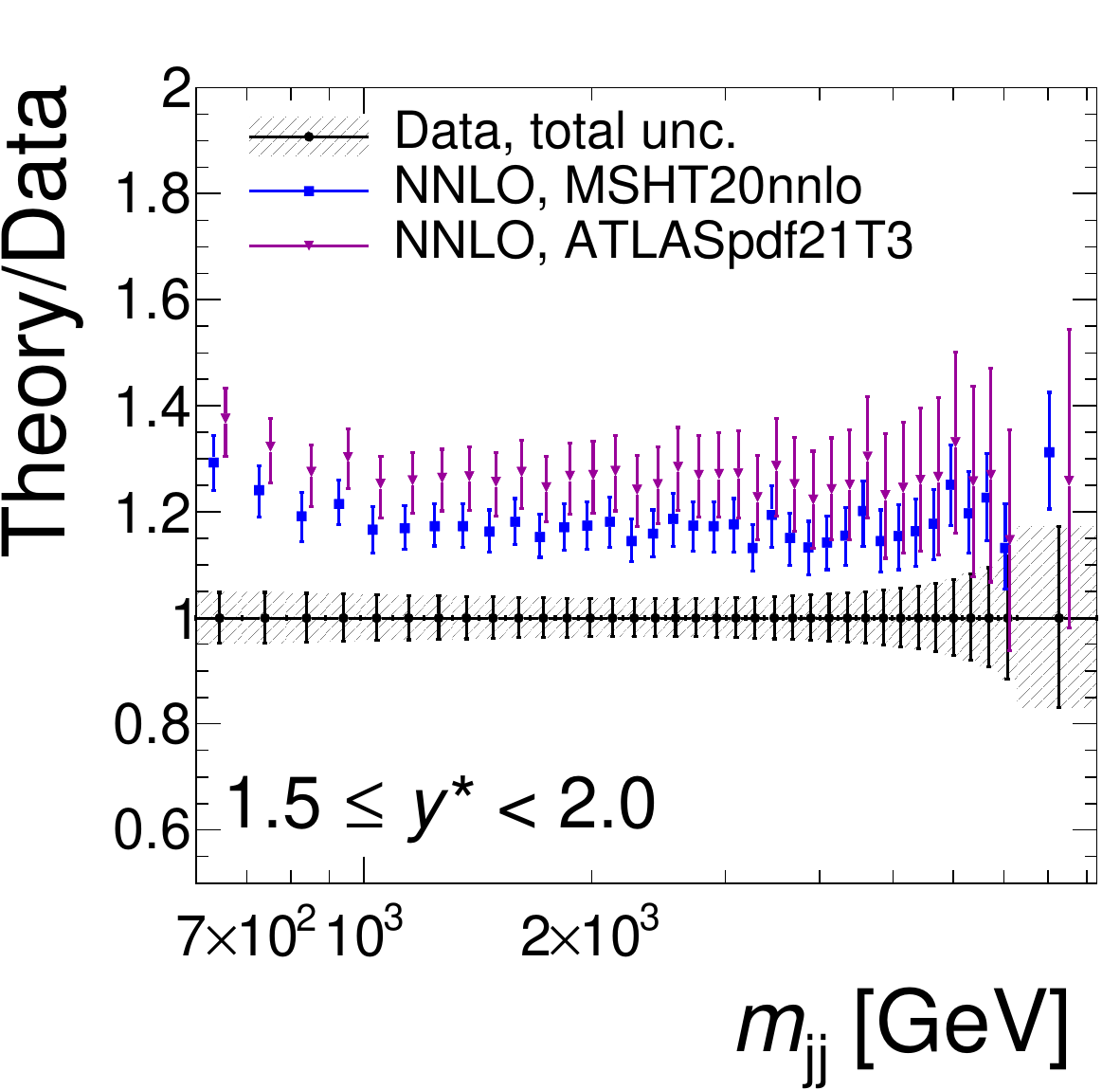} &
\includegraphics[width=0.31\textwidth]{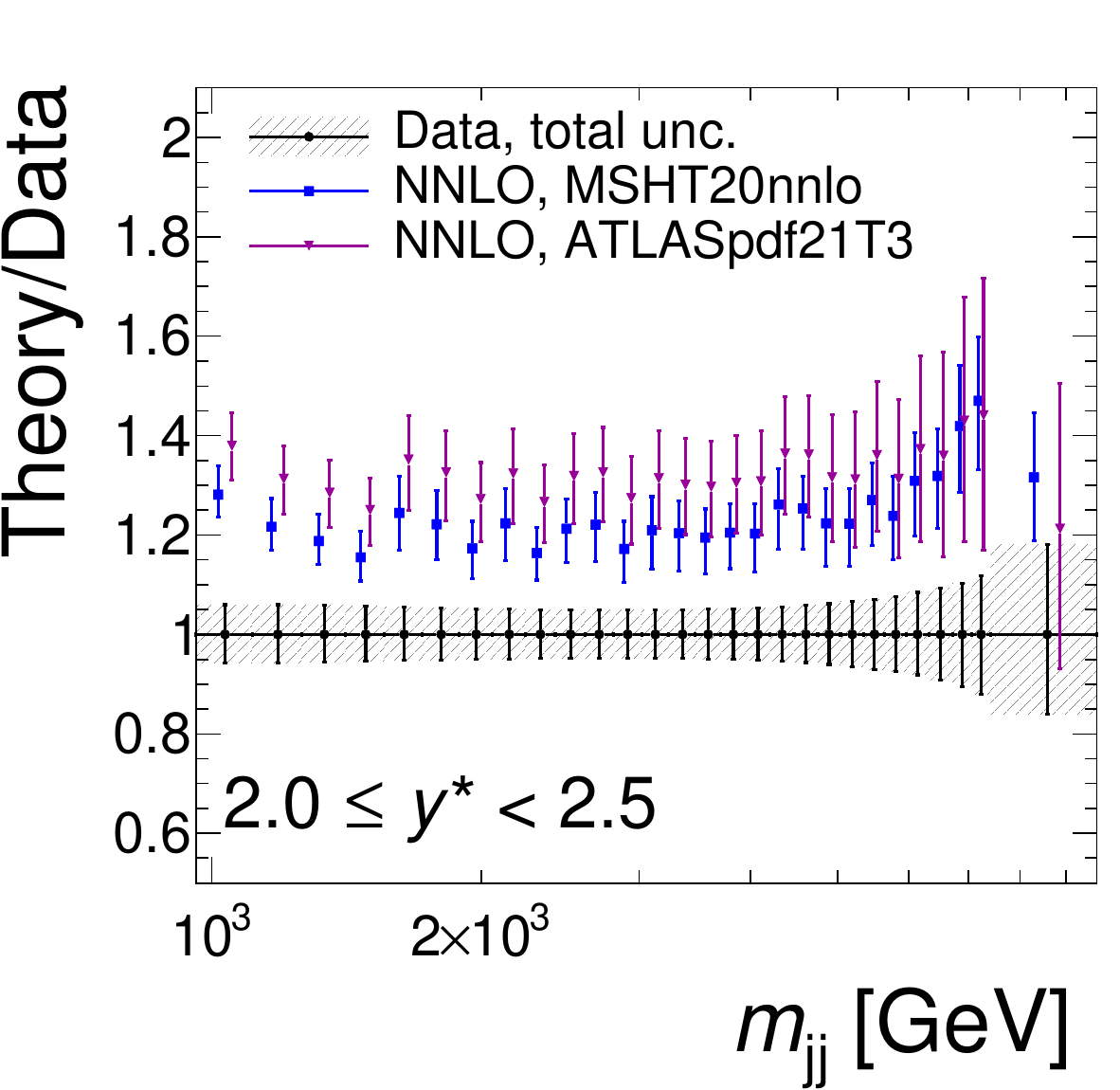} &
\includegraphics[width=0.31\textwidth]{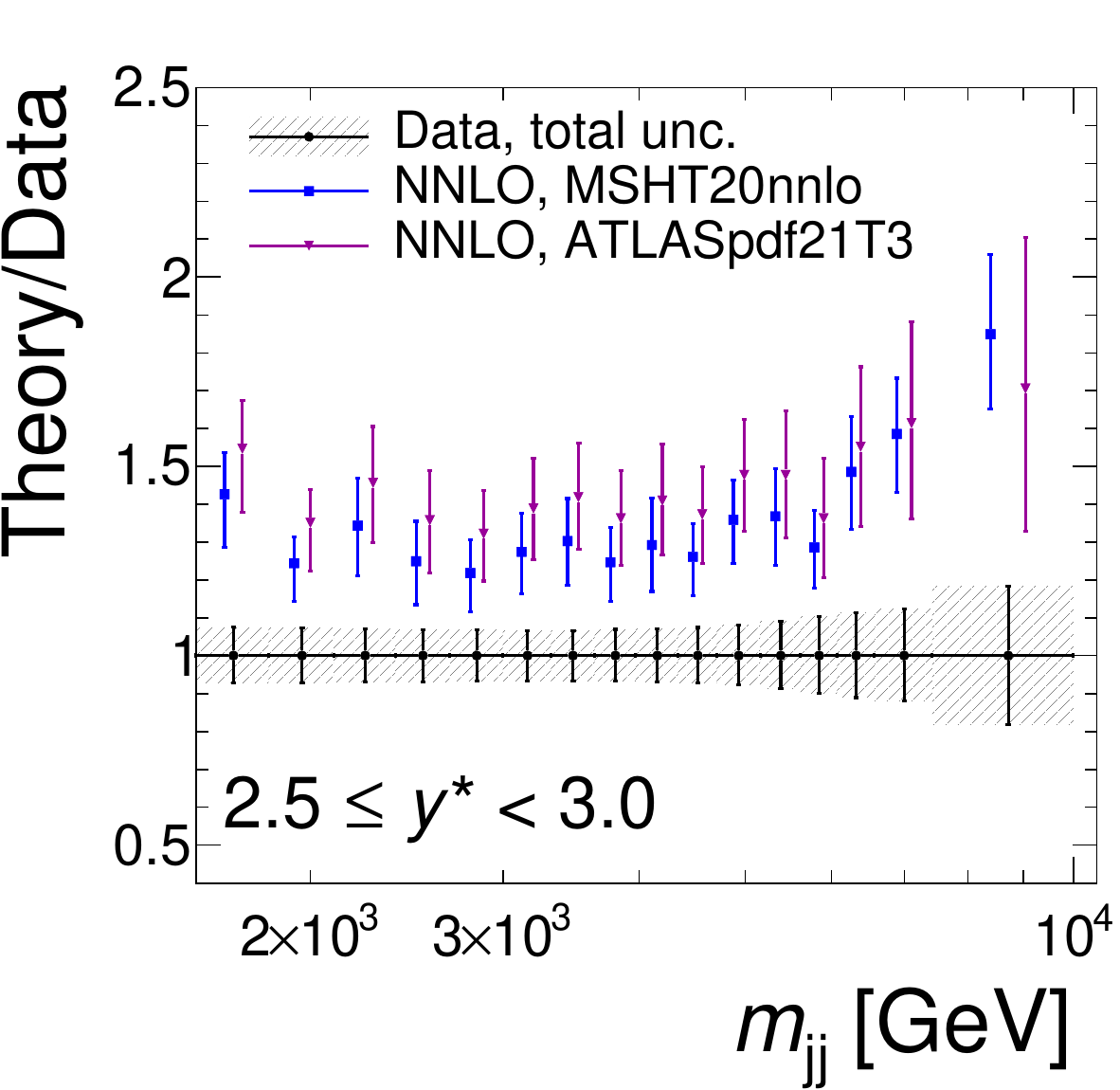} \\
\multicolumn{3}{c}{\includegraphics[width=1.0\textwidth]{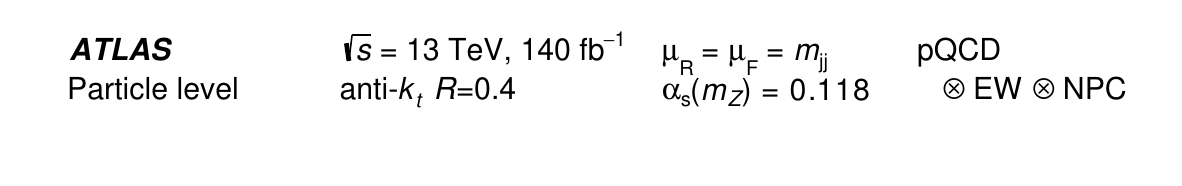}}  \\
\end{tabular}
\caption{   Theory-to-data ratio for the double-differential cross-section as a function of the invariant dijet mass \mjj in six equidistant \ystar bins for \antikt $R=0.4$ jets at the particle level.
The sub-figures compare \NNLO pQCD calculations with different PDF sets (\MSHT, \ATLASPDF{21T3}) to the ATLAS data.
The pQCD full colour predictions via the NNLOJET programme include renormalisation and factorisation scales of $\muRmuF = \mjj$.
The pQCD calculations are corrected by electroweak and non-perturbative corrections to obtain the complete predictions at the particle level.
The theory is compared with the data unfolded to the particle level using the IDS method with one iteration.
Coloured vertical line errors indicate the total systematic uncertainties in pQCD calculations, including the effect of non-perturbative corrections. The prediction using the \ATLASPDF{21T3} PDF set does not include $\alpha_s$ variation as a source of theoretical systematic uncertainty.
Vertical band errors around $1.0$ indicate total experimental uncertainty in the data.
}
\label{fig:DataToTheory_pdf2_mjj_yStar__MJJ_SCALE}
\end{figure}

\begin{figure}[tbp]
\centering
\begin{tabular}{lll}
\includegraphics[width=0.31\textwidth]{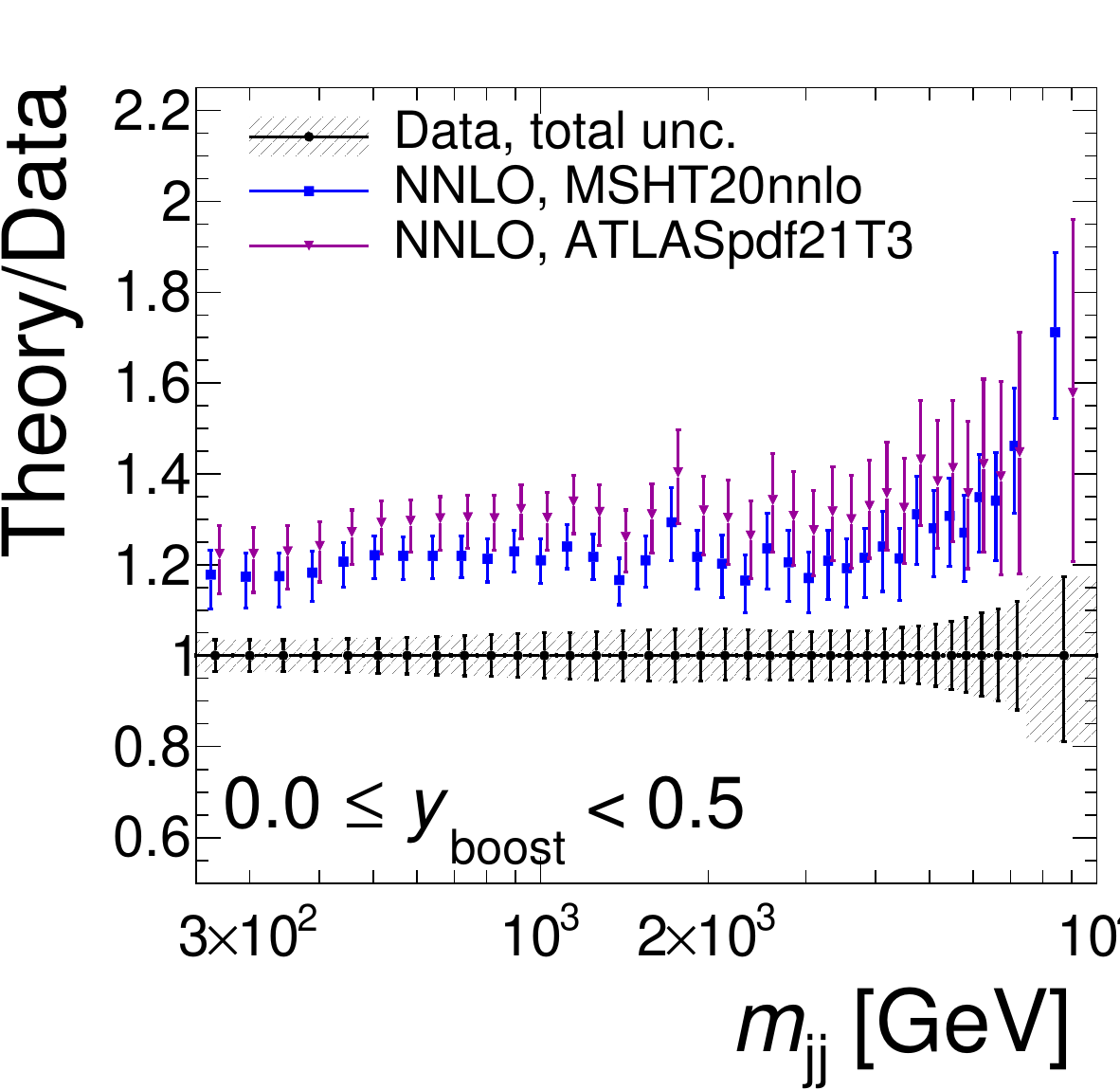} &
\includegraphics[width=0.31\textwidth]{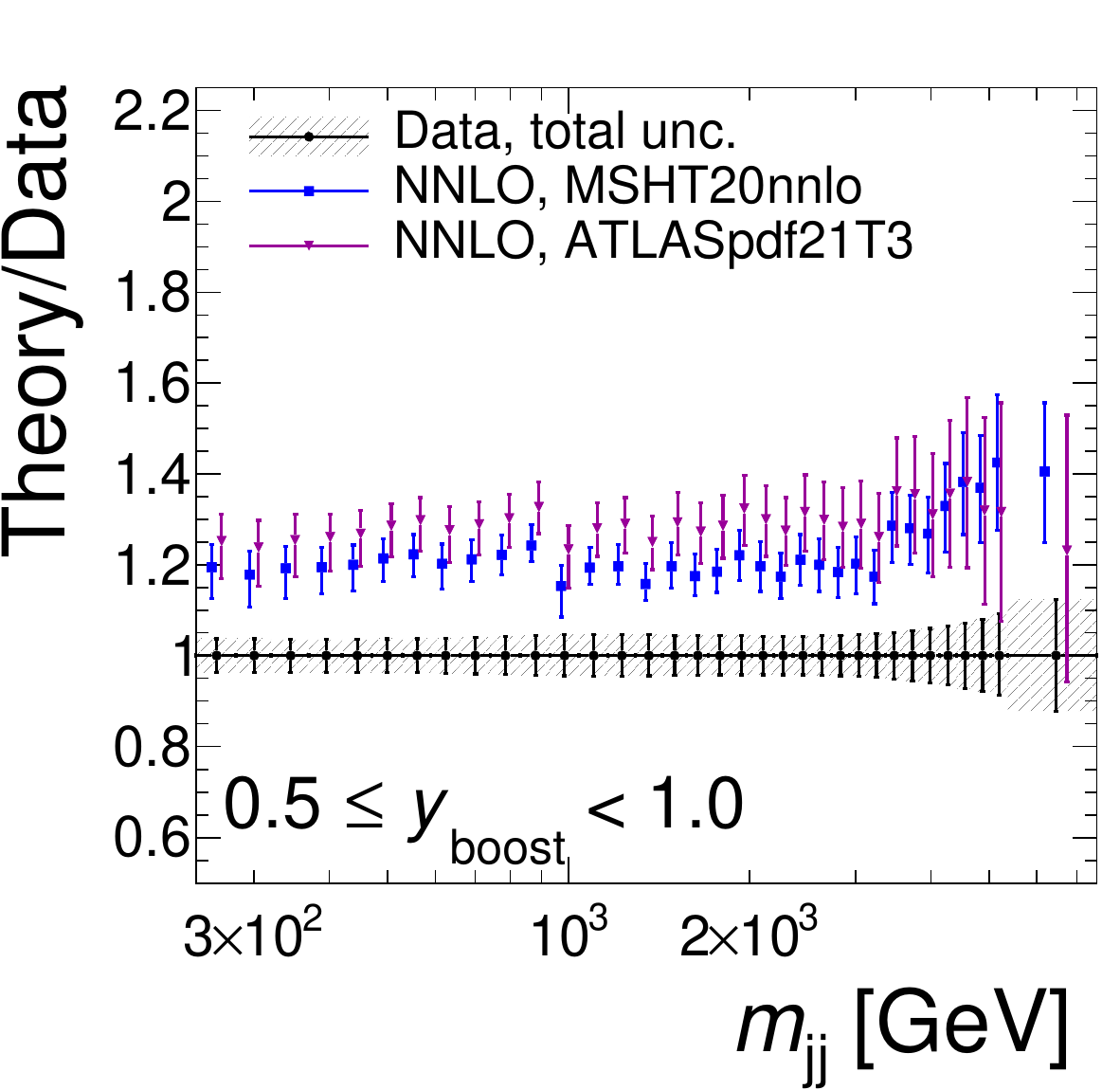} &
\includegraphics[width=0.31\textwidth]{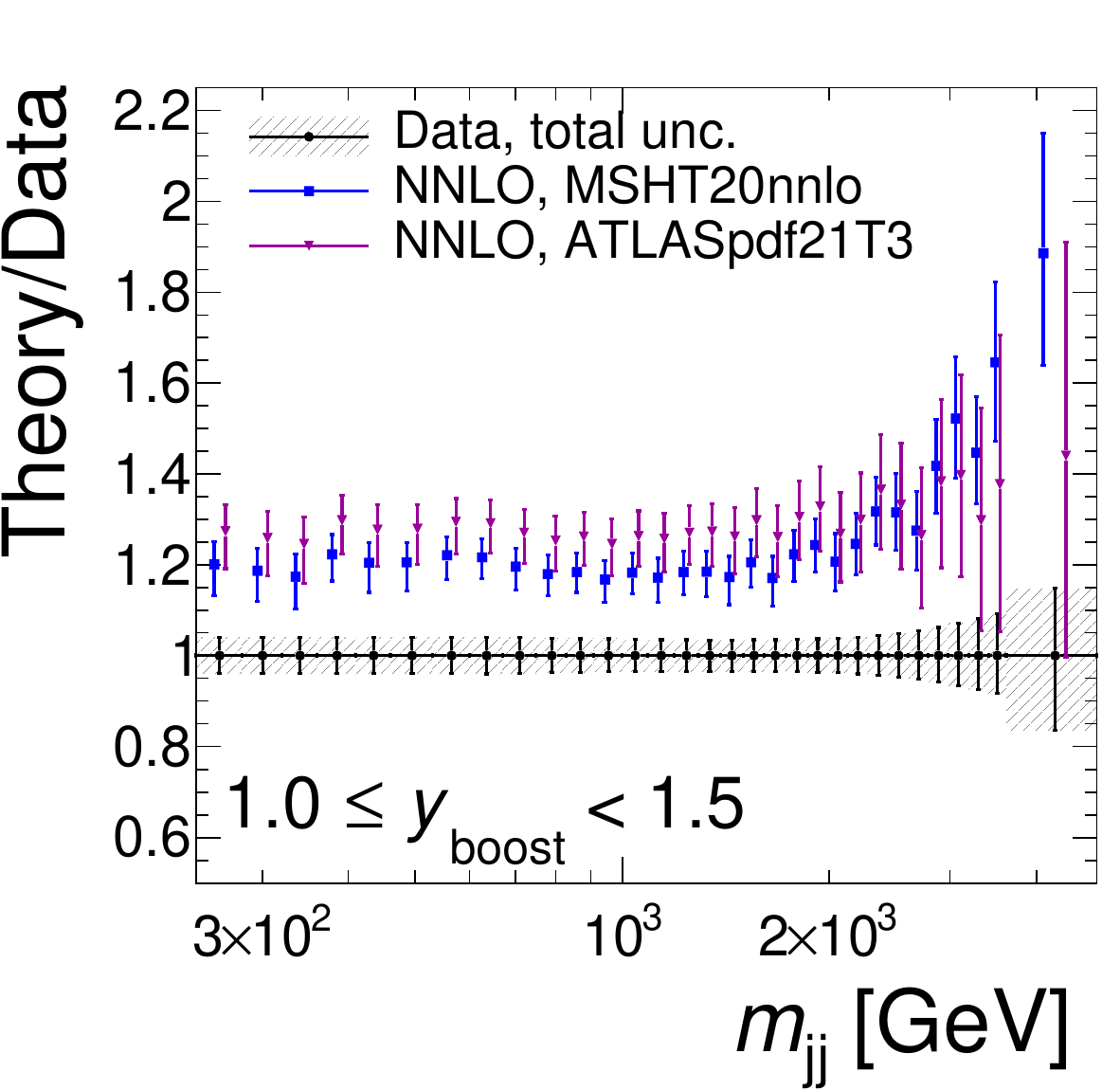} \\
\includegraphics[width=0.31\textwidth]{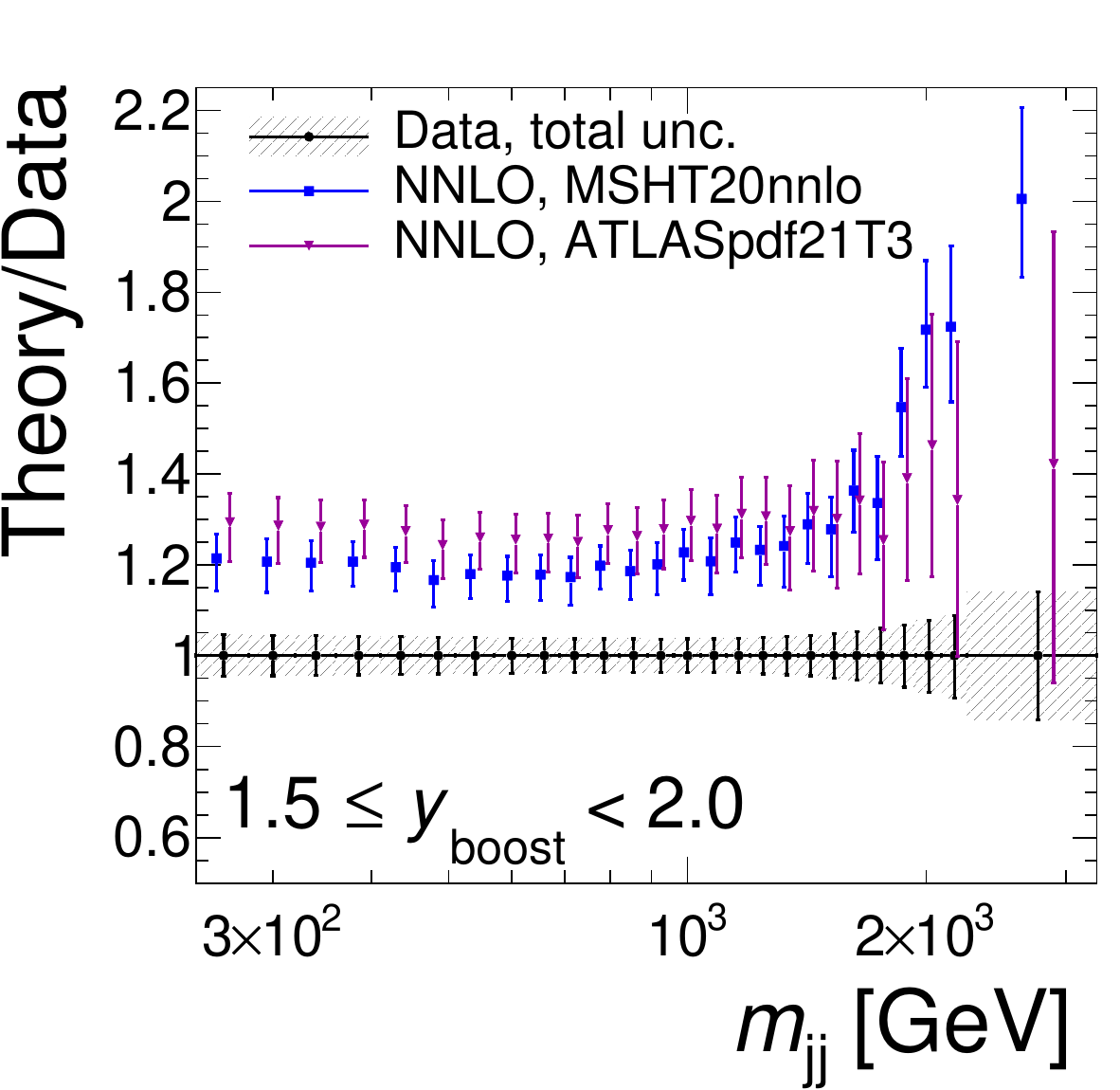} &
\includegraphics[width=0.31\textwidth]{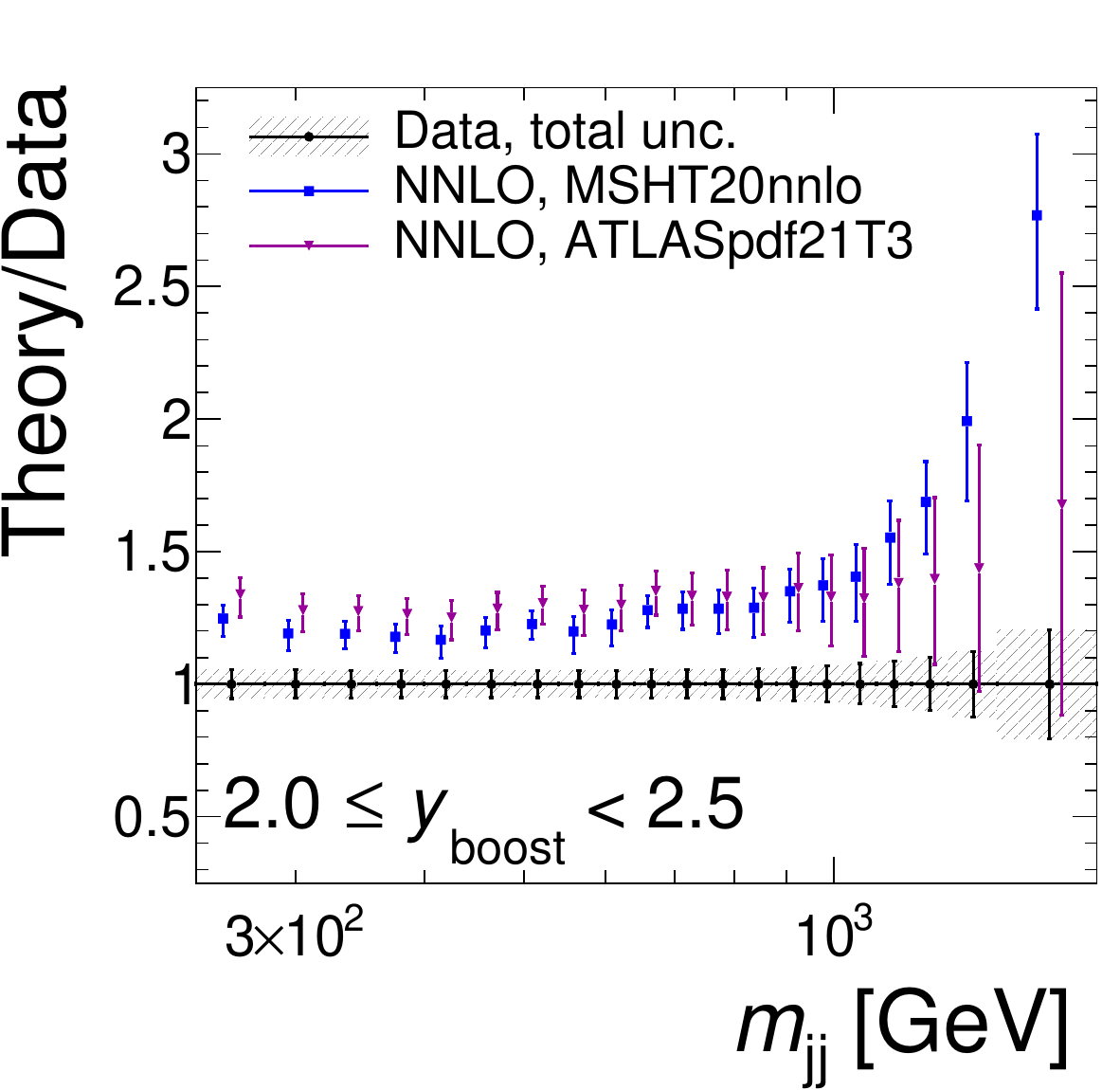} &
\includegraphics[width=0.31\textwidth]{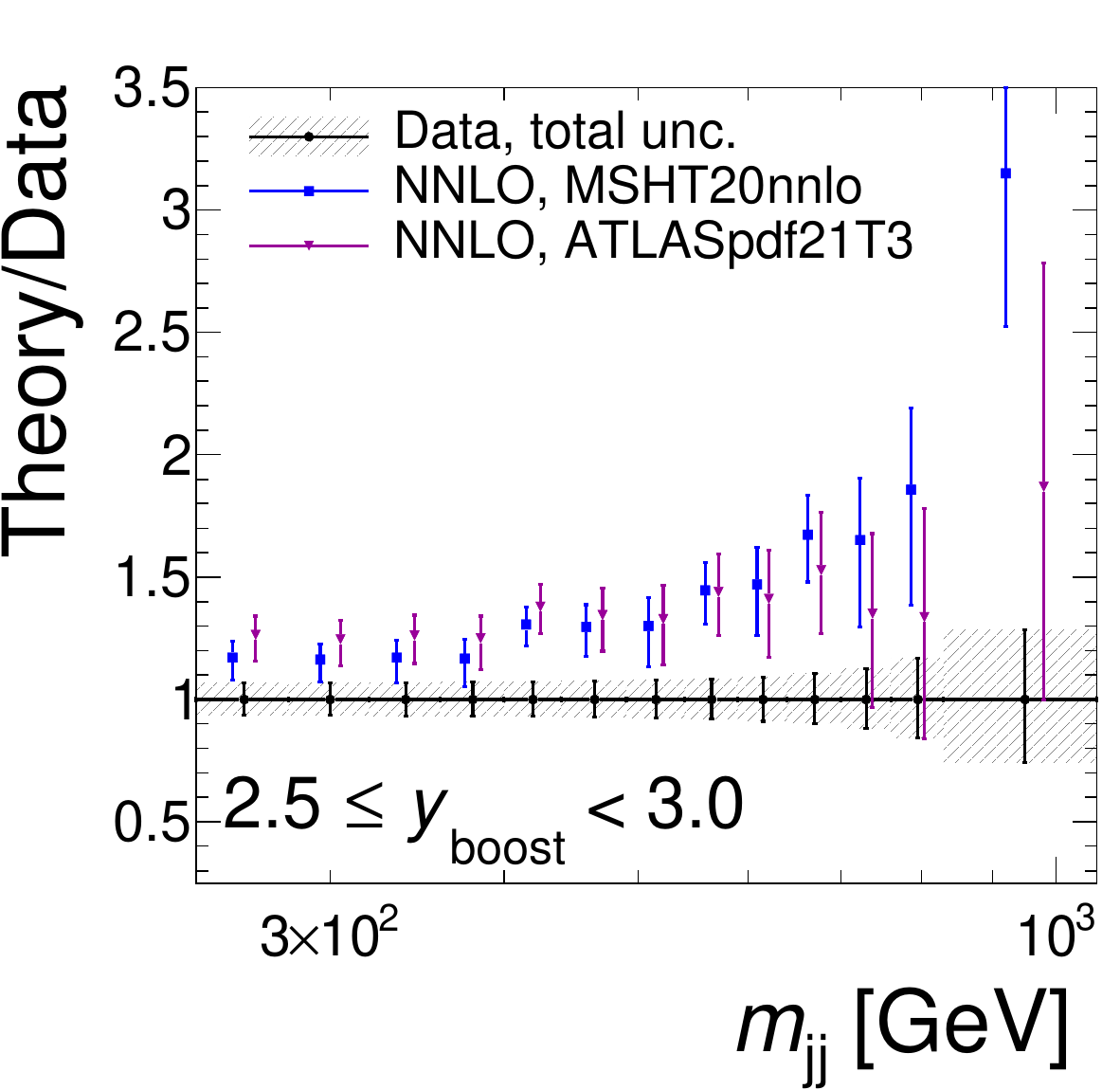} \\
\multicolumn{3}{c}{\includegraphics[width=1.0\textwidth]{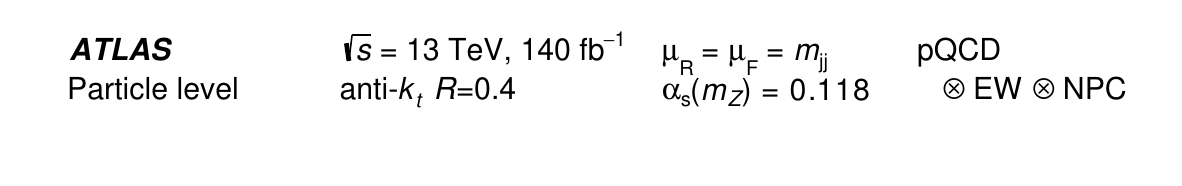}} \\

\end{tabular}
\caption{   Theory-to-data ratio for the double-differential cross-section as a function of the invariant dijet mass \mjj in six equidistant \yboost bins for \antikt $R=0.4$
jets at the particle level.
The sub-figures compare \NNLO pQCD calculations with different PDF sets (\MSHT, \ATLASPDF{21T3}) to the ATLAS data.
The pQCD full colour predictions via the NNLOJET programme include renormalisation and factorisation scales of $\muRmuF = \mjj$.
The pQCD calculations are corrected by electroweak and non-perturbative corrections to obtain the complete predictions at the particle level.
The theory is compared with the data unfolded to the particle level using the IDS method with two iterations.
Coloured vertical line errors indicate the total systematic uncertainties in pQCD calculations, including the effect of non-perturbative corrections. The prediction using the \ATLASPDF{21T3} PDF set does not include $\alpha_s$ variation as a source of theoretical systematic uncertainty.
Vertical band errors around $1.0$ indicate total experimental uncertainty in the data.
}
\label{fig:DataToTheory_pdf2_mjj_yBoost__MJJ_SCALE}
\end{figure}

\FloatBarrier
\newpage
A quantitative assessment of the agreement between the measured dijet cross-section and theoretical predictions requires a statistical approach. To evaluate the consistency of the data with theoretical expectations, a $\chi^{2}$ test is employed as described in Ref.~\cite{STDM-2012-03}.

The $\chi^{2}$ test provides a measure of the overall compatibility between the experimental results and the predictions, including the correlations, accounting for statistical and (possibly asymmetric)~systematic uncertainties. The individual experimental and theory uncertainty components are treated as mutually uncorrelated, while full correlation is assumed across the (\mjj, \ystar) or (\mjj, \yboost) bins, respectively. Statistical correlations across these bins are evaluated using covariance matrices derived from 1000 pseudo experiments, in which both the data and MC simulation are fluctuated according to Poisson statistics. From the theory perspective, separate components for scale variations, PDF eigenvectors, non-perturbative corrections, and the strong coupling constant are implemented. The PDF eigenvectors for \NNPDF{} sets are obtained from a covariance matrix derived based on the pseudo experiments.

The $\chi^2$ tests are performed to evaluate the sensitivity of the measurement to electroweak and non-perturbative corrections using a fixed-order \NNLO pQCD prediction with the \CT{18NNLO} PDF set. For the (\mjj,~\ystar) measurement, a strong sensitivity to electroweak effects is observed in global $\chi^2/\text{ndf}$ ($329/237$ with vs.\ $396/237$ without). In contrast, the impact of non-perturbative corrections is small ($329/237$ with vs.\ $338/237$ without). For the (\mjj,~\yboost) measurement, no sensitivity is observed to either electroweak nor non-perturbative effects, as the $\chi^2/\mathrm{ndf}$ remains nearly unchanged ($183/160$ with vs.\ $184/160$ without electroweak corrections, and $184/160$ with vs.\ $184/160$ without non-perturbative corrections).

The $\chi^2$ tests are also used to perform quantitative comparisons between the data and different PDF sets (\CT{18}, \NNPDF{3.1}, \NNPDF{4.0}, \MSHT, and \ATLASPDF{21T3}) across both individual and global $\ystar$ and $\yboost$ bins, as shown in \Cref{tab:results:chi2_yStar,tab:results:chi2_yBoost}. The results indicate good agreement in the individual $\ystar$ and $\yboost$ bins. Overall, higher $\chi^2$ values are observed in the (\mjj,~\ystar) measurement compared with (\mjj,~\yboost); this trend is more pronounced in the global $\chi^2/\text{ndf}$. While the global results in $\yboost$ remain acceptable, the global $\chi^2/\text{ndf}$ in $\ystar$ corresponds to a low $p$-value of about $10^{-6}$. This low $p$-value in (\mjj,~\ystar) is primarily driven by residual bin-to-bin fluctuations in the pQCD predictions, arising from incomplete cancellations between large positive and negative integrals of the real–real and real–virtual pQCD contributions.

The lowest global $\chi^2/\text{ndf}$ is obtained with \ATLASPDF{21T3} in both $(\mjj,\ystar)$ and $(\mjj,\yboost)$, at $288/237$ and $166/160$, respectively. Because \ATLASPDF{21T3} does not yet provide a published $\alpha_s$ uncertainty, this component is omitted here; its inclusion would be expected to lower the $\chi^2$ further.

For the other PDF sets, the global $\chi^2/\text{ndf}$ values saturate at higher values—around $330/237$ in $(\mjj,\ystar)$ and $180/160$ in $(\mjj,\yboost)$—and are mutually consistent. The second-best global $\chi^2$ value in $(\mjj,\ystar)$ is found for \CT{18NNLO} with $329/237$, while in $(\mjj,\yboost)$ it is \NNPDF{3.1NNLO} with $178/160$.

\begin{table}[htbp]
\centering
\caption{Evaluated $\chi^2/\text{ndf}$ for the dijet cross-section in the individual and global \ystar bins taking into account \NNLO pQCD predictions using \CT{18}, \NNPDF{3.1}, \NNPDF{4.0}, \MSHT, and \ATLASPDF{21T3} PDF sets, and $\muRmuF=\mjj$ corrected for electroweak and non-perturbative corrections to the unfolded data. The \ATLASPDF{21T3} does not include an $\alpha_s$ uncertainty, as it is not published.}

\begin{tabular}{c|ccccc}
\hline
& \multicolumn{5}{c}{$\chi^2/\text{ndf}$} \\
& \footnotesize{\CT{18NNLO}} & \footnotesize{\NNPDF{3.1NNLO}} & \footnotesize{\NNPDF{4.0NNLO}} & \footnotesize{\MSHT{NNLO}} & \footnotesize{\ATLASPDF{21T3} } \\
\hline
$\ystar < 0.5$          & 58/58 & 68/58 & 59/58 & 61/58 & 58/58 \\
$0.5 \leq \ystar < 1.0$ & 62/54 & 70/54 & 63/54 & 67/54 & 59/54 \\
$1.0 \leq \ystar < 1.5$ & 56/45 & 67/45 & 63/45 & 65/45 & 58/45 \\
$1.5 \leq \ystar < 2.0$ & 51/36 & 59/36 & 57/36 & 54/36 & 63/36 \\
$2.0 \leq \ystar < 2.5$ & 30/28 & 41/28 & 36/28 & 34/28 & 39/28 \\
$2.5 \leq \ystar < 3.0$ & 17/16 & 21/16 & 17/16 & 21/16 & 19/16 \\
\hline
$0.0 \leq \ystar < 3.0$ & 329/237 & 343/237 & 342/237 & 347/237 & 288/237 \\
\hline
\end{tabular}

\label{tab:results:chi2_yStar}
\end{table}

\begin{table}[htbp]
\centering
\caption{Evaluated $\chi^2/\text{ndf}$ for the dijet cross-section in the individual and global \yboost bins taking into account \NNLO pQCD predictions using \CT{18}, \NNPDF{3.1}, \NNPDF{4.0}, \MSHT, and \ATLASPDF{21T3} PDF sets, and $\muRmuF=\mjj$ corrected for electroweak and non-perturbative corrections to the unfolded data. The \ATLASPDF{21T3} does not include an $\alpha_s$ uncertainty, as it is not published.}
\begin{tabular}{c|ccccc}
\hline
& \multicolumn{5}{c}{$\chi^2/\text{ndf}$} \\
& \footnotesize{\CT{18NNLO}} & \footnotesize{\NNPDF{3.1NNLO}} & \footnotesize{\NNPDF{4.0NNLO}} & \footnotesize{\MSHT{NNLO}} & \footnotesize{\ATLASPDF{21T3} } \\
\hline
$\yboost < 0.5$          & 33/36 & 48/36 & 37/36 & 43/36 & 44/36 \\
$0.5 \leq \yboost < 1.0$ & 51/34 & 55/34 & 53/34 & 55/34 & 54/34 \\
$1.0 \leq \yboost < 1.5$ & 41/31 & 40/31 & 39/31 & 47/31 & 47/31 \\
$1.5 \leq \yboost < 2.0$ & 26/26 & 27/26 & 26/26 & 31/26 & 35/26 \\
$2.0 \leq \yboost < 2.5$ & 24/20 & 24/20 & 21/20 & 28/20 & 31/20 \\
$2.5 \leq \yboost < 3.0$ & 15/13 & 15/13 & 14/13 & 20/13 & 16/13 \\
\hline
$0.0 \leq \yboost < 3.0$ & 183/160 & 178/160 & 185/160 & 195/160 & 166/160 \\
\hline
\end{tabular}
\label{tab:results:chi2_yBoost}
\end{table}

The $\chi^2/\text{ndf}$ values remain relatively good despite the observed differences between theory and data in \Cref{fig:DataToTheory_mjj_yStar_NNLO_CT18__MJJ_SCALE,fig:DataToTheory_mjj_yBoost_NNLO_CT18__MJJ_SCALE,fig:DataToTheory_pdf1_mjj_yStar__MJJ_SCALE,fig:DataToTheory_pdf1_mjj_yBoost__MJJ_SCALE,fig:DataToTheory_pdf2_mjj_yStar__MJJ_SCALE,fig:DataToTheory_pdf2_mjj_yBoost__MJJ_SCALE}. This behaviour arises because the $\chi^2$ test is largely insensitive to overall normalisation shifts, which can be absorbed due to the bin-to-bin correlations of the systematic uncertainties.
Indeed, small coherent shifts of a few nuisance parameters are sufficient to account for the systematic differences observed between data and theory. The sum of the penalty terms induced in the $\chi^2$ by these shifts remains small compared with the number of bins. Consequently, the $\chi^2$ values are driven primarily by local bin-to-bin fluctuations rather than by global offsets.
\FloatBarrier


%

\FloatBarrier

\section{Conclusion}
Inclusive, double-differential dijet cross-sections have been measured as a function of $(\mjj, \ystar)$ and $(\mjj, \yboost)$ using particle flow \antikt jets with a radius parameter of $R=0.4$. The measurements were performed in $pp$ collisions at a centre-of-mass energy of $13~\TeV$ using the  $140$~\ifb data collected by the ATLAS experiment during the LHC Run-2 phase between 2015 and 2018. The results incorporate the most detailed and up-to-date breakdown of systematic uncertainties currently available in the ATLAS experiment.

The measured distributions are unfolded to the particle level and compared with several theoretical predictions, including state-of-the-art full colour \NNLO pQCD calculations from NNLOJET, with additional electroweak and non-perturbative corrections applied. A qualitative comparison between the measured data and the theoretical calculations is performed using both \NLO and \NNLO matrix elements and various PDF sets.

The \NNLO theoretical predictions tend to overestimate the data by about 15--20\%, with the best qualitative agreement seen for the \NNPDF{4.0NNLO} PDF set. In contrast, the quantitative results of the $\chi^2$ test, which accounts for correlations, show the best results (lowest global $\chi^2/\text{ndf}$) for \ATLASPDF{21T3} in both $(\mjj,\ystar)$ and  $(\mjj,\yboost)$. The other PDF sets give slightly higher and mutually comparable $\chi^2/\text{ndf}$ values. The second-best $\chi^2/\text{ndf}$ is obtained with \CT{18NNLO} in $(\mjj,\ystar)$ and with \NNPDF{3.1NNLO} in $(\mjj,\yboost)$. Overall, better qualitative and quantitative results are found in (\mjj, \yboost) compared with (\mjj, \ystar).

\FloatBarrier


%
%
%
%
%
\section*{Acknowledgements}
%

%

%
%

%
%

We thank CERN for the very successful operation of the LHC and its injectors, as well as the support staff at
CERN and at our institutions worldwide without whom ATLAS could not be operated efficiently.

The crucial computing support from all WLCG partners is acknowledged gratefully, in particular from CERN, the ATLAS Tier-1 facilities at TRIUMF/SFU (Canada), NDGF (Denmark, Norway, Sweden), CC-IN2P3 (France), KIT/GridKA (Germany), INFN-CNAF (Italy), NL-T1 (Netherlands), PIC (Spain), RAL (UK) and BNL (USA), the Tier-2 facilities worldwide and large non-WLCG resource providers. Major contributors of computing resources are listed in Ref.~\cite{ATL-SOFT-PUB-2025-001}.

We gratefully acknowledge the support of ANPCyT, Argentina; YerPhI, Armenia; ARC, Australia; BMWFW and FWF, Austria; ANAS, Azerbaijan; CNPq and FAPESP, Brazil; NSERC, NRC and CFI, Canada; CERN; ANID, Chile; CAS, MOST and NSFC, China; Minciencias, Colombia; MEYS CR, Czech Republic; DNRF and DNSRC, Denmark; IN2P3-CNRS and CEA-DRF/IRFU, France; SRNSFG, Georgia; BMFTR, HGF and MPG, Germany; GSRI, Greece; RGC and Hong Kong SAR, China; ICHEP and Academy of Sciences and Humanities, Israel; INFN, Italy; MEXT and JSPS, Japan; CNRST, Morocco; NWO, Netherlands; RCN, Norway; MNiSW, Poland; FCT, Portugal; MNE/IFA, Romania; MSTDI, Serbia; MSSR, Slovakia; ARIS and MVZI, Slovenia; DSI/NRF, South Africa; MICIU/AEI, Spain; SRC and Wallenberg Foundation, Sweden; SERI, SNSF and Cantons of Bern and Geneva, Switzerland; NSTC, Taipei; TENMAK, T\"urkiye; STFC/UKRI, United Kingdom; DOE and NSF, United States of America.

Individual groups and members have received support from BCKDF, CANARIE, CRC and DRAC, Canada; CERN-CZ, FORTE and PRIMUS, Czech Republic; COST, ERC, ERDF, Horizon 2020 and Marie Sk{\l}odowska-Curie Actions, European Union; Investissements d'Avenir Labex, Investissements d'Avenir Idex and ANR, France; DFG and AvH Foundation, Germany; Herakleitos, Thales and Aristeia programmes co-financed by EU-ESF and the Greek NSRF, Greece; BSF-NSF and MINERVA, Israel; NCN and NAWA, Poland; La Caixa Banking Foundation, CERCA Programme Generalitat de Catalunya and PROMETEO and GenT Programmes Generalitat Valenciana, Spain; G\"{o}ran Gustafssons Stiftelse, Sweden; The Royal Society and Leverhulme Trust, United Kingdom.

In addition, individual members wish to acknowledge support from Chile: Agencia Nacional de Investigaci\'on y Desarrollo (FONDECYT 1230812, FONDECYT 1240864, Fondecyt 3240661, Fondecyt Regular 1240721); China: Chinese Ministry of Science and Technology (MOST-2023YFA1605700, MOST-2023YFA1609300), National Natural Science Foundation of China (NSFC - 12175119, NSFC 12275265); Czech Republic: Czech Science Foundation (GACR - 24-11373S), Ministry of Education Youth and Sports (ERC-CZ-LL2327, FORTE CZ.02.01.01/00/22\_008/0004632), PRIMUS Research Programme (PRIMUS/21/SCI/017); EU: H2020 European Research Council (ERC - 101002463); European Union: European Research Council (BARD No. 101116429, ERC - 948254, ERC 101089007), European Regional Development Fund (HE COFUND GA No.101081355, ERDF), European Union, Future Artificial Intelligence Research (FAIR-NextGenerationEU PE00000013); France: Agence Nationale de la Recherche (ANR-21-CE31-0013, ANR-21-CE31-0022, ANR-22-EDIR-0002, ANR-24-CE31-0504-01); Germany: Deutsche Forschungsgemeinschaft (DFG - 469666862, DFG - CR 312/5-2); China: Research Grants Council (GRF); Italy: Ministero dell'Università e della Ricerca (NextGenEU 153D23001490006 M4C2.1.1, NextGenEU I53D23000820006 M4C2.1.1, NextGenEU I53D23001490006 M4C2.1.1, SOE2024\_0000023); Japan: Japan Society for the Promotion of Science (JSPS KAKENHI JP22H01227, JSPS KAKENHI JP22H04944, JSPS KAKENHI JP22KK0227, JSPS KAKENHI JP24K23939, JSPS KAKENHI JP24KK0251, JSPS KAKENHI JP25H00650, JSPS KAKENHI JP25H01291, JSPS KAKENHI JP25K01023); Norway: Research Council of Norway (RCN-314472); Poland: Ministry of Science and Higher Education (IDUB AGH, POB8, D4 no 9722), Polish National Science Centre (NCN 2021/42/E/ST2/00350, NCN OPUS 2023/51/B/ST2/02507, NCN UMO-2019/34/E/ST2/00393, UMO-2022/47/O/ST2/00148, UMO-2023/49/B/ST2/04085, UMO-2023/51/B/ST2/00920, UMO-2024/53/N/ST2/00869); Portugal: Foundation for Science and Technology (FCT); Spain: Generalitat Valenciana (ASFAE/2022/008), Ministry of Science and Innovation (RYC2019-028510-I, RYC2020-030254-I, RYC2021-031273-I, RYC2022-038164-I), Ministerio de Ciencia, Innovación y Universidades/Agencia Estatal de Investigaci\'on (PID2022-142604OB-C22); Sweden: Carl Trygger Foundation (Carl Trygger Foundation CTS 22:2312), Swedish Research Council (Swedish Research Council 2023-04654, VR 2021-03651, VR 2022-03845, VR 2022-04683, VR 2023-03403, VR 2024-05451), Knut and Alice Wallenberg Foundation (KAW 2018.0458, KAW 2022.0358, KAW 2023.0366); Switzerland: Swiss National Science Foundation (SNSF - PCEFP2\_194658); United Kingdom: The Binks Trust, Royal Society (NIF-R1-231091); United States of America: U.S. Department of Energy (ECA DE-AC02-76SF00515), John Templeton Foundation (John Templeton Foundation 63206), Neubauer Family Foundation.

%
%


%
%
\appendix
\printbibliography
\clearpage
\input{atlas_authlist}
%

%
%

%

%
%
%
%

\FloatBarrier
\end{document}

%% file: atlas_authlist.tex
 
\begin{flushleft}
\hypersetup{urlcolor=black}
{\Large The ATLAS Collaboration}

\bigskip

\AtlasOrcid[0000-0002-6665-4934]{G.~Aad}$^\textrm{\scriptsize 103}$,
\AtlasOrcid[0000-0001-7616-1554]{E.~Aakvaag}$^\textrm{\scriptsize 17}$,
\AtlasOrcid[0000-0002-5888-2734]{B.~Abbott}$^\textrm{\scriptsize 122}$,
\AtlasOrcid[0000-0002-0287-5869]{S.~Abdelhameed}$^\textrm{\scriptsize 118a}$,
\AtlasOrcid[0000-0002-1002-1652]{K.~Abeling}$^\textrm{\scriptsize 55}$,
\AtlasOrcid[0000-0001-5763-2760]{N.J.~Abicht}$^\textrm{\scriptsize 49}$,
\AtlasOrcid[0000-0002-8496-9294]{S.H.~Abidi}$^\textrm{\scriptsize 30}$,
\AtlasOrcid[0009-0003-6578-220X]{M.~Aboelela}$^\textrm{\scriptsize 45}$,
\AtlasOrcid[0000-0002-9987-2292]{A.~Aboulhorma}$^\textrm{\scriptsize 36e}$,
\AtlasOrcid[0000-0001-5329-6640]{H.~Abramowicz}$^\textrm{\scriptsize 156}$
% ...
\bigskip
\\

% ...
$^{17}$Department for Physics and Technology, University of Bergen, Bergen; Norway.\\
% ...
$^{30}$Physics Department, Brookhaven National Laboratory, Upton NY; United States of America.\\
% ...
$^{36}$$^{(a)}$Facult\'e des Sciences Ain Chock, Universit\'e Hassan II de Casablanca;$^{(b)}$Facult\'{e} des Sciences, Universit\'{e} Ibn-Tofail, K\'{e}nitra;$^{(c)}$Facult\'e des Sciences Semlalia, Universit\'e Cadi Ayyad, LPHEA-Marrakech;$^{(d)}$LPMR, Facult\'e des Sciences, Universit\'e Mohamed Premier, Oujda;$^{(e)}$Facult\'e des sciences, Universit\'e Mohammed V, Rabat;$^{(f)}$Institute of Applied Physics, Mohammed VI Polytechnic University, Ben Guerir; Morocco.\\
% ...
$^{45}$Physics Department, Southern Methodist University, Dallas TX; United States of America.\\
% ...
$^{49}$Fakult\"{a}t Physik, Technische Universit{\"a}t Dortmund, Dortmund; Germany.\\
% ...
$^{55}$II. Physikalisches Institut, Georg-August-Universit\"{a}t G\"ottingen, G\"ottingen; Germany.\\
% ...
$^{103}$CPPM, Aix-Marseille Universit\'e, CNRS/IN2P3, Marseille; France.\\
% ...
$^{118}$$^{(a)}$New York University Abu Dhabi, Abu Dhabi;$^{(b)}$United Arab Emirates University, Al Ain; United Arab Emirates.\\
% ...
$^{122}$Homer L. Dodge Department of Physics and Astronomy, University of Oklahoma, Norman OK; United States of America.\\
% ...
$^{156}$Raymond and Beverly Sackler School of Physics and Astronomy, Tel Aviv University, Tel Aviv; Israel.\\
% ...
$^{a}$ Also at Affiliated with an institute formerly covered by a cooperation agreement with CERN.\\
$^{b}$ Also at An-Najah National University, Nablus; Palestine.\\
$^{c}$ Also at Borough of Manhattan Community College, City University of New York, New York NY; United States of America.\\
$^{d}$ Also at Center for Interdisciplinary Research and Innovation (CIRI-AUTH), Thessaloniki; Greece.\\
$^{e}$ Also at Centre of Physics of the Universities of Minho and Porto (CF-UM-UP); Portugal.\\
$^{f}$ Also at CERN, Geneva; Switzerland.\\
% ...

\end{flushleft}
